\documentclass[twocolumn,floatfix,prb,aps,showpacs]{revtex4-1}

\usepackage{graphicx,amsmath,amssymb,color}
\usepackage{nicefrac}
\usepackage[titletoc,title]{appendix}

\usepackage{subfigure}

\usepackage{psfrag}
\usepackage{bm,color}
\usepackage{array,multirow}
\usepackage{booktabs}

\newcommand{\be}{\begin{equation}}
\newcommand{\ee}{\end{equation}}

\newcommand{\ba}{\begin{eqnarray}}
\newcommand{\ea}{\end{eqnarray}}

\renewcommand{\vec}[1]{{\textbf{\textit{#1}}}}

\begin{document}

\title{Tunnel transport and interlayer excitons in bilayer fractional quantum Hall systems}

\author{Yuhe Zhang and J. K. Jain}

\affiliation{Department of Physics, The Pennsylvania State University, University Park, Pennsylvania, 16802, USA}

\author{J. P. Eisenstein} 

\affiliation{Institute of Quantum Information and Matter,
Department of Physics, California Institute of Technology, Pasadena, California 91125, USA}
   
\date{\today}

\begin{abstract}
In a bilayer system consisting of a composite-fermion Fermi sea in each layer, the tunnel current is exponentially suppressed at zero bias, followed by a strong peak at a finite bias voltage $V_{\rm max}$. This behavior, which is qualitatively different from that observed for the electron Fermi sea, provides fundamental insight into the strongly correlated non-Fermi liquid nature of the CF Fermi sea and, in particular, offers a window into the short-distance high-energy physics of this highly non-trivial state. We identify the exciton responsible for the peak current and provide a quantitative account of the value of $V_{\rm max}$. The excitonic attraction is shown to be quantitatively significant, and its variation accounts for the increase of $V_{\rm max}$ with the application of an in-plane magnetic field. We also estimate the critical Zeeman energy where transition occurs from a fully spin polarized composite fermion Fermi sea to a partially spin polarized one, carefully incorporating corrections due to finite width and Landau level mixing, and find it to be in satisfactory agreement with the Zeeman energy where a qualitative change has been observed for the onset bias voltage [Eisenstein {\em et al.}, Phys. Rev. B {\bf 94}, 125409 (2016)]. For fractional quantum Hall states, we predict a substantial discontinuous jump in $V_{\rm max}$ when the system undergoes a transition from a fully spin polarized state to a spin singlet or a partially spin polarized state.
\end{abstract}

\maketitle

\section{Introduction}

Much attention on bilayer systems in a high magnetic field has focused on the emergence of an excitonic superfluid at total filling factor $\nu_{\rm T}=1$ \cite{Eisenstein14,Spielman00, Kellogg02, Kellogg04, Tutuc04, Wiersma04, Halperin83, MacDonald90, Wen92c}, where the electrons in one layer become strongly correlated with the holes of the other layer to produce an excitonic superfluid that exhibits remarkable phenomena. We will be concerned in this article with the situation when the distance between the two layers is sufficiently large  to preclude excitonic superfluidity, but small enough that tunnel transport is feasible. In this regime, each layer presumably consists of a Fermi sea of composite fermions \cite{Jain89,Halperin93,Jain07}. Experimental studies of the tunnel transport during the last two decades \cite{Ashoori90, Eisenstein92b, Brown94, Eisenstein95, Eisenstein09, Eisenstein16} have revealed many interesting features. (i) The tunnel current is exponentially suppressed at zero bias. (ii) The tunnel current exhibits a strong maximum at a certain bias voltage denoted $V_{\rm max}$. (iii)  $V_{\rm max}$ increases under the application of an additional parallel magnetic field. (iv) $V_{\rm max}$  does not exhibit a qualitative change when the spin polarization of the composite fermion (CF) Fermi sea\cite{Halperin93} decreases from its maximal value. (v) The onset of the tunnel transport is sensitive to the spin polarization of the CF Fermi sea, and shifts to lower bias voltages as the spin polarization of the CF Fermi sea decreases. While tunnel transport has been experimentally studied in most detail when each layer is in the compressible $1/2$ state, many of the above mentioned features are not particularly sensitive to the filling factor~\cite{Eisenstein92b, Brown94, Eisenstein09}. (vi) A double peaked structure is observed at $\nu=3/2$ and nearby filling factors in the range $4/3 \leq \nu \leq 5/3$ \cite{Eisenstein09}; in contrast, a single peak is observed at $\nu=1/2$ and vicinity, although there is evidence for a split peak at $\nu\approx 2/3$ \cite{Eisenstein92b}. 

These experimental observations, which are dramatically different from those at zero magnetic field, provide a unique experimental window into the strongly correlated nature of the CF Fermi sea and fractional quantum Hall (FQH) states. In particular, the interlayer tunneling experiments in general involve high energy excitations of the FQH state, and thus probe physics beyond what is accessible through many other measured quantities, e.g. transport gaps, which relate only to low energy excitations.  The problem of interlayer tunneling in the FQH regime has been theoretically addressed by numerical diagonalization~\cite{He93b, Haussmann96}, using a Chern-Simons theory \cite{Hatsugai93}, and also by treating the state either classically~\cite{Efros93, Levitov97} or as a Wigner crystal~\cite{Johansson93}. While these studies capture certain features of the phenomenology, the physical nature of the excitation responsible for the peak current has not been clarified, and a detailed quantitative comparison between theory and  experimental data has been lacking. We report on progress in this direction and address many, though not all, of the phenomenological observations listed in the preceding paragraph.

Tunneling of an electron from one layer to another is essentially a spectroscopic probe of the interlayer exciton, whose energy $E_{\rm ex}$ consists of three parts: the energy required to add an electron to a FQH state; the energy required to create a hole in a FQH state; and the interlayer interaction energy between the electron and hole excitations. We write $E_{\rm ex}$ as
\be
E_{\rm ex} =(E_{\rm e}-E_{\rm gs}) + (E_{\rm h}-E_{\rm gs}) + E_{\rm e-h} 
\label{Eex}
\ee
where $E_{\rm gs}$ is the ground state energy of electrons in one layer, $E_{\rm e}$ ($E_{\rm h}$) is the energy of the state with one electron added to (removed from) the ground state, and $E_{\rm e-h}$ is the attractive interaction between the tunneled electron and the hole left behind.  
Of course, an electron or a hole can be added into a continuum of excited states, producing excitons with a continuum of exciton energies. 

The tunnel transport probes high energy physics of the FQH state or the CF Fermi sea, because the low energy spectrum does not contain any object with the quantum numbers of an electron or a hole. In this sense, bilayer tunneling provides information distinct from the activation gap deduced from the temperature dependence of the resistance, which corresponds to the lowest energy charged excitations.

\begin{figure}
\resizebox{0.45\textwidth}{!}{\includegraphics{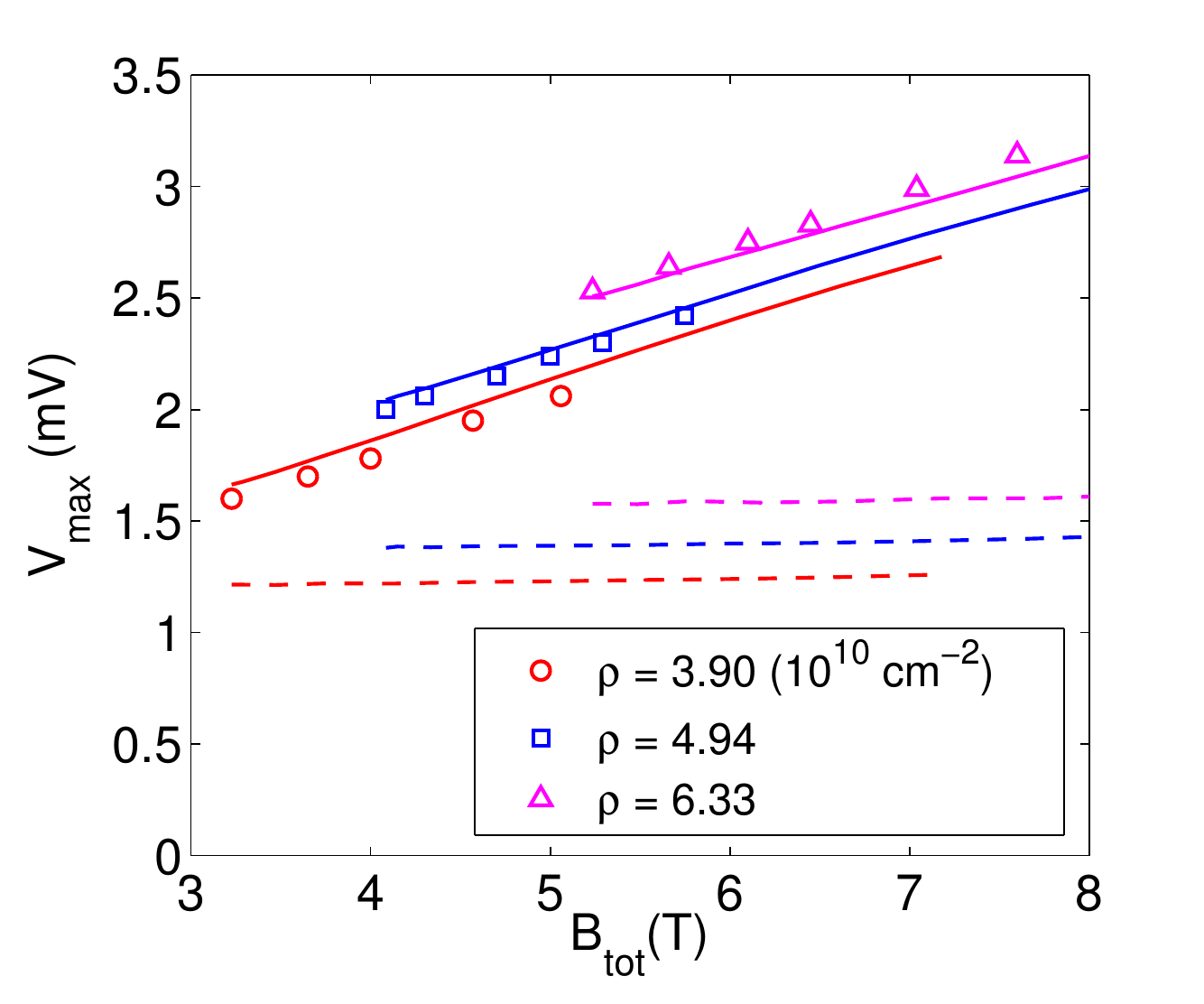}} 
\caption{The symbols show, for three densities $\rho$, the experimental bias voltage $V_{\text{max}}$ that produces the peak tunnel current, taken from Eisenstein {\em et al.}~\cite{Eisenstein16}, which studies a system of two quantum wells of width $w=18$ nm separated by $d=28$ nm (center to center). The solid lines and dashed lines depict the theoretical energies of the hard exciton and soft exciton, respectively. The total magnetic field is given by $B_{\rm tot}=\sqrt{B_{\parallel}^2+B_{\perp}^2}$, where the normal component $B_{\perp}$ is equal to the leftmost depicted value. The theory contains no adjustable parameters. Further details are given in the main text.
}
\label{expcompare0}
\end{figure}

We consider below two types of excitons. The first is that in which an electron (hole) is added to the state by application of a lowest-Landau-level (LLL) projected local creation (annihilation) operator.  We label this exciton a ``hard exciton" because its electron and hole components occupy the smallest wave packets that can be created in the LLL within the background of the correlated CF state. This is the object with the largest tunneling amplitude, and thus should correspond to the maximum current. We determine all three contributions to the exciton energy in a microscopic calculation. The attractive interaction between the exciton makes a substantial correction to the total energy, reducing it by a factor of $\sim 2$ for typical experimental parameters. An elegant way of singling out the contribution of the excitonic attraction energy $E_{\rm e-h}$, which depends in a complicated manner on both the density profiles of the electron and the hole and the interlayer separation, is through the application of a parallel magnetic field. Such a field provides a momentum boost to the tunneled electron, producing an interlayer exciton for which the electron and the hole are laterally displaced (by an amount that depends on the magnitude of the parallel magnetic field), thus reducing the magnitude of the excitonic attraction. The measurement of the interlayer exciton energy under the influence of a parallel magnetic field is equivalent to measuring its energy-wave vector dispersion.

The primary result of the comparison between theoretical calculations and the experiments of Ref.~\onlinecite{Eisenstein16} is shown in Fig.~\ref{expcompare0} (details given below).  The comparison shows that the energy of the hard exciton does indeed nicely correlate with the interlayer chemical potential difference $eV_{\rm max}$ that produces the maximum current, and also accurately captures the observed dependence on the parallel magnetic field.

It is indeed possible to add an electron and a hole into lower energy states, which take advantage of the correlations of the background FQH state. As an illustration we consider another exciton, called the ``soft" exciton, made of a soft electron and a soft hole. The soft electron is the lowest energy state that has the quantum numbers of an electron, represented as a bound complex of $(2n\pm 1)$ fractionally charged CF quasiparticles for the $n/(2n\pm 1)$ FQH state. Similarly, a soft hole is represented as a bound complex of $(2n\pm 1)$ CF quasiholes. The internal CF structure of the soft electron or soft hole is determined uniquely within the CF theory. In Fig.~\ref{expcompare0} we also show the energy of the soft exciton as a function of density and $B_{\rm tot}$. Because the soft exciton is of very large size, its energy is largely insensitive to the parallel magnetic field. The comparison with experiment in Fig.~\ref{expcompare0} shows that the soft exciton is not relevant to the peak current. We do not see any signature of the soft exciton in the experimental data, which we attribute to the smallness of the tunneling matrix element for this rather complex object. 

\begin{figure}
\resizebox{0.4\textwidth}{!}{\includegraphics{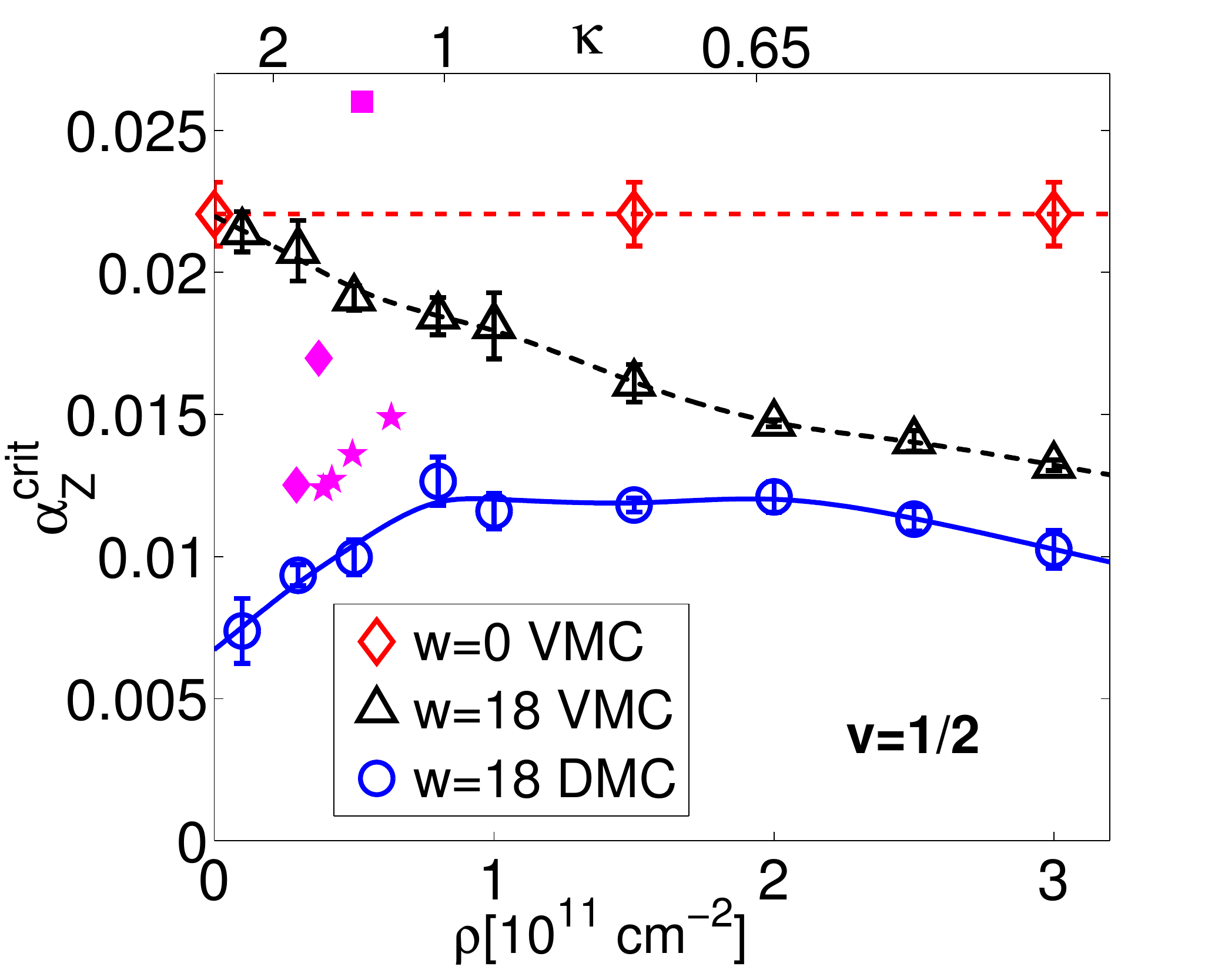}} 
\vspace{-2mm}
\caption{The critical Zeeman energy $\alpha^{\rm crit}_{\rm Z}=E^{\rm crit}_{\rm Z}/(e^2/\epsilon l)$ above which the $\nu = 1/2$ CF Fermi sea is fully spin polarized.  
The energies are quoted in units of $e^2/\epsilon l$ where $\epsilon$ is the dielectric function of the host semiconductor and $l=\sqrt{\hbar c/eB_{\perp}}$ is the magnetic length. The red dashed line is the variational Monte Carlo (VMC) result for a zero width system with no LL mixing; the dashed black line includes corrections due to finite width but not LL mixing; and the solid blue line depicts diffusion Monte Carlo (DMC) result that include corrections due to both LL mixing and finite width.  The latter two results are obtained for a quantum well of width 18nm. The experimental data are taken from Eisenstein {\em et al.}\cite{Eisenstein16} (magenta stars), Finck {\em et al.}\cite{Finck10} (magenta diamonds), and Giudici {\em et al.}\cite{Giudici08} (magenta square). All results are for a quantum well of width $w=18$nm.  The lower axis shows the electron density, whereas the LL mixing parameter $\kappa=(e^2/\epsilon l)/(\hbar \omega_c)$ is shown on top, assuming parameters appropriate for GaAs.
}
\label{spintransition0}
\end{figure}

We also revisit the issue of the spin polarization of the CF Fermi sea, specifically the determination of the critical Zeeman energy above which the CF Fermi sea is fully spin polarized. This is motivated by the recent experiment of Eisenstein {\em et al.} \cite{Eisenstein16} where they find a change in the behavior of the onset tunneling gap as a function of the parallel magnetic field, which they interpret as transition into a fully spin polarized state. An earlier calculation by Park and Jain \cite{Park98} had estimated the critical Zeeman energy for the CF Fermi sea but did not take into account corrections due to finite quantum well width and Landau level (LL) mixing. Using a fixed phase diffusion Monte Carlo method, we incorporate both of these corrections and find, as shown in Fig.~\ref{spintransition0} (details given below), that the theoretical critical Zeeman energy is reduced by roughly a factor of 2, bringing theory into better agreement with the experiment of Eisenstein {\em et al.}\cite{Eisenstein16}

Finally, we predict that the exciton energy has a substantial dependence on the spin polarization of the state. For example, as seen in Fig.~\ref{Eexw0}, our calculations show that the energy of the exciton jumps up by a factor of $\sim 2$ at $\nu=2/5$ when the system goes from a fully polarized state into a spin singlet state. This increase can be attributed to the fact that for the spin singlet state the electron and the hole are more spatially localized than for the fully spin polarized state (because the Pauli repulsion is less effective in the spin singlet state), thus enhancing $E_{\rm e}$ and $E_{\rm h}$.  The bilayer tunneling experiments may thus provide a new method for studying spin polarization phase transitions in FQH effect. 

We provide in this article a quantitative account of the observations (i) - (iv) listed in the leading paragraph of this article. We are not able to obtain a quantitative understanding of the small gap that marks the onset of transport in the I-V plot, nor of its dependence on the spin polarization of the state, although we do make speculations for the underlying physics. We also do not understand the origin of peak splitting at and near $\nu=3/2$.

The plan of the paper is as follows. In Section \ref{SecII} we define the hard and the soft excitons and discuss their relevance to the tunnel current. In Section \ref{SecIII} we show theoretically calculated values and compare with experiment. In Section \ref{SecIV} we calculate the critical Zeeman energy beyond which the CF Fermi sea becomes fully spin polarized and compare it to experiments. The paper is concluded in Section \ref{SecV}.

\section{Tunneling and interlayer excitons}
\label{SecII}

\subsection{Interlayer tunnel current}

We consider the tunneling Hamiltonian
\be
H_{\rm tunnel}\sim {\bar \Psi}^{\dagger}_L(0)  {\bar \Psi}_R(0) + h. c.
\ee
where ${\bar \Psi}_R(\vec{r})$ is the LLL-projected electron annihilation operator on the right layer and ${\bar \Psi}^{\dagger}_L(\vec{r})$ is a LLL-projected electron creation operator on the left layer. We have assumed that tunneling from a given point occurs to a point directly across, which has the highest tunneling amplitude. (In the presence of an additional in-plane magnetic field, tunneling occurs to a laterally displaced point, as discussed in more detail below.) We have also assumed that the system is translationally invariant, and therefore the tunnel amplitude does not depend on the position. The use of the LLL-projected operators is appropriate when the energies of interest are small compared to the cyclotron energy, so the higher LLs are not relevant. Following the standard many body methods \cite{Giuliani08, Mahan00}, the tunnel current at voltage $V$ is given by
\be
I(V)\sim \sum_{\alpha} |\langle \Psi_{\rm ex}^{\alpha} | {\bar \Psi}_L^{\dagger}(0) {\bar \Psi}_R(0) | \Psi_{0} \rangle |^2 \delta(E_{\rm ex}^{\alpha}-eV)
\label{IV}
\ee
where $\Psi_{0}$ is the bilayer ground state. The sum is over all interlayer exciton eigenstates $\Psi_{\rm ex}^{\alpha}$ labeled by $\alpha$, which involve a transfer of an electron from one layer to the other.  The exciton energy $E_{\rm ex}^{\alpha}$ is defined relative to the bilayer ground state energy and includes intra- as well as inter-layer interaction. We   also set the temperature to zero, which is a good approximation given that the temperatures in the relevant experiments are much smaller than the energies of interest. 

If one assumes that the interlayer interaction is negligible, then the above expression can be cast into a perhaps more familiar form:
\be
I(V)\sim \int_0^{eV} dE A_L^{>}(E) A_R^{<}(E-eV)
\ee
where the spectral functions for each individual layer are defined as $A^{>}(E)=\sum_m \langle m | \Psi^{\dagger}(0)|0\rangle |^2 \delta(E-E_m)$ and $A^{<}(E)=\sum_n \langle n | \Psi(0)|0\rangle |^2 \delta(E+E_n)$, where $|n\rangle$ are the eigenstates of the single layer system with $N-1$ particles, $|m\rangle$ are the eigenstates of the single layer system with $N+1$ particles, and $E_n$ and $E_m$ are their energies measured with respect to the ground state energy of the $N$ particle system. Eq.~\ref{IV} is more useful when the attractive energy between the electron and the hole in the two layers produced due to  tunneling is not negligible (as is seen to be the case below).

From Eq.~\ref{IV}, it is clear that at a voltage $V$, the interlayer excitonic states which have $E_{\rm ex}=eV$ and a non-zero overlap with ${\bar \Psi}_L^{\dagger}(0) {\bar \Psi}_R(0) | \Psi_{0} \rangle$ contribute to the tunnel current. For a Landau Fermi liquid, the interacting ground state is not explicitly known and the calculation of the relevant matrix elements and exciton energies proceeds through the standard perturbative treatment of the interaction. Such a perturbative treatment may be performed for the FQH effect as well within the Chern-Simons formulation, but that formulation is valid only for low-energy long-wave length physics whereas, as seen below, the interlayer tunneling probes short-distance high-energy behavior. Fortunately, the explicit knowledge of accurate wave functions for the ground states of various incompressible states and the CF Fermi sea allows us to make progress. While an evaluation of the full line shape of the I-V curve is a complicated task within our approach, we are able to identify the exciton responsible for the peak current and give a quantitative account of the phenomenology associated with it. 

Below we consider two specific (interlayer) excitons.  The so-called ``hard" exciton, defined below, is identified with the peak current. The ``soft" exciton represents a low energy exciton in which the electron and the hole are represented as complexes of excited composite fermion particles or holes. 

In what follows below, we shall assume the system is in a regime where the ground state does not involve interlayer correlations. In other words, we assume that the FQH / CF Fermi sea state in each layer is unaffected by the presence of the other layer. (See Refs.~\onlinecite{Halperin83,Scarola01b} for bilayer FQH states that involve interlayer correlations.) We will also concentrate on incompressible FQH states, because these are easier to deal with theoretically than the 1/2 CF Fermi sea, and approach the CF Fermi sea along the sequence $\nu=n/(2n+1)$. Our analysis of the CF Fermi sea is aided by our finding below that $E_{\rm ex}$ is not particularly sensitive to the filling factor. 

The evaluation of Eq.~\ref{IV} by the standard perturbative methods of many particle theory is not feasible, as the physics of the FQH state is non-perturbative. Fortunately, we have an excellent quantitative understanding of the various FQH states as well as the 1/2 state through the CF theory, which will allow us to perform detailed microscopic calculations. 

\subsection{Hard exciton}

\begin{figure*}
\resizebox{0.26\textwidth}{!}{\includegraphics{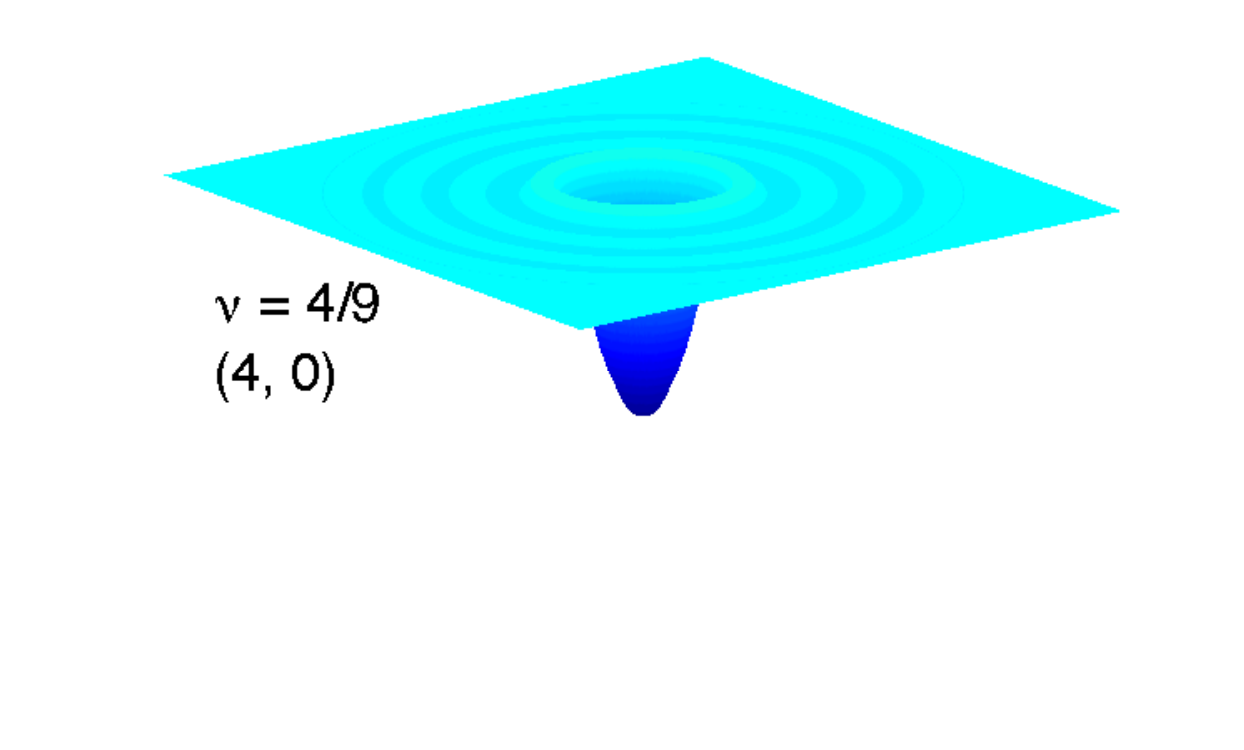}}
\hspace{-5mm}
\resizebox{0.26\textwidth}{!}{\includegraphics{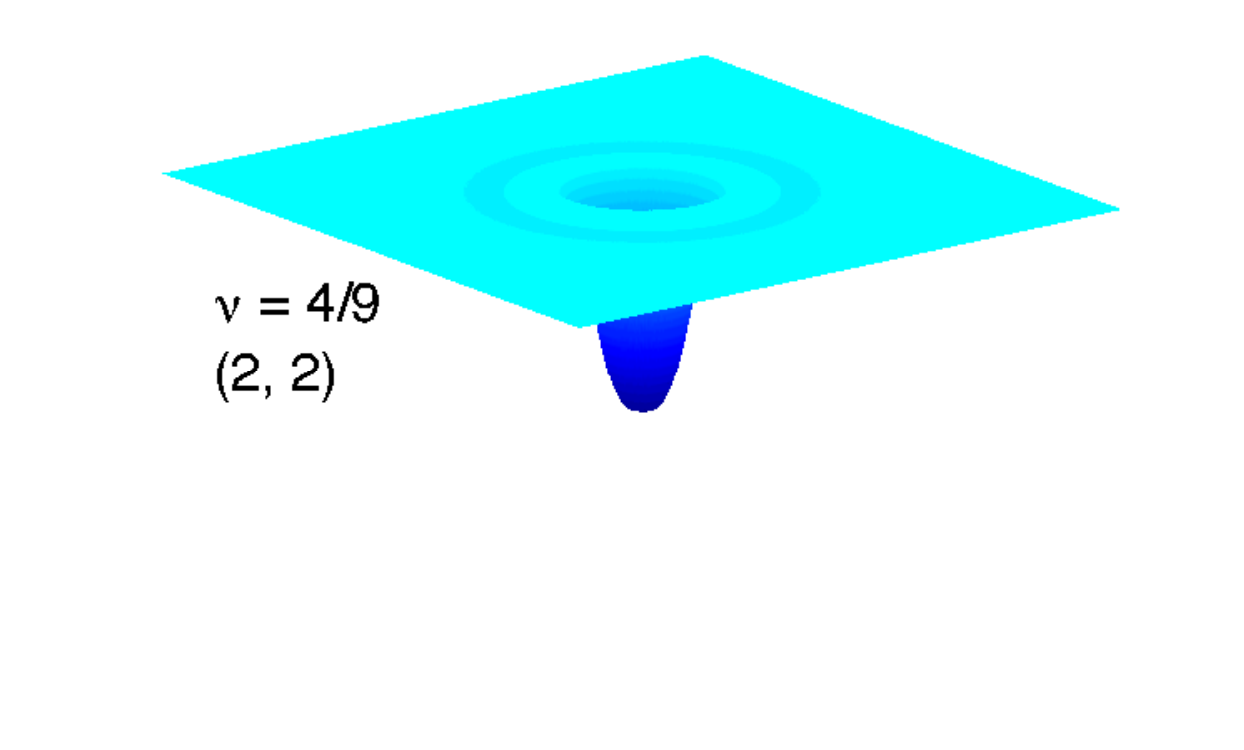}}
\hspace{-5mm}
\resizebox{0.26\textwidth}{!}{\includegraphics{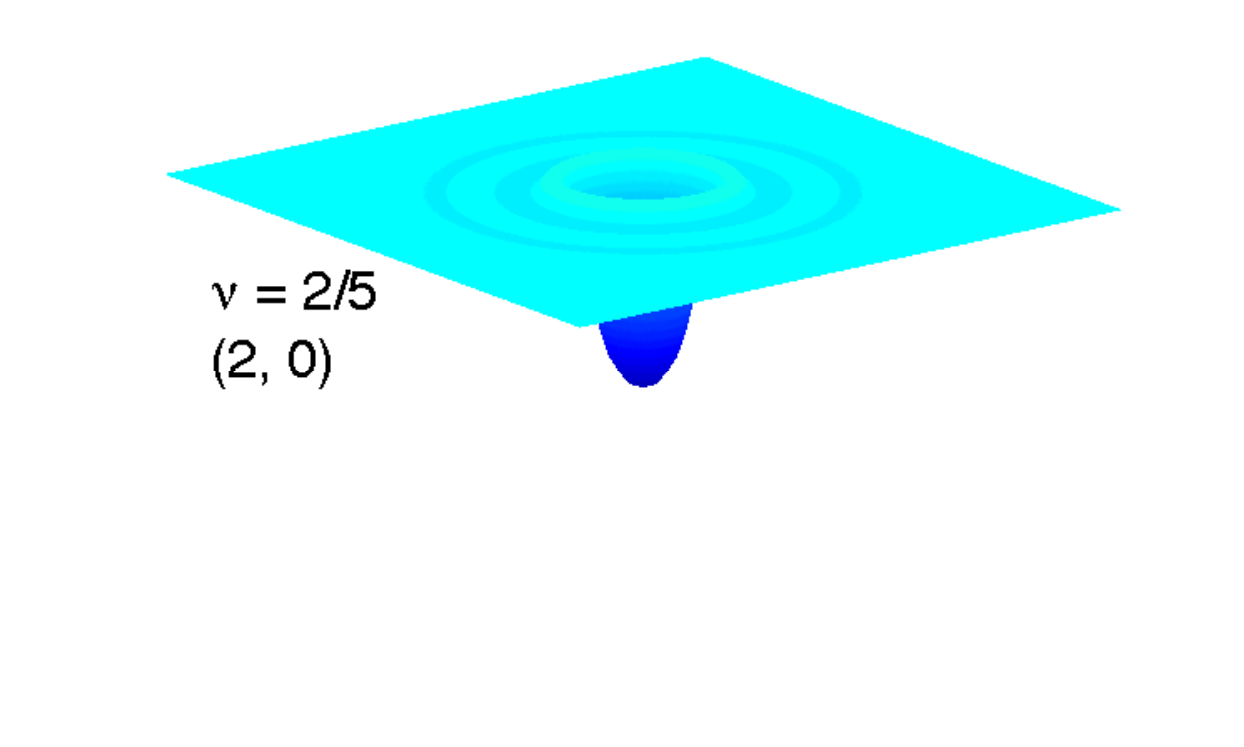}}
\hspace{-5mm}
\resizebox{0.26\textwidth}{!}{\includegraphics{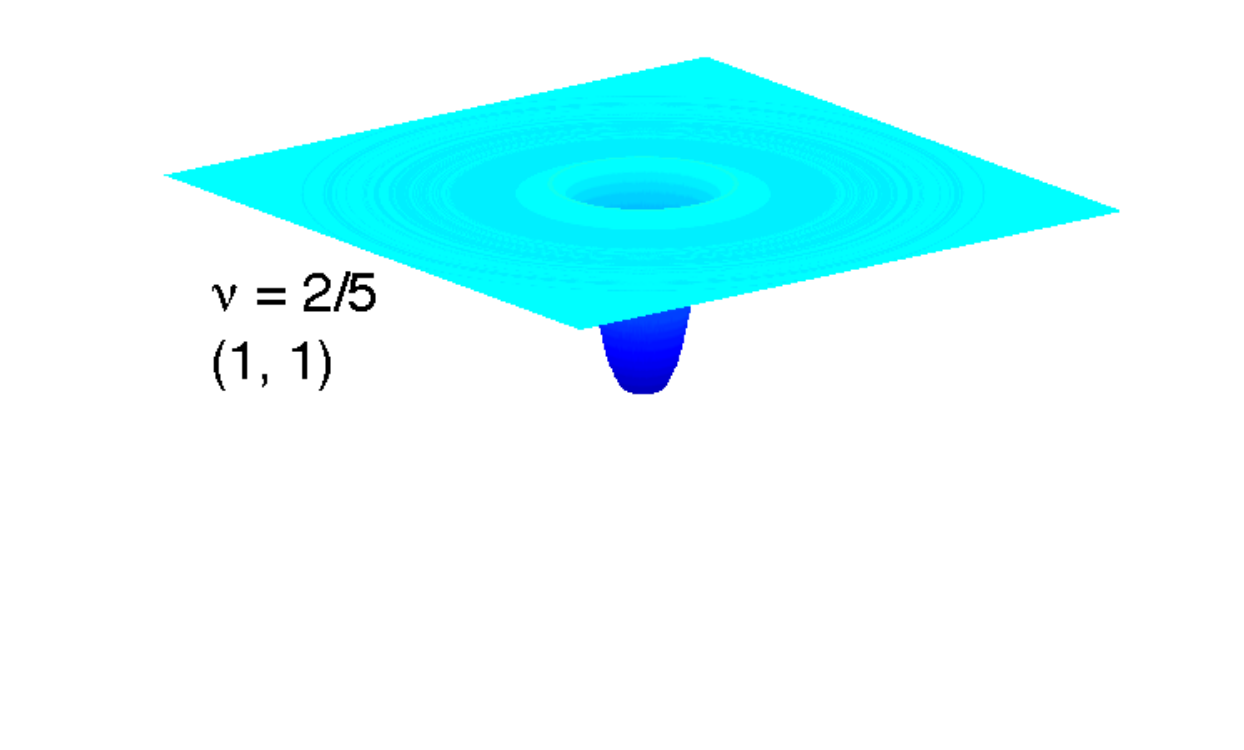}}\\
\vspace{-10mm}
\resizebox{0.26\textwidth}{!}{\includegraphics{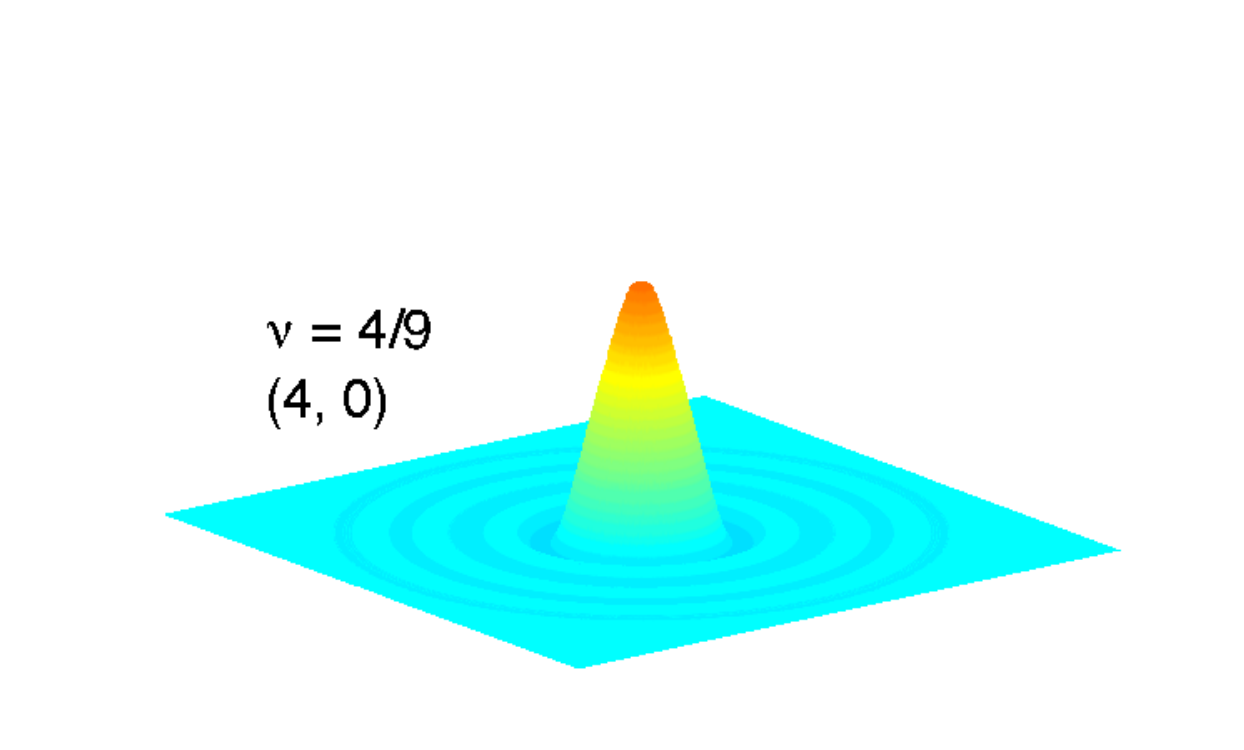}}
\hspace{-5mm}
\resizebox{0.26\textwidth}{!}{\includegraphics{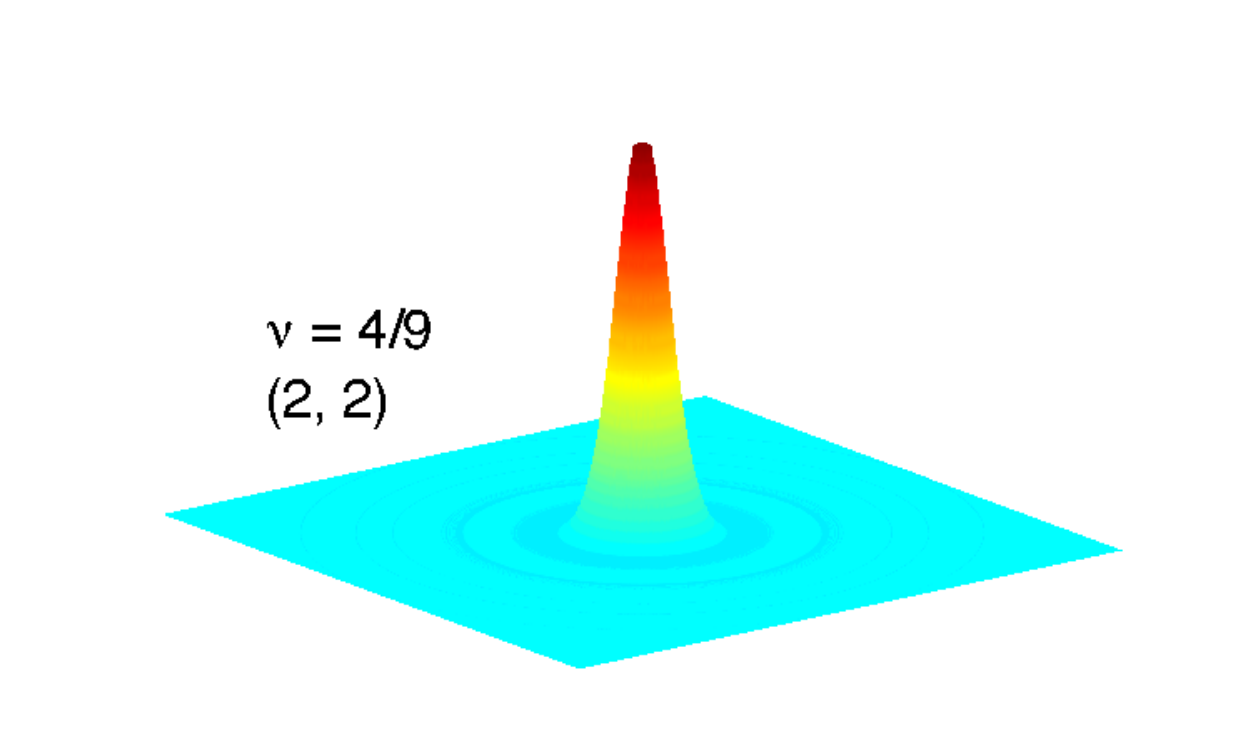}}
\hspace{-5mm}
\resizebox{0.26\textwidth}{!}{\includegraphics{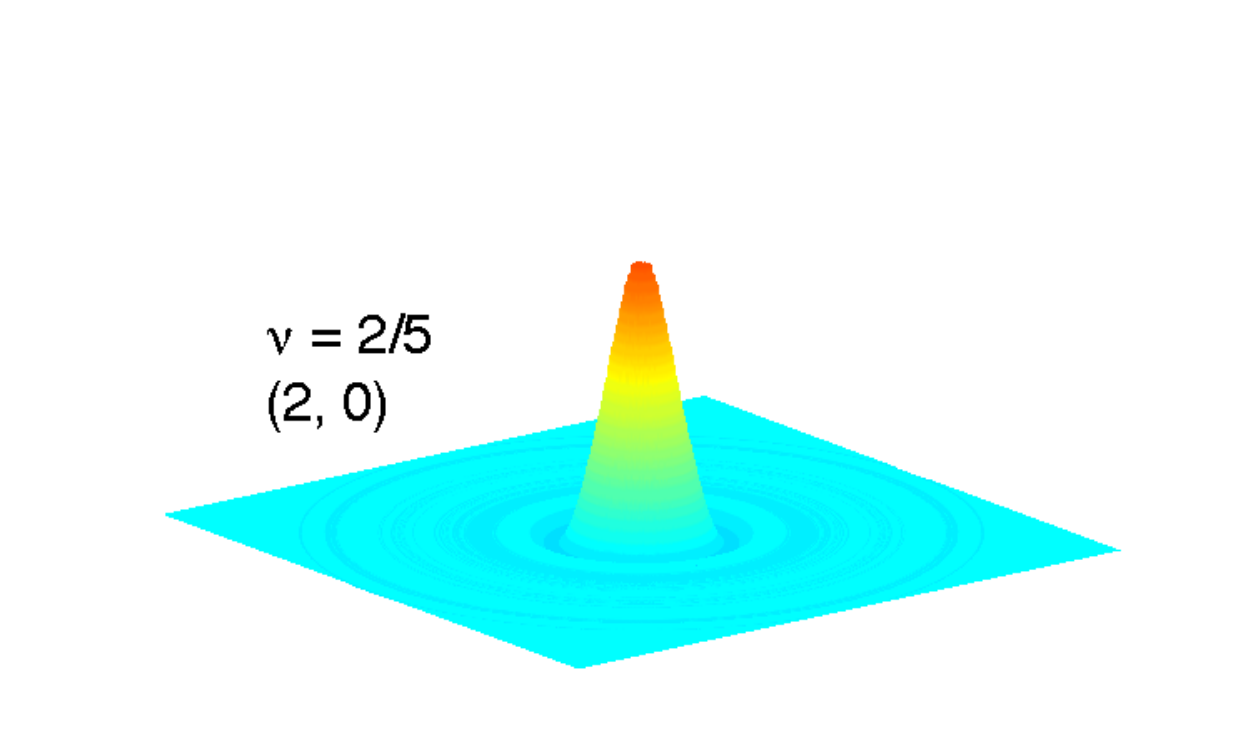}}
\hspace{-5mm}
\resizebox{0.26\textwidth}{!}{\includegraphics{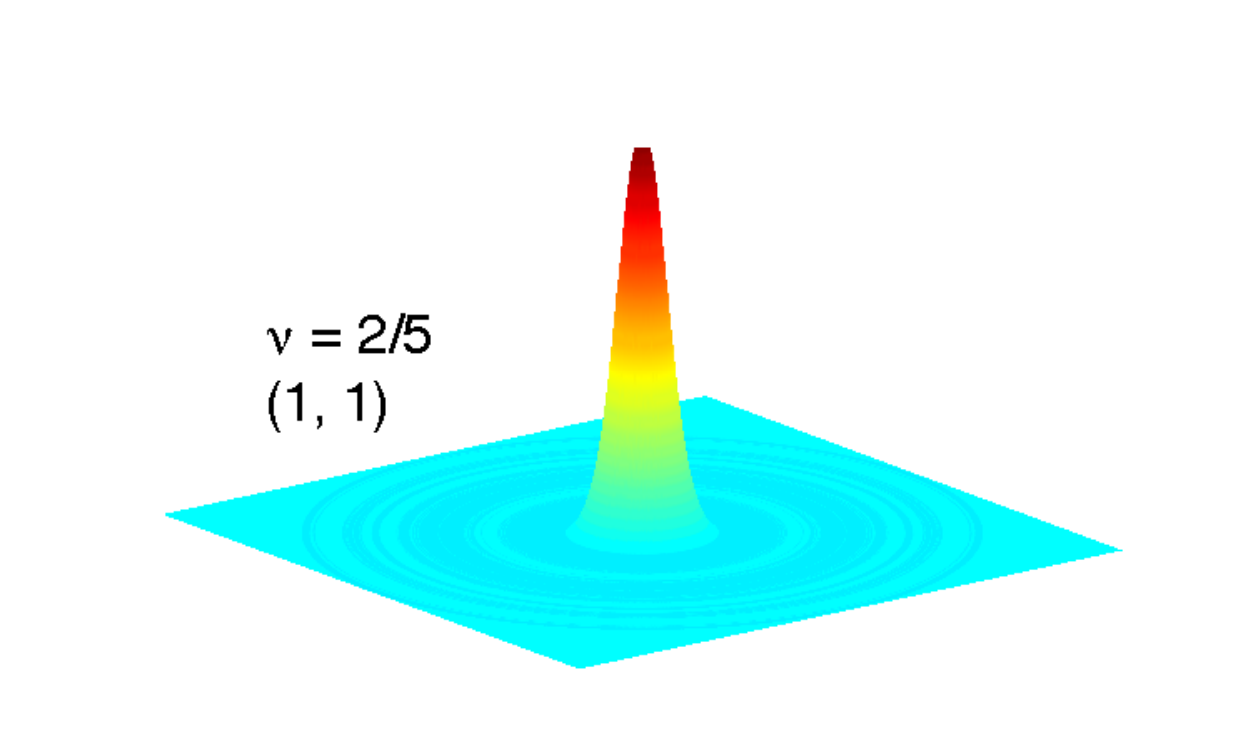}}\\
\resizebox{0.25\textwidth}{!}{\includegraphics{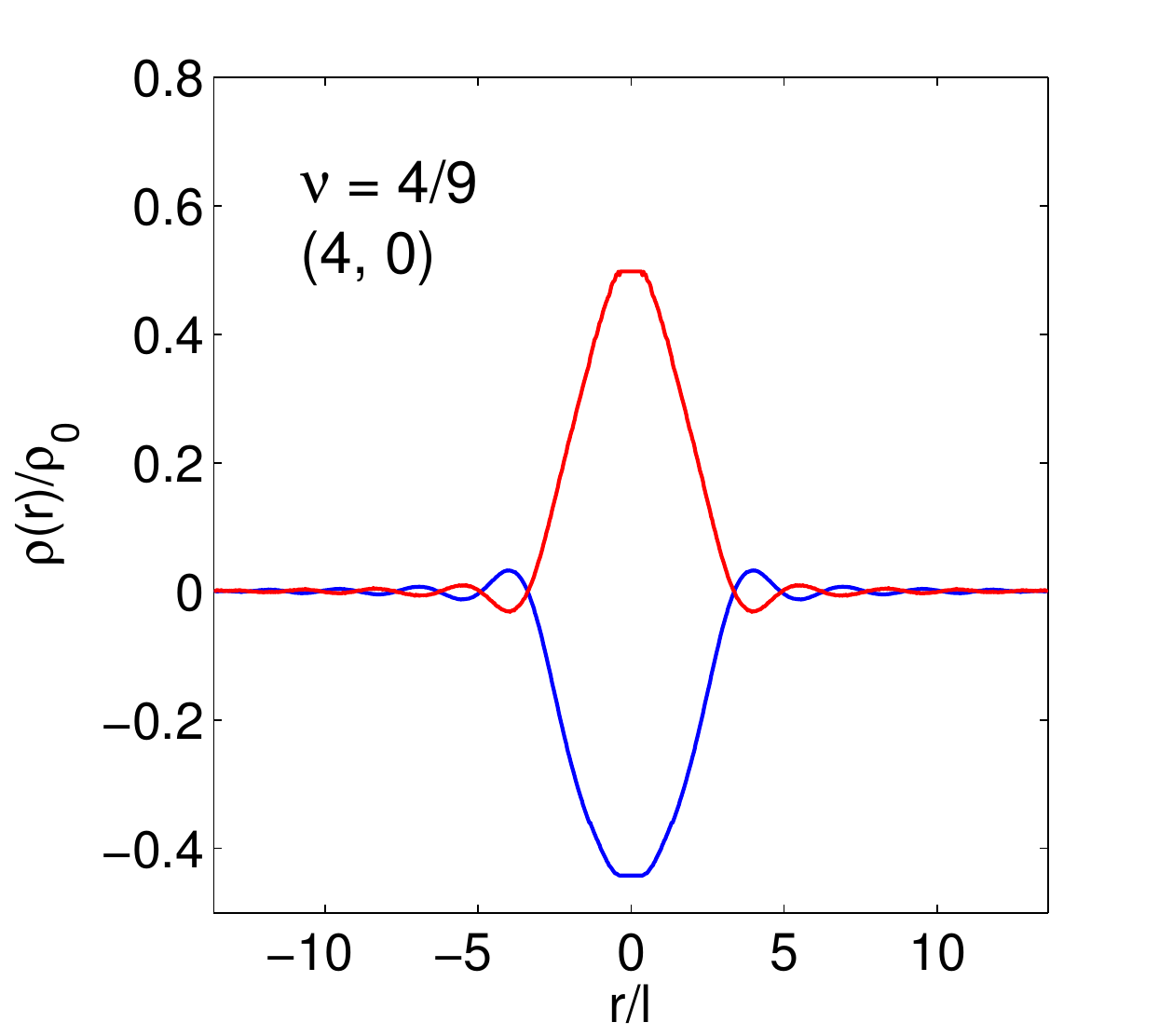}}
\hspace{-3mm}
\resizebox{0.25\textwidth}{!}{\includegraphics{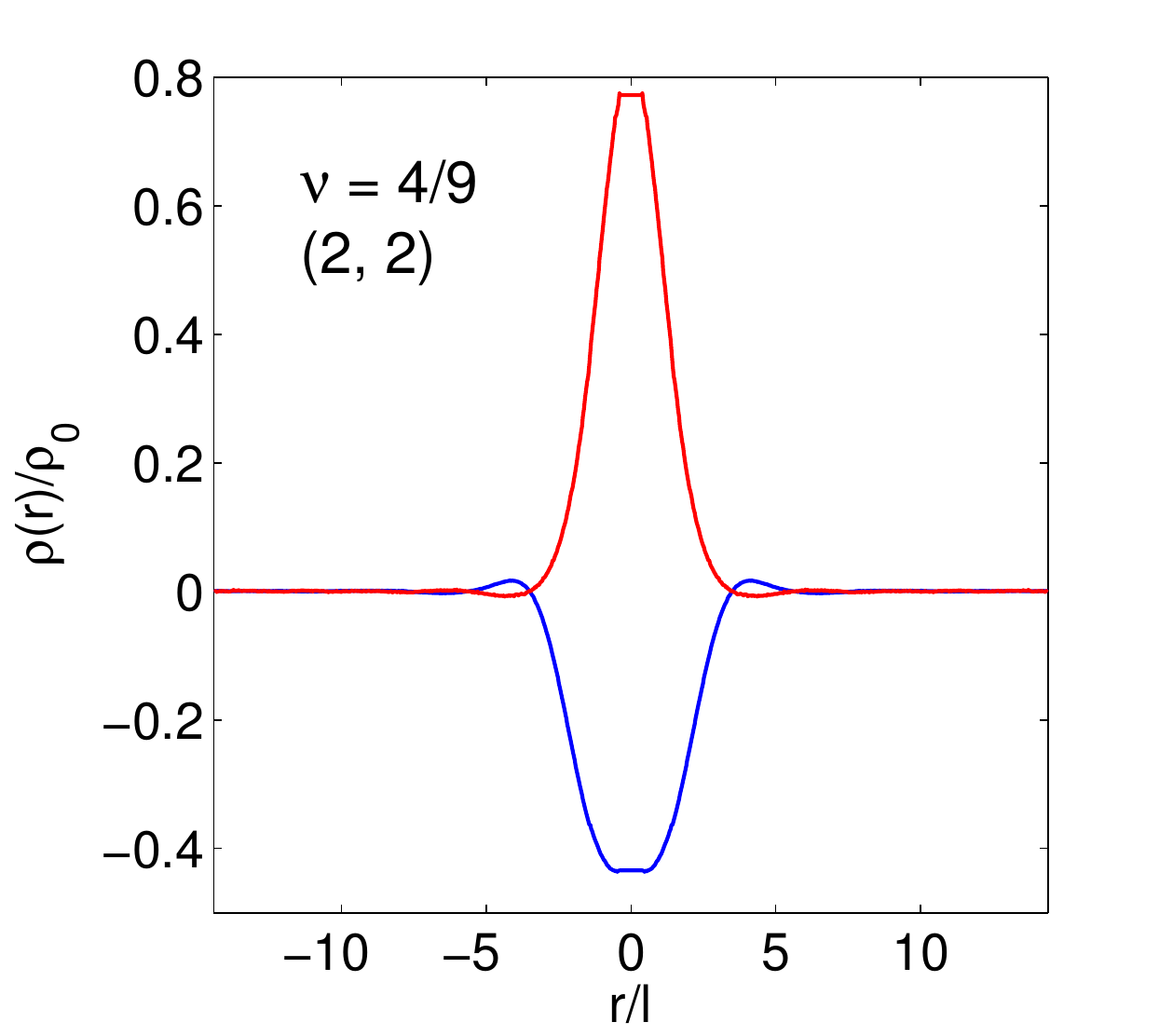}}
\hspace{-3mm}
\resizebox{0.25\textwidth}{!}{\includegraphics{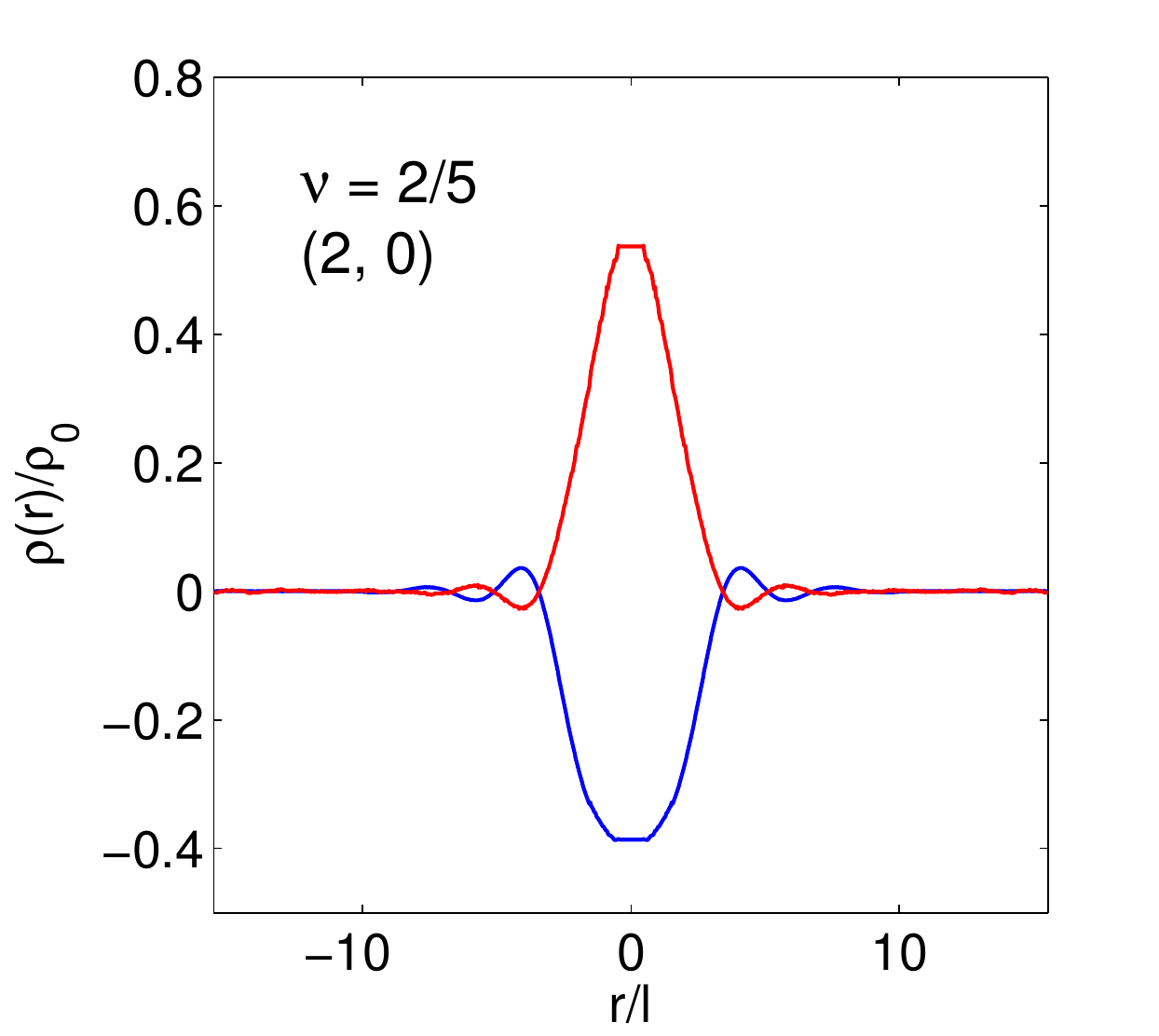}}
\hspace{-3mm}
\resizebox{0.25\textwidth}{!}{\includegraphics{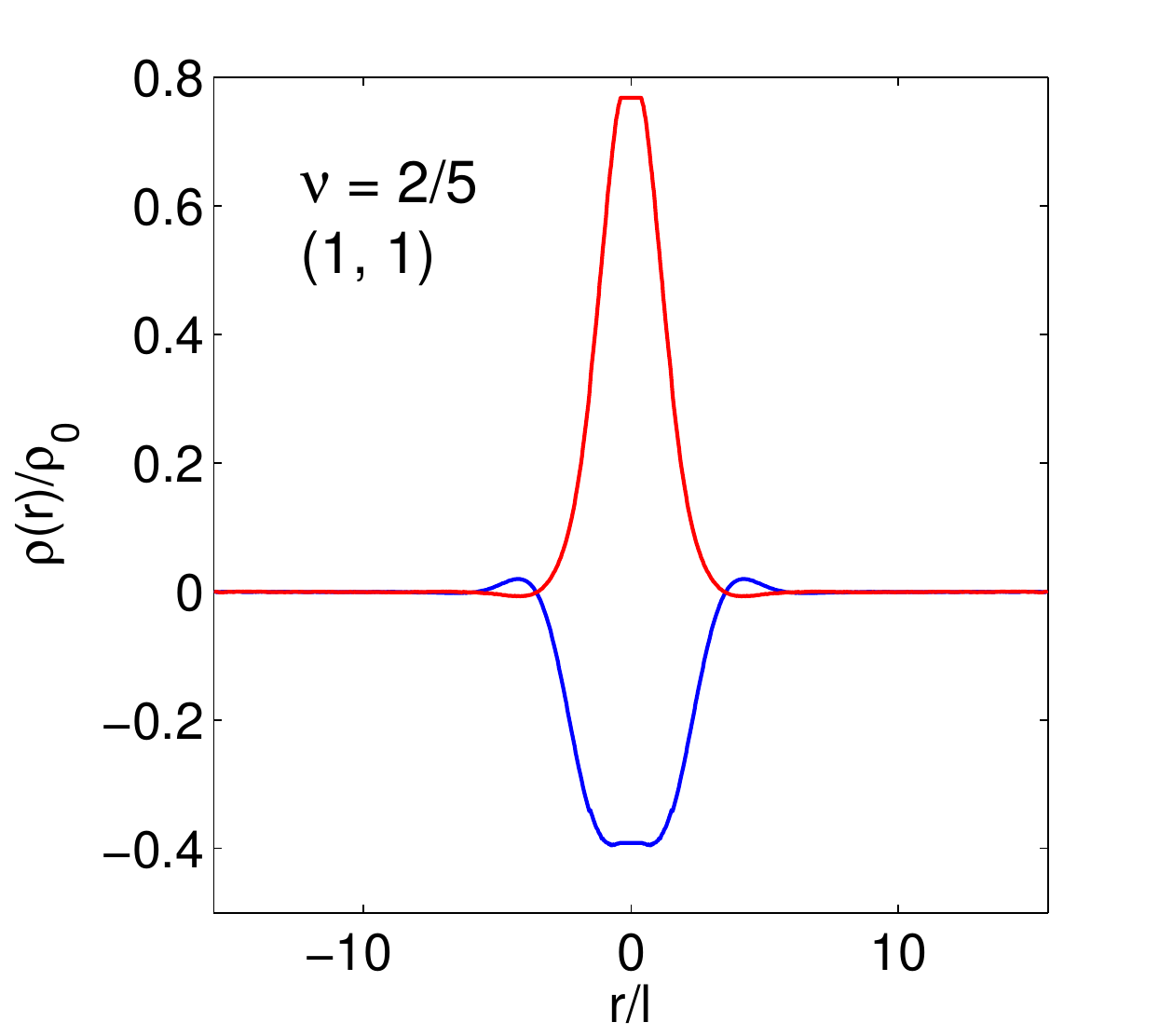}}
\vspace{-5mm}
\caption{Lower panels show the density profiles in unit of $\rho_{0} = 1/(2\pi\ell^{2})$ as a function of the cord distance $r/l$ for hard electrons (red lines) and hard holes (blue lines). The results are shown for filling factors $\nu=2/5$ and $4/9$, where we consider both the fully spin polarized states labeled $(2,0)$ and $(4,0)$, and the spin singlet states labeled $(1,1)$ and $(2,2)$.
The system sizes are $N=44-50$. The corresponding 3D density profiles (assuming planar geometry) of the hard hole and hard electron are shown in the top and the middle panels. 
}
\label{harddensity}
\end{figure*}

We define the interlayer exciton ${\bar \Psi}_L^{\dagger}(0) {\bar \Psi}_R(0) | \Psi_{0} \rangle$ as the ``hard" exciton. It consists of two parts. The operator ${\bar \Psi}_L^{\dagger}(0)$ creates in the left layer an electron which is uncorrelated with the background state except for Pauli exclusion.  This is the smallest size object in the background of the given ground state that has the quantum numbers of an electron. Hence the adjective ``hard."  In the disk geometry, the wave function of the hard electron at the origin is given by 
\be
{\bar \Psi}^{\dagger}(\vec{r}=\vec{0}) | \Psi_{0} \rangle \sim c_0^{\dagger} | \Psi_{0} \rangle
\ee
where $c_0^{\dagger}$ creates an electron in the state with angular momentum zero. A hard hole is similarly given by $c_0 | \Psi_{0} \rangle$. 

If the hard exciton ${\bar \Psi}_L^{\dagger}(0) {\bar \Psi}_R(0) | \Psi_{0} \rangle$ were an eigenstate, then the tunnel coupling in Eq.~\ref{IV} would be the largest for the hard exciton. In general, one may expect largest matrix elements for tunneling into eigenstates with energy close to the hard exciton.  We therefore find it natural to identify the energy of the hard exciton with the voltage $V_{\rm max}$ at the peak current. This identification is supported below by a detailed, quantitative comparison between theory and experiment. 

We shall use for our calculations the spherical geometry~\cite{Haldane83} where electrons are confined to move on the surface of a sphere with radius $R$. A magnetic monopole of strength $Q$ is located at the center, producing a total flux of $2Q\phi_{0}$ and a radial magnetic field $B=2Q\phi_{0}/4\pi R^{2}$. The Hamiltonian is 
\begin{equation}
H=\frac{1}{2m_{\rm b}}\sum_{i}{[-i\hbar\nabla_{i}+e{\bf A}({\bf r}_{i})]^{2} + V(\mathcal{R})},
\label{Hamiltonian}
\end{equation}
where the vector potential is ${\bf A}=-\frac{\hbar c Q}{eR_{0}} \cot \theta \bm{\hat{\phi}}$ in the Haldane gauge. The single-particle eigenstates of this Hamiltonian are described by the monopole harmonics $Y_{Q,l,m}$ where $l = |Q|, |Q|+1, ...$ is the orbital angular momentum and $m = -l, -l+1, ..., l$ is the $z$ component of the orbital angular momentum. Different angular momentum shells are the LLs. Ignoring spin, the degeneracy of each LL is equal to $(2l+1)$, increasing by 2 for each successive shell.

The electron creation operator in the spherical geometry is given by 
\be
{\bar \Psi}^{\dagger} (\Omega) = \sum_{m} Y^{*}_{QQm}(\Omega) c^{\dagger}_{QQm}
\ee
where $\Omega$ is the position of the added electron and $Y_{QQm}$ is the LLL single-particle wave function
\be
Y_{QQm}(\Omega) = \left[ N_{Q} (2Q, Q-m) \right]^{1/2} v^{Q-m} u^{Q+m}
\ee
with spinor coordinates $u = \cos\frac{\theta}{2} e^{i\phi/2}$, $v = \sin\frac{\theta}{2} e^{-i\phi/2}$, and $N_{Q} = (2Q+1)/4\pi$. For simplicity, we add an electron at the north pole of the sphere, which is denoted as $\Omega = {\bar \Omega}$ ($u=1$, $v=0$). Now creation operator simplified to $\sqrt{N_{Q}} c^{\dagger}_{QQQ}$. Application of $c^{\dagger}_{QQQ}$ to a spinless ground state $| \Psi_{0}\rangle$ leads to the (un-normalized) wave function for the hard electron
\be
\Psi_{\rm e}^{\text{hard}}(\Omega_{1}, ..., \Omega_{N+1}) = A[Y_{QQQ}(\Omega_{N+1}) \Psi_{0}(\Omega_{1}, ..., \Omega_{N})],
\label{hard-e}
\ee
where A denotes antisymmetrization over all the coordinates. For a spinful state, the above antisymmetrization should operate only on the coordinates with the same spin as the added electron. Since we start with the ground state with $L=0$ and add a electron with $l = m = Q$, this hard electron state has a total angular momentum $L = |M| = Q$ where $M$ is the $z$ component of the orbital angular momentum.

A hard hole at the north pole is created similarly by application of the electron annihilation operator ${\bar \Psi} ({\bar \Omega}) = \sqrt{N_{Q}} c_{QQQ}$. The wave function is obtained by replacing one of the coordinates with the north pole coordinate ${\bar \Omega}$
\be
\Psi_{h}^{\text{hard}}(\Omega_{1}, ..., \Omega_{N-1}) = \Psi_{0}(\Omega_{1}, ..., \Omega_{N-1},\bar{ \Omega}).
\label{hard-h}
\ee
Note that for a spinful state, the coordinate being replaced should have the same type of spin as the hard electron as we assume spin is conserved during tunneling. The hard hole state also has $L = |M| = Q$. Fig. \ref{harddensity} shows the density profiles of hard electrons and hard holes for different spinful states at $\nu = 4/9$ and $2/5$.

To calculate the energy of the hard exciton, we need the ground state wave function $\Psi_{0}$. We will calculate various quantities within the CF theory\cite{Jain89,Jain07}, which maps the interacting electrons at filling $\nu$ to non-interacting CFs (bound states of one electron and $2p$ vortices) at filling $\nu^{*}$, where $\nu$ and $\nu^*$ are related by $\nu=\nu^{*}/(2p\nu^{*} \pm1)$. FQH effect at $\nu=n/(2pn\pm1)$ is explained as the integer quantum Hall effect of CFs at $\nu^{*}=n$. We will in general consider spinful electrons, and take $n=n_{\uparrow}+n_{\downarrow}$, where $n_{\uparrow}$ is the number of filled spin-up $\Lambda$ levels (CF LLs), and $n_{\downarrow}$ is the number of filled spin-down $\Lambda$ levels. The wave function for the FQH ground state (suppressing the spin part) is given by\cite{Jain89,Wu93,Jain07}
\begin{equation}
\begin{aligned}
\Psi_{n/(2pn\pm 1)}  = \mathcal{P}_{\text{LLL}} \Phi_{\pm n_{\uparrow}} \Phi_{\pm n_{\downarrow}} \Phi_1^{2p} .
\label{CFWF}
\end{aligned}
\end{equation}
Here $\Phi_n$ is the wave function for $n$ filled Landau levels of independent fermions, $\Phi_{-n}\equiv [\Phi_n]^*$, and $\mathcal{P}_{\text{LLL}}$ denotes LLL projection. We label the spinful states as $(n_{\uparrow},n_{\downarrow})$.   In the spherical geometry, a system with $N$ particles at monopole strength $Q$ reduces to composite fermions at a reduced effective monopole strength $Q^{*} = Q - p(N-1)$; the wave functions $\Phi_{\pm n_{\uparrow}}$ and $\Phi_{\pm n_{\downarrow}}$ at the right-hand side of Eq. (\ref{CFWF}) correspond to $Q^*$. From the standard CF theory, a relation between $Q^{*}$, $n_{\uparrow}$, $n_{\downarrow}$ and the particle numbers of each spin ($N_{\uparrow}$ and $N_{\downarrow}$) is derived as
\be
Q^{*} = (N_{\uparrow} - n_{\uparrow}^{2})/2n_{\uparrow} = (N_{\downarrow} - n_{\downarrow}^{2})/2n_{\downarrow}.
\ee
With this, we can write down the wave functions $\Phi_{\pm n_{\uparrow}}(\Omega_{1}, \Omega_{2}, ..., \Omega_{N_{\uparrow}})$ and $\Phi_{\pm n_{\downarrow}} (\Omega_{N_{\uparrow} +1}, ..., \Omega_{N}) $ at $Q^{*}$, and perform the LLL projection in spherical geometry~\cite{Jain97,Jain97b,Jain07} for Eq. (\ref{CFWF}). We can then evaluate the Coulomb energy of a ground state from Eq. (\ref{CFWF}) and the energies of the hard exciton from Eqs. (\ref{hard-e}) (\ref{hard-h}) using Monte Carlo method~\cite{Foulkes01}.

\subsection{Soft exciton}

\begin{figure}
\resizebox{0.4\textwidth}{!}{\includegraphics{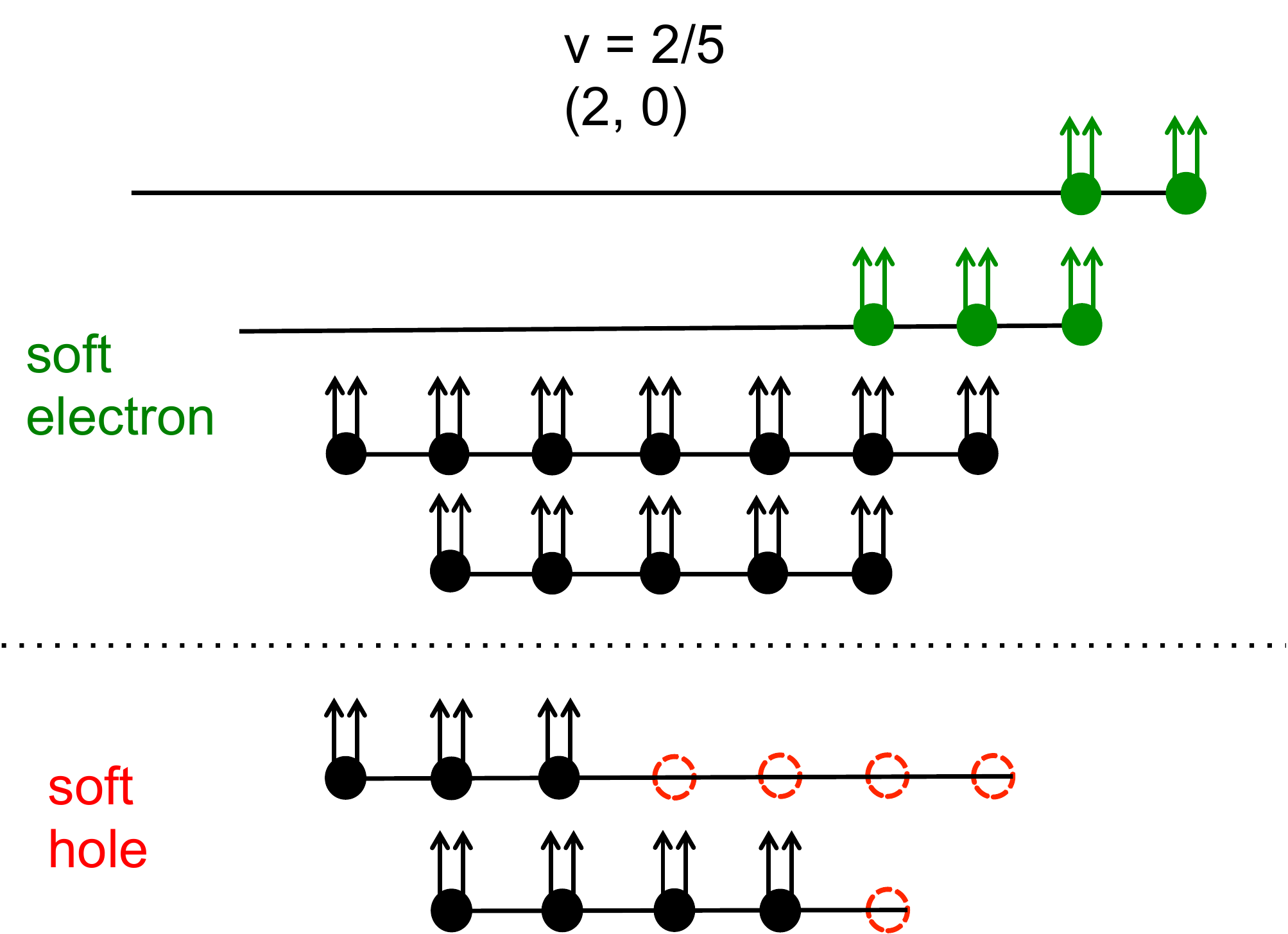}}\\
\caption{Schematic picture for the soft electron and the soft hole in terms of CF occupation for the fully spin polarized state at $\nu=2/5$. The composite fermions colored with green make a soft electron, whereas the CF holes colored in red make a soft hole.  A total of 5 excited CF particles or holes are needed to give a total charge of magnitude 1. These are the lowest energy excitations with the quantum numbers of an electron and a hole. We have depicted a finite system for illustration, but the structure remains the same for larger systems.
}
\label{CFoccupation}
\end{figure}

\begin{figure}
\resizebox{0.49\textwidth}{!}{\includegraphics{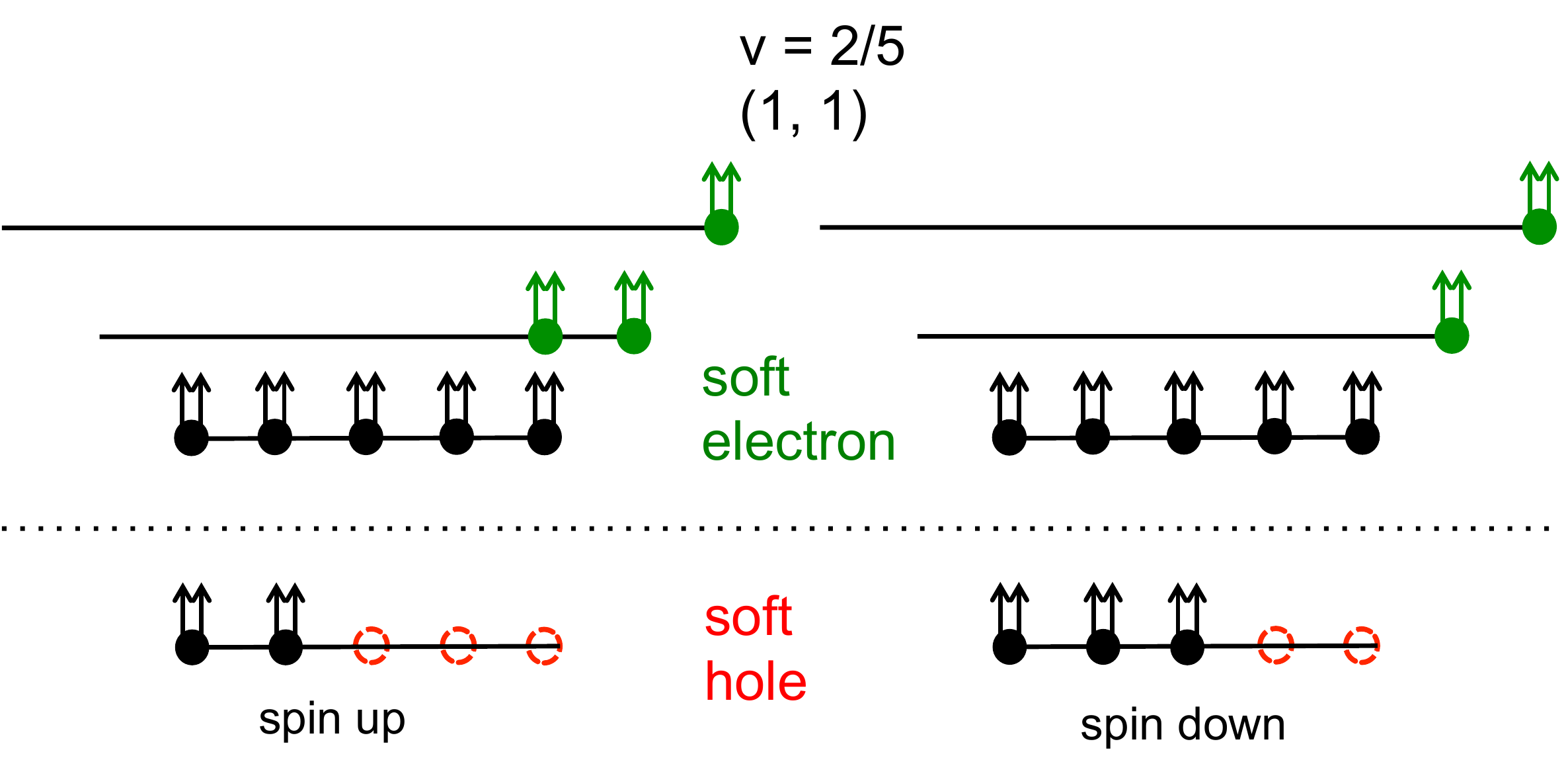}}\\
\caption{Schematic picture for the soft electron and the soft hole of the spin singlet state at $\nu=2/5$. The composite fermions colored with green make a soft electron, whereas the CF holes colored in red make a soft hole. As in Fig.~\ref{CFoccupation} we have a total of 5 excited CF particles or holes, but 3 of them with spin up and 2 with spin down, to produce a net spin 1/2. 
}
\label{CFoccupation2}
\end{figure}

\begin{figure*}
\resizebox{0.26\textwidth}{!}{\includegraphics{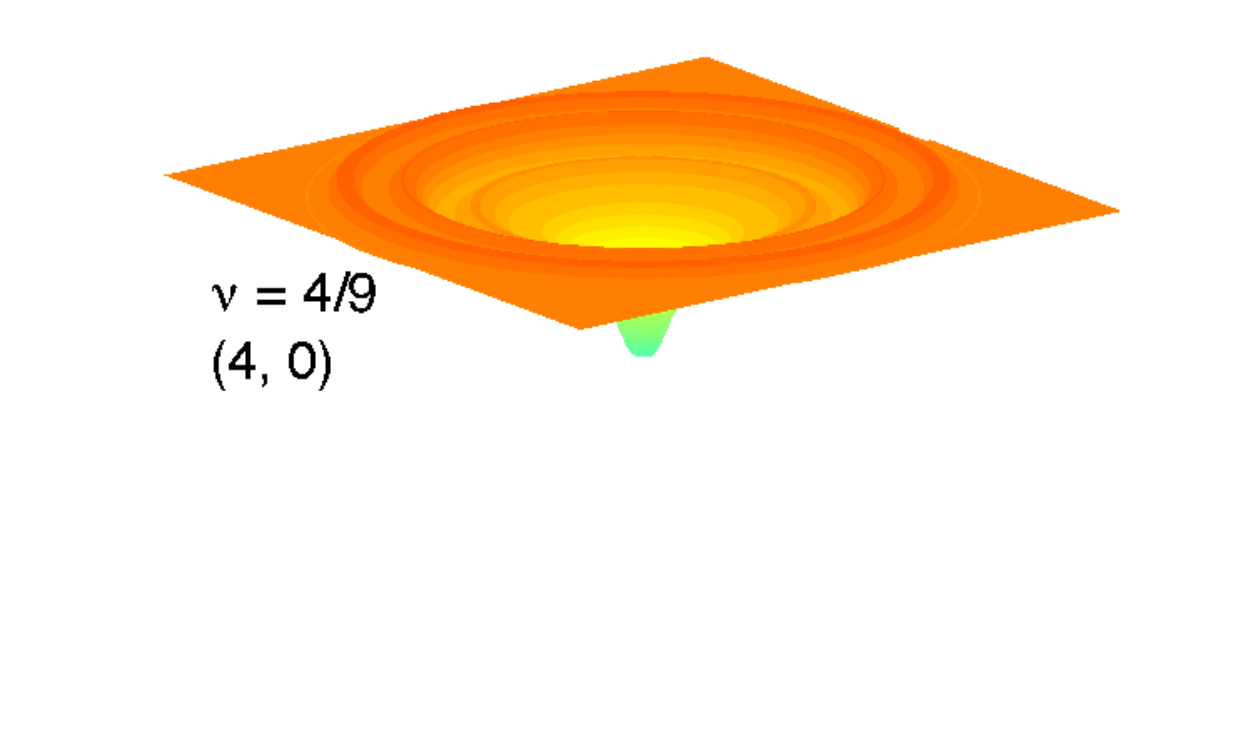}}
\hspace{-5mm}
\resizebox{0.26\textwidth}{!}{\includegraphics{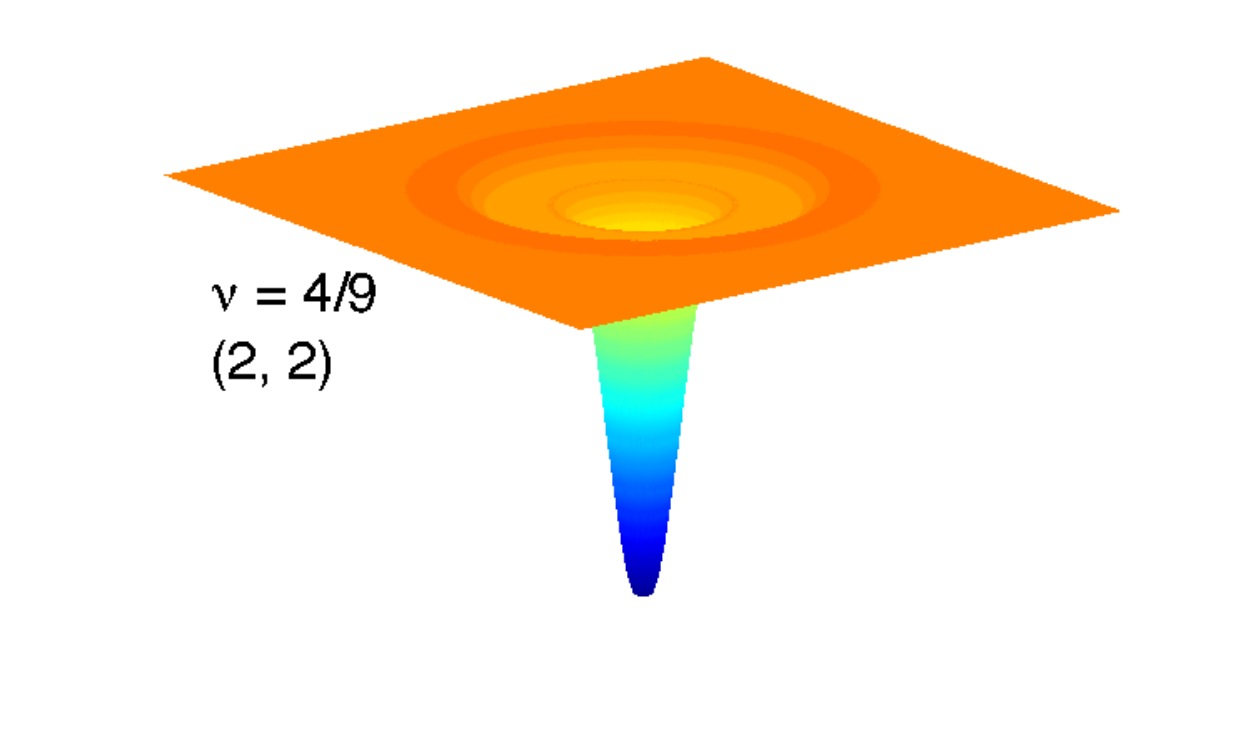}}
\hspace{-5mm}
\resizebox{0.26\textwidth}{!}{\includegraphics{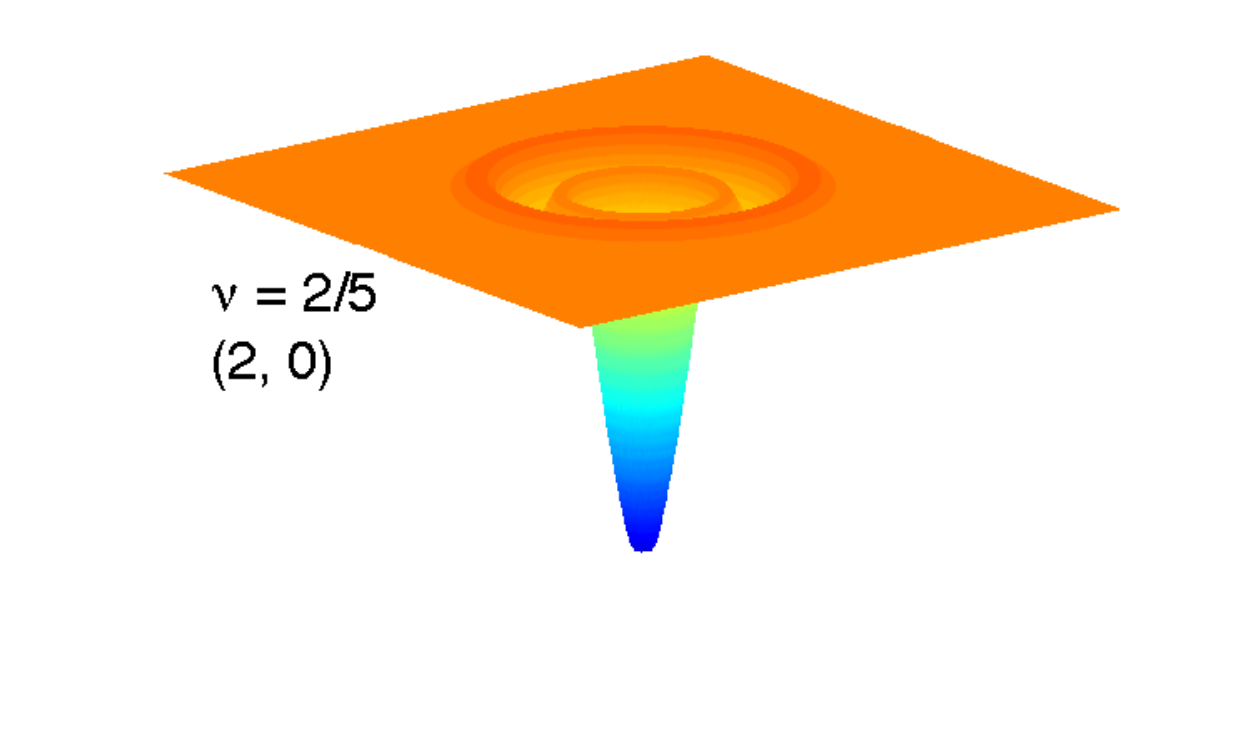}}
\hspace{-5mm}
\resizebox{0.26\textwidth}{!}{\includegraphics{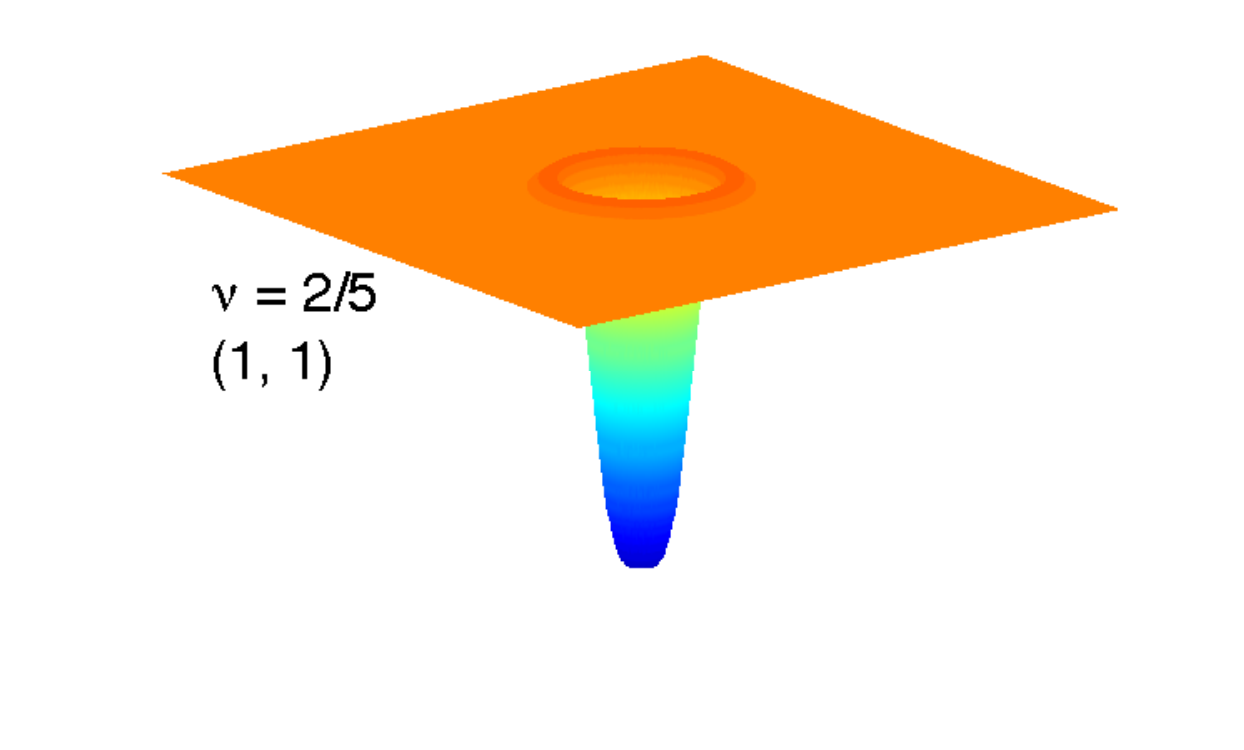}}\\
\vspace{-6mm}
\resizebox{0.26\textwidth}{!}{\includegraphics{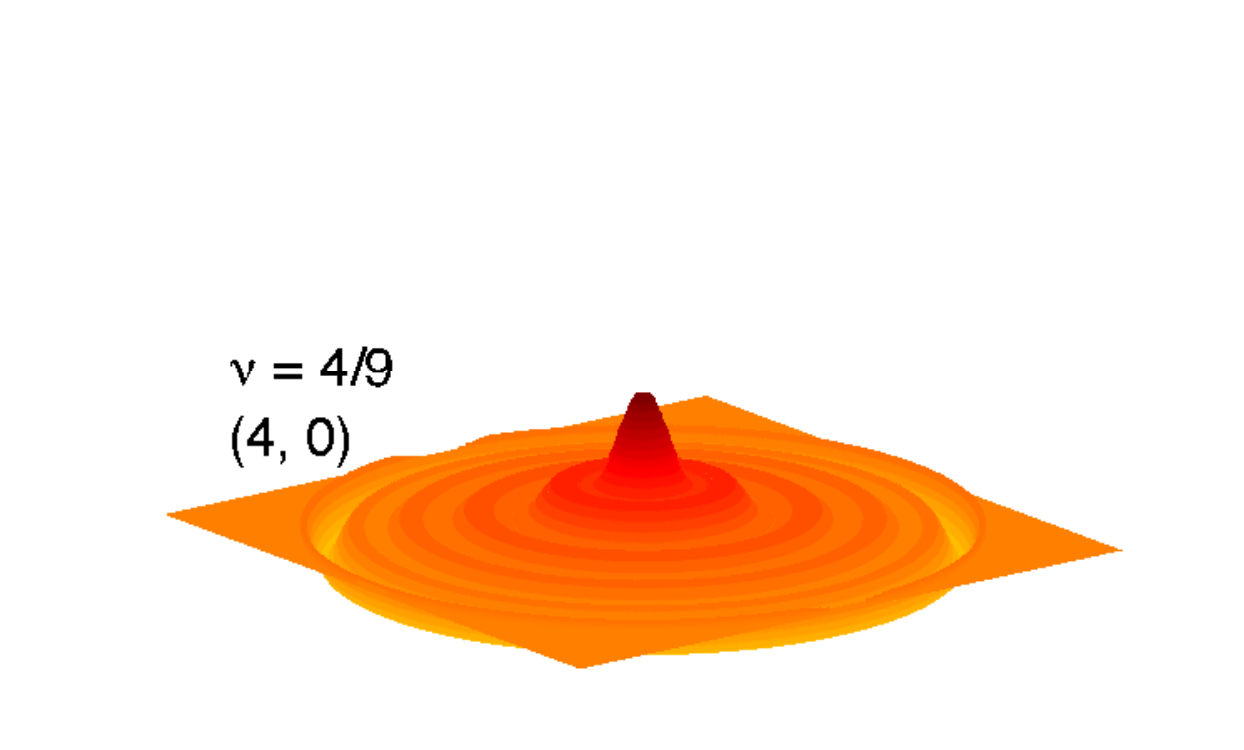}}
\hspace{-5mm}
\resizebox{0.26\textwidth}{!}{\includegraphics{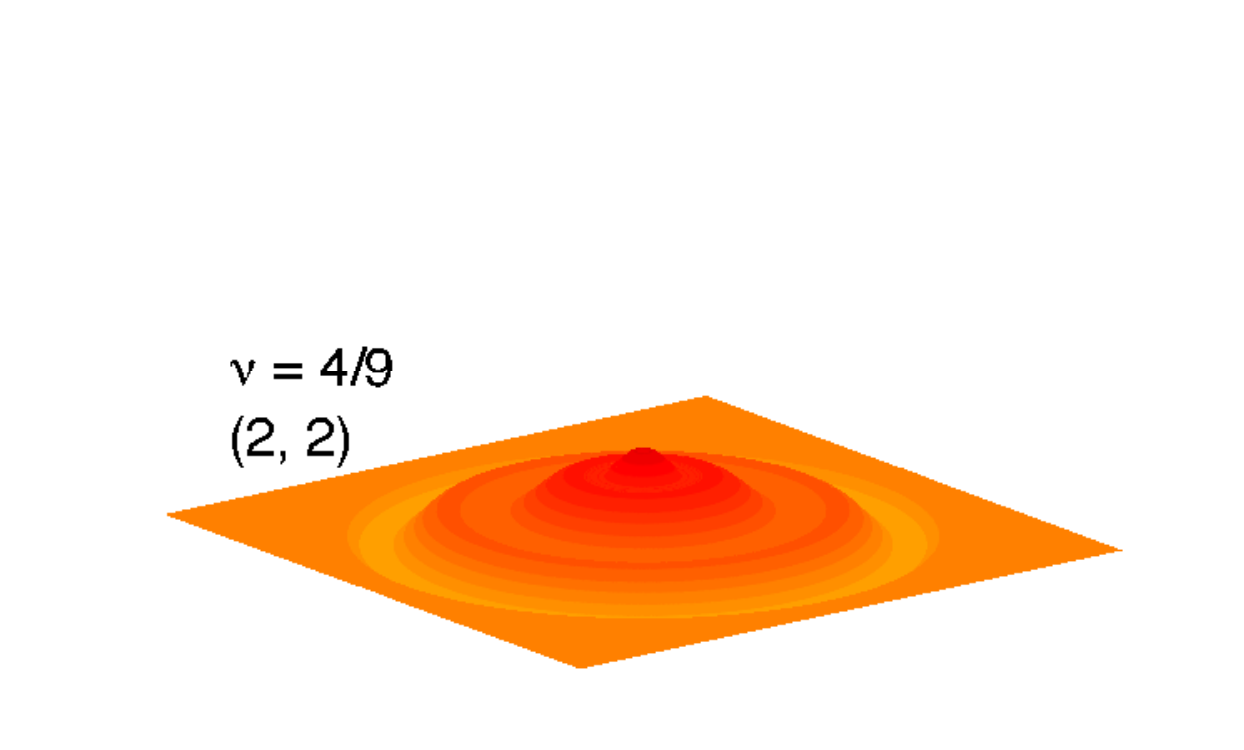}}
\hspace{-5mm}
\resizebox{0.26\textwidth}{!}{\includegraphics{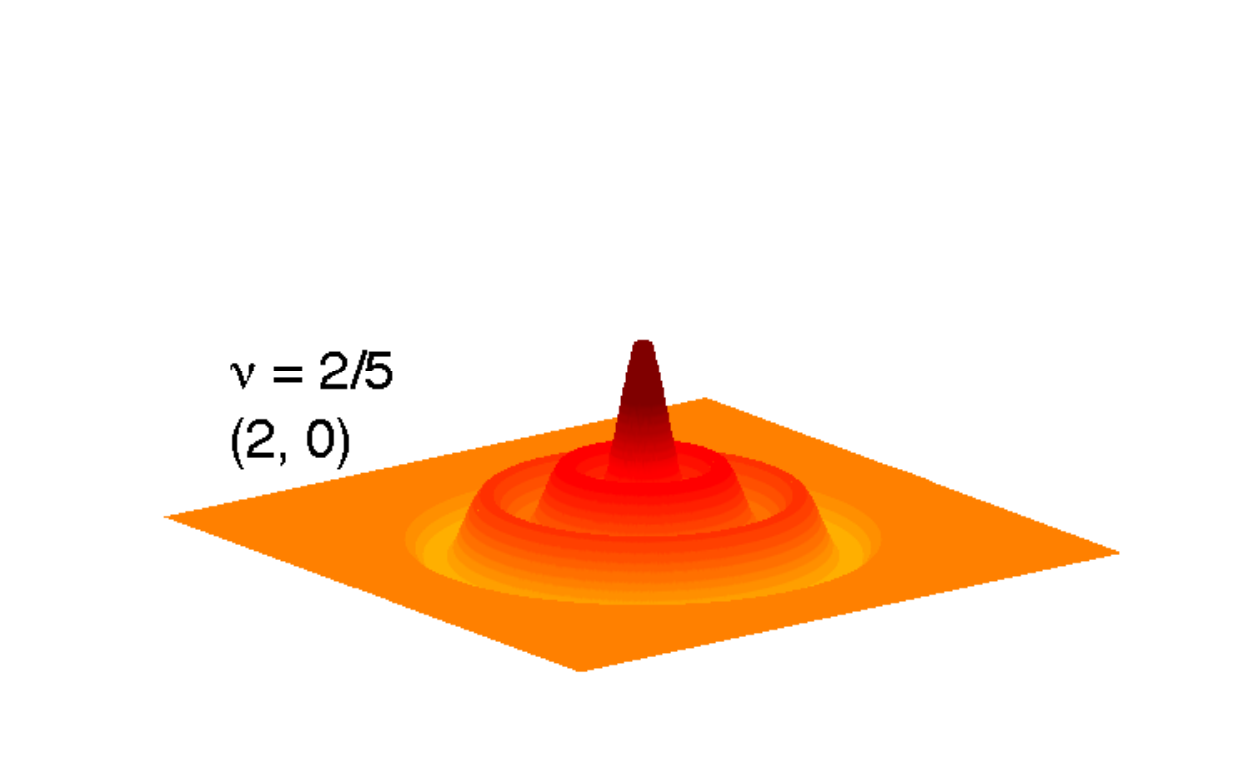}}
\hspace{-5mm}
\resizebox{0.26\textwidth}{!}{\includegraphics{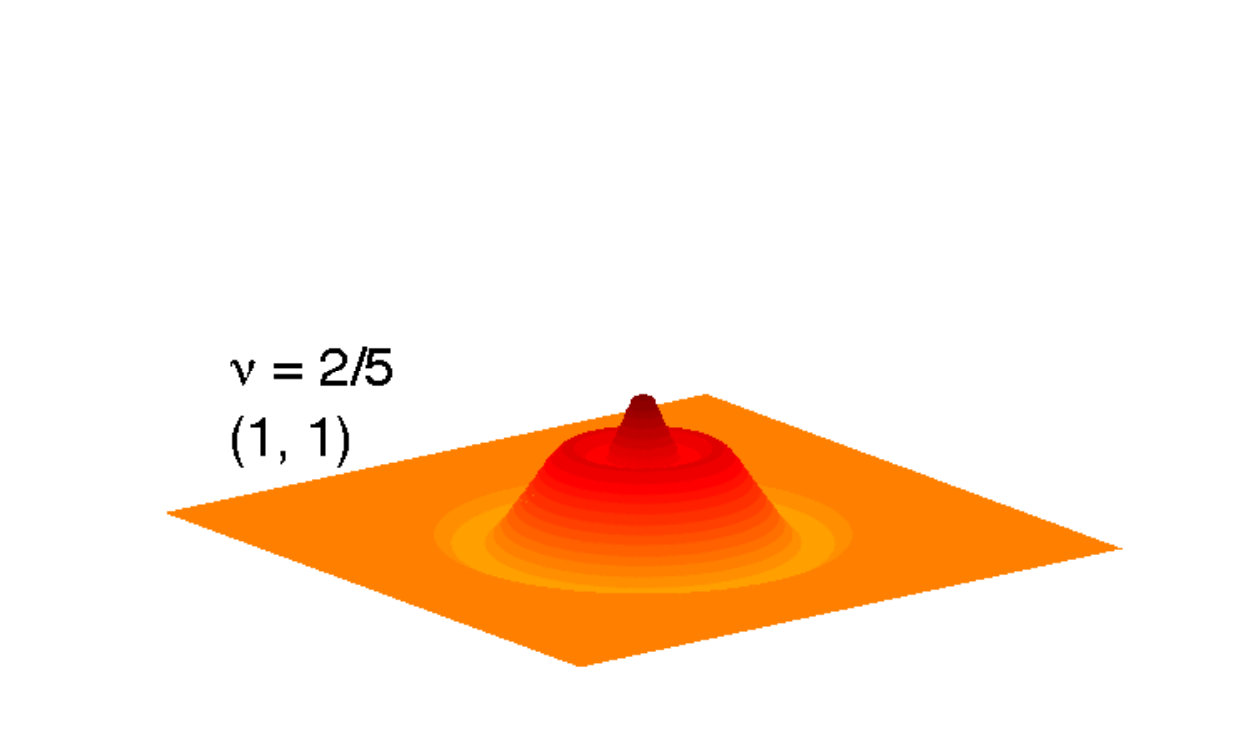}}\\
\resizebox{0.25\textwidth}{!}{\includegraphics{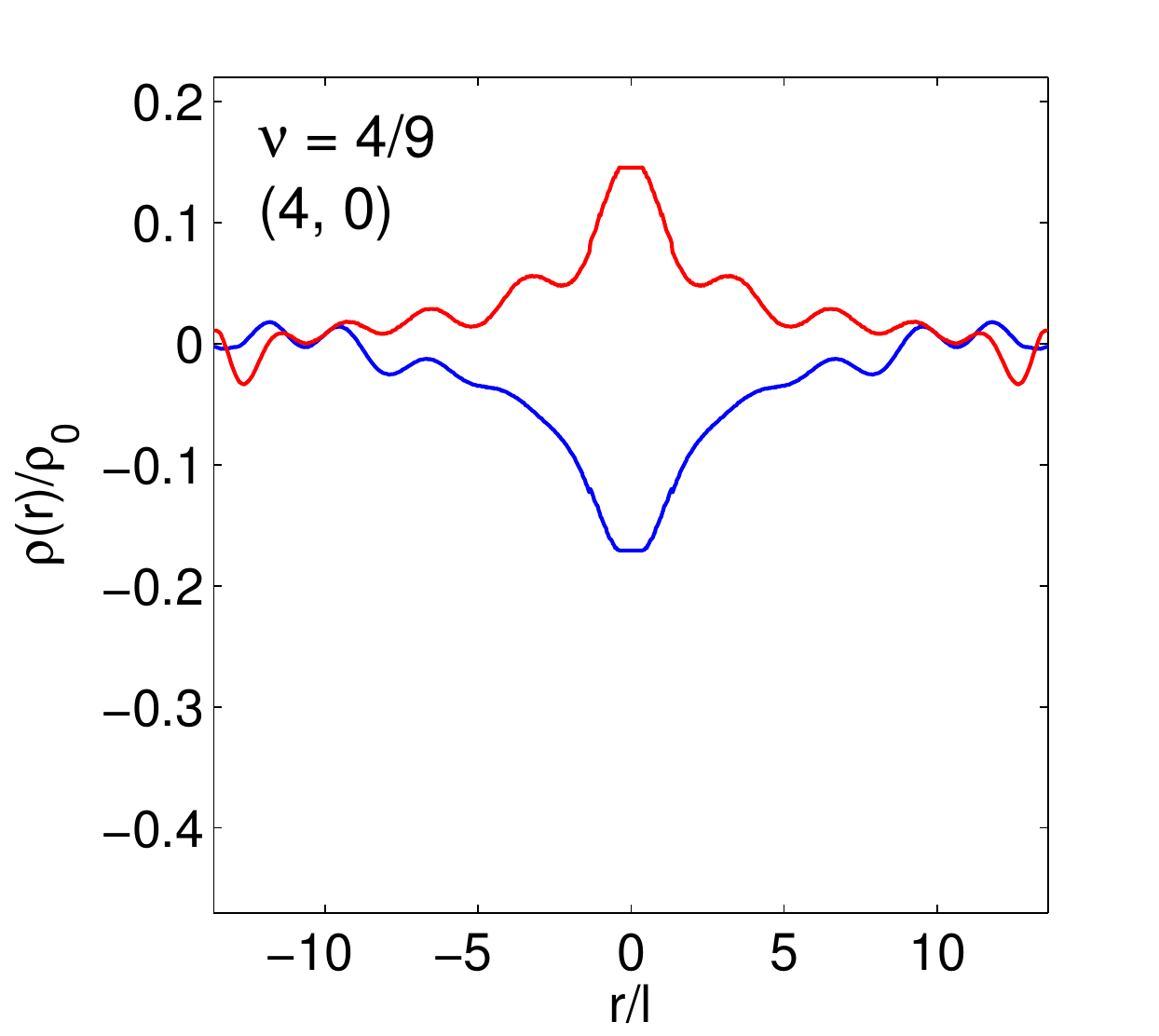}}
\hspace{-3mm}
\resizebox{0.25\textwidth}{!}{\includegraphics{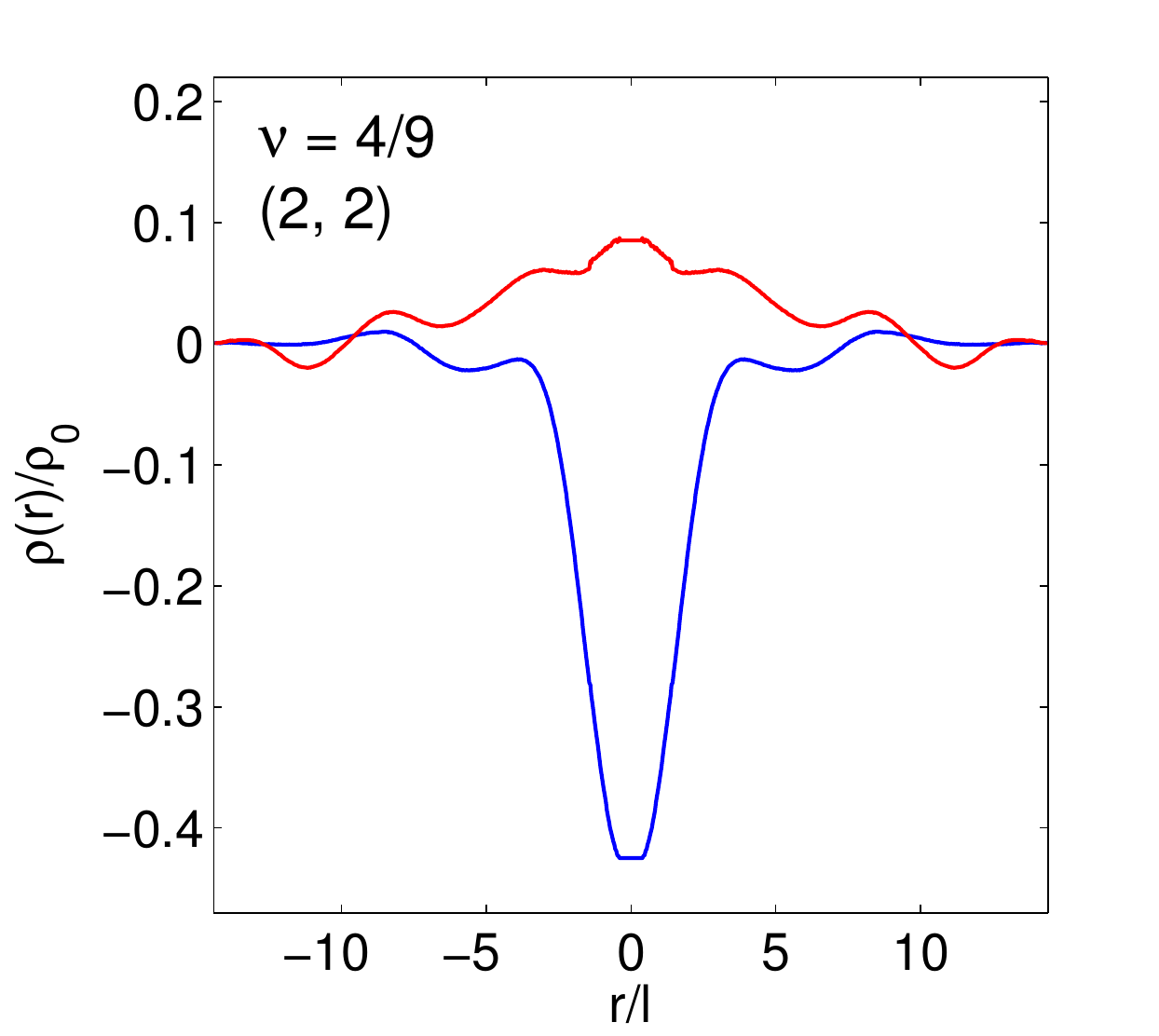}}
\hspace{-3mm}
\resizebox{0.25\textwidth}{!}{\includegraphics{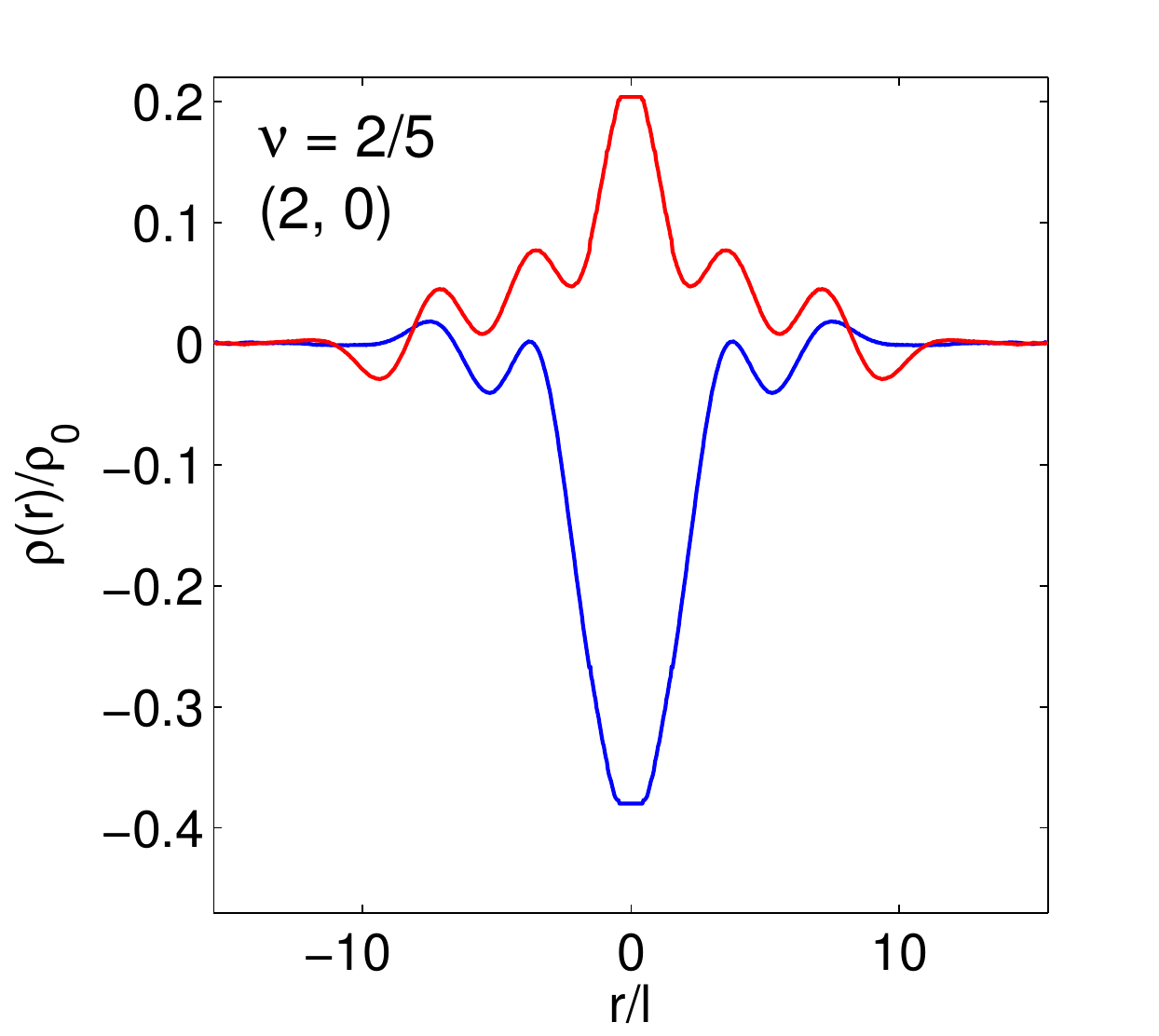}}
\hspace{-3mm}
\resizebox{0.25\textwidth}{!}{\includegraphics{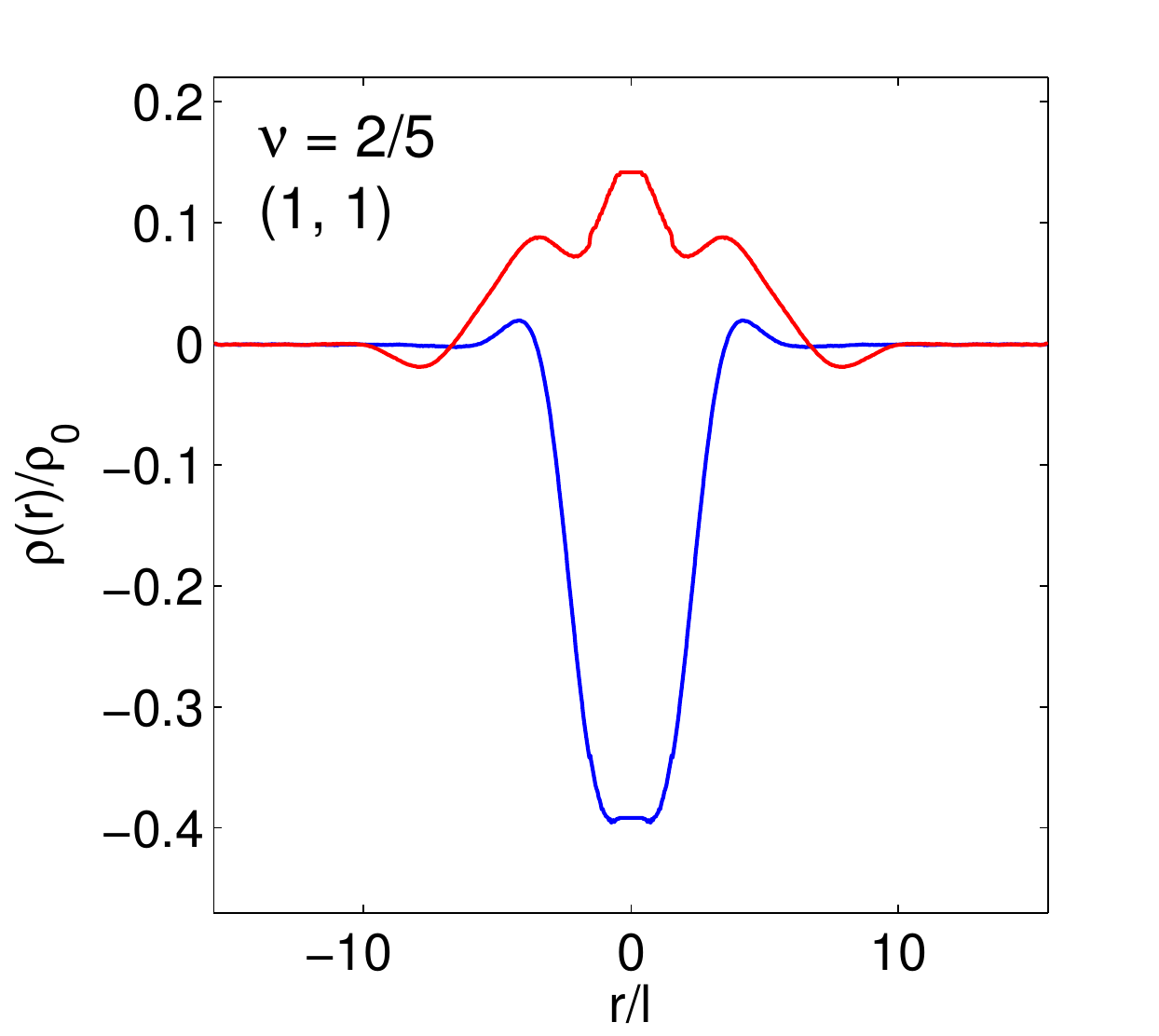}}
\vspace{-5mm}
\caption{Density profiles for soft holes (top row) and soft electrons (middle row) for fully spin polarized and spin singlet states at $\nu=2/5$ and $\nu=4/9$. These correspond to bound complexes of composite fermions, as shown in Figs.~\ref{CFoccupation}-\ref{CFoccupation2} for $\nu=2/5$. The bottom panels show the line plots for the densities.
}
\label{softdensity}
\end{figure*}

We now ask what is the lowest energy exciton that has a non-zero matrix element with ${\bar \Psi}_L^{\dagger}(0) {\bar \Psi}_R(0) | \Psi_{0} \rangle$. We can identify this exciton within the CF theory, to the extent that we can neglect the inter-layer interaction energy of the electron and the hole. 

The lowest energy excitations of a single layer FQH state are excited composite fermions or the CF holes they leave behind. It is possible to construct a low energy ``electron" from a combination of such excited composite fermions\cite{Peterson05}. Of relevance to the current problem is the excitation that has the same quantum numbers as the hard electron. Specifically, the excitation should have the same total angular momentum $L = |M| = Q$ as the hard electron. Because the excited composite fermions carry a fractional local charge equal to $1/(2pn+ 1)$ of an electron charge, one needs to consider a collection of $2pn+1$ composite fermions in excited $\Lambda$Ls to produce an excitation with the charge of an electron. Ref.~\onlinecite{Peterson05} studied the problem of how these excited composite fermions arrange themselves in various $\Lambda$Ls to produce such an excitation, and showed that the lowest energy state can be identified uniquely. We call this lowest-energy excitation a ``soft'' electron. For partially spin polarized states, the spin-up soft electron consists of $2pn_{\uparrow}+1$ spin-up and $2pn_{\downarrow}$ spin-down composite fermions in the excited $\Lambda$Ls; the $\Lambda$L occupations of composite fermions can again be determined uniquely for the lowest energy state. Specifically, for $p=1$, a soft spin-up electron consists of $(n_{\uparrow} + 1)$ and $n_{\uparrow}$ spin-up composite fermions in the lowest two unoccupied spin-up $\Lambda$Ls (namely $n_{\uparrow}$-th and $(n_{\uparrow} + 1)$-th spin-up $\Lambda$Ls), and $n_{\downarrow}$ spin-down CFs in each of the lowest two unoccupied spin-down $\Lambda$Ls, with all the composite fermions occupying the largest available $m$ orbitals in each $\Lambda$L. A spin-up ``soft'' hole can be similarly defined. It consists of $(n_{\uparrow} + 2)$ and $(n_{\uparrow} - 1)$ CF holes in the top two occupied spin-up $\Lambda$Ls, and $(n_{\uparrow} + 1)$ and $(n_{\uparrow} - 1)$ CF holes in the top two occupied spin-down $\Lambda$Ls, again in the largest $m$ orbitals. Figs.~\ref{CFoccupation}-\ref{CFoccupation2} show some examples of the lowest-energy CF complexes corresponding to soft electron and soft hole. 

While the soft excitons are the lowest energy excitons which have a non-zero matrix element with the hard exciton ${\bar \Psi}_L^{\dagger}(0) {\bar \Psi}_R(0) | \Psi_{0} \rangle$, they are much more spread out than the hard excitons (see Figs.~\ref{harddensity}, \ref{softdensity}). As a result, they are expected to have much lower tunneling amplitude, especially for states $n/(2n+1)$ for large $n$. This has been confirmed by explicit calculation\cite{Peterson05} which shows that the overlaps of a hard hole and a soft hole for $\nu=1/3$, 2/5, 3/7 and 4/9 are $\sim$ 1.0, 0.52, 0.08, 0.015, whereas the overlaps of a hard electron and a soft electron for $\nu=1/3$, 2/5 and 3/7 are $\sim$ 0.3, 0.03, and 0.005 in the thermodynamic limit (all numbers are for fully spin polarized states).  As an interesting aside, even though the energy $E_{\rm e}+E_{\rm h}$ for the soft exciton is lower than that of the hard exciton, addition of an electron-hole interaction term $E_{\rm e-h}$ can reverse their ordering, because $E_{\rm e-h}$ is more negative for the hard exciton than for the soft exciton. Please see the next section for the detailed definition of $E_{\rm e}$, $E_{\rm h}$ and $E_{\rm e-h}$.

If we do not insist on angular momentum conservation during tunneling (which is strictly valid only in the absence of disorder), then an even lower energy exciton becomes available consisting of $2pn+1$ far separated quasiparticle-quasihole excitons. We believe it to be unlikely that the electron tunneling term in the Hamiltonian would couple to such excitons in a significant fashion, and therefore do not consider them.

\section{Exciton energy: Calculation and comparison with experiment}
\label{SecIII}

As noted in the introduction, the  exciton energy $E_{\rm ex}$ is a sum of three parts: 
\be
E_{\rm ex} =(E_{\rm e}-E_{\rm gs}) + (E_{\rm h}-E_{\rm gs}) + E_{\rm e-h} \equiv \Delta + E_{\rm e-h}
\label{Eex}
\ee
where $E_{\rm e}$ / $E_{\rm h}$ are the energies of the state with an additional electron / hole, and $E_{\rm e-h}$ is attractive interaction between them.  We have defined the `bare' gap $\Delta$, namely the exciton energy without including the interaction between the electron and the hole. Given the density profiles of an electron [$\rho_{\rm e}(r)$] and a hole [$\rho_{h}(r)$] at the center of a disk, the interaction term can be evaluated as
\be
E_{\rm e-h} = \frac{e^{2}}{\epsilon} \int{ \frac{\rho_{\rm e}(|{\bf r}_{1}|) \rho_{h}(|{\bf r}_{2}|) }{\sqrt{({\bf r}_{1} - {\bf r}_{2})^{2} + d^{2}}} d{\bf r}_{1} d{\bf r}_{2}}
\ee
where $d$ is the distance between the two electron-gas layers and $\epsilon$ is the dielectric constant of the material. 

{\em Parallel magnetic field}: The above equation is appropriate when the electron tunnels perpendicularly across the barrier. When a parallel magnetic field $B_{||}$ is added to a pre-existing perpendicular field $B_{\perp}$, the tunneling electron acquires a ``momentum boost'' $\hbar q$ due to the Lorentz force associated with $B_{||}$, with $q = edB_{||}/\hbar$. This momentum boost causes the electron to tunnel in a non-perpendicular direction,  leading to a lateral shift in the location of the tunneled electron. Since the single particle wave function in Landau gauge is centered at $y = k_{x}\ell^{2}$, where $\ell$ is the magnetic length, the shift distance can be calculated as $s = q\ell^{2} = d\frac{B_{||}}{B_{\perp}}$. Therefore the interaction term is modified to
\be
E_{\rm e-h} = \frac{e^{2}}{\epsilon} \int{ \frac{\rho_{\rm e}(|({\bf r}_{1} - {\bf s})|) \rho_{h}(|{\bf r}_{2}|) }{\sqrt{({\bf r}_{1} - {\bf r}_{2})^{2} + d^{2}}} d{\bf r}_{1} d{\bf r}_{2}}.
\ee
In Sec. III, we will show that this $E_{\rm e-h}$ dependence on $B_{||}$ quantitatively explains the experimental finding that $V_{\rm max}$ shifts to higher bias voltages with increasing parallel magnetic field.

To evaluate the exciton energy $E_{\rm ex}$ in Eq. (\ref{Eex}), we calculate the energies $E_{gs}$, $E_{\rm e}$, $E_{\rm h}$ as well as the density profiles using the microscopic theory of composite fermion. We use the standard LLL projection method \cite{Jain97,Jain97b} and evaluate various integrals using the Monte Carlo method. We also assume that the wave functions of the hard and soft electron and hole are not modified significantly due to the interlayer interaction.

FQH experiments are generally performed on GaAs-${\mathrm{Al}}_{\mathrm{x}}$${\mathrm{Ga}}_{1\mathrm{-}\mathrm{x}}$As heterojunctions and quantum wells. These structures have nonzero transverse width, which can lead to quantitative changes to observables. In our numerical computation, we consider an effective two-dimensional interaction evaluated from the transverse wave function $\xi(z)$:
\begin{equation}
V^{\text{eff}}(r) = \frac{e^{2}}{\epsilon} \int dz_{1} \int dz_{2} \frac{|\xi(z_{1})|^{2} |\xi(z_{2})|^{2}}{[r^{2} + (z_{1} - z_{2})^{2}]^{1/2}},
\end{equation}
where $z_{1}$ and $z_{2}$ denote the coordinates in the transverse direction, and $r =\sqrt{ (x_{1} - x_{2})^{2}+ (y_{1} - y_{2})^{2}}$ is the distance on the 2D plane. $V^{\text{eff}}(r)$ approaches the ideal 2D interaction $e^{2}/\epsilon r$ at long distances, but is softened at short distances. We obtain  $\xi(z)$ by solving the 1D Schr\"odinger and Poisson equations self-consistently~\cite{Ortalano97} for a zero-magnetic-field system with different geometries and charge densities. The local density approximation~\cite{Ortalano97} is used.

We also neglect the effect of the parallel magnetic field on the transverse wave function in what follows below. The justification is that for the experimental parameters of interest here, the finite width corrections are actually small, changing the exciton energies by only a small amount ($\sim$10\%), presumably because of the relatively small quantum well width $w=18$nm and small densities. This suggests that the changes in the transverse wave function due to any parallel magnetic field will not cause significant correction to the calculated energies. There is another effect due to a parallel magnetic field, namely that the electron mass becomes anisotropic (e.g. see Ref.~\onlinecite{Mueed15a}), thereby breaking rotational symmetry. This leads to excitations that are not exactly circularly symmetric. Experiments have shown that the effect of parallel magnetic fields is relatively small for composite fermions than for electrons\cite{Kamburov14}, and theoretical calculations (e.g. see Ref.~\onlinecite{Balram16} and references therein) show that the change in excitation energies is small.

In the evaluation of the $E_{\rm ex}$, we neglect the effect of LL mixing, because it does not affect the excitations gaps substantially~\cite{Melik-Alaverdian95, Melik-Alaverdian97, Scarola00}. In contrast, it has been found \cite{Zhang16} that LL mixing can cause substantial quantitative correction to the critical Zeeman energies where transitions between differently spin polarized states occur. We will show below that the Zeeman energy below which the CF Fermi sea ceases to be fully spin polarized also is affected by LL mixing.

We also assume spin is conserved during tunneling. For a partially polarized state such as $(n_{\uparrow}, n_{\downarrow}) = (3, 1)$ ($\nu=4/9$), we will consider the two different cases in which the tunneling electron belongs to the majority and minority spin species.

\begin{figure*}
\resizebox{0.335\textwidth}{!}{\includegraphics{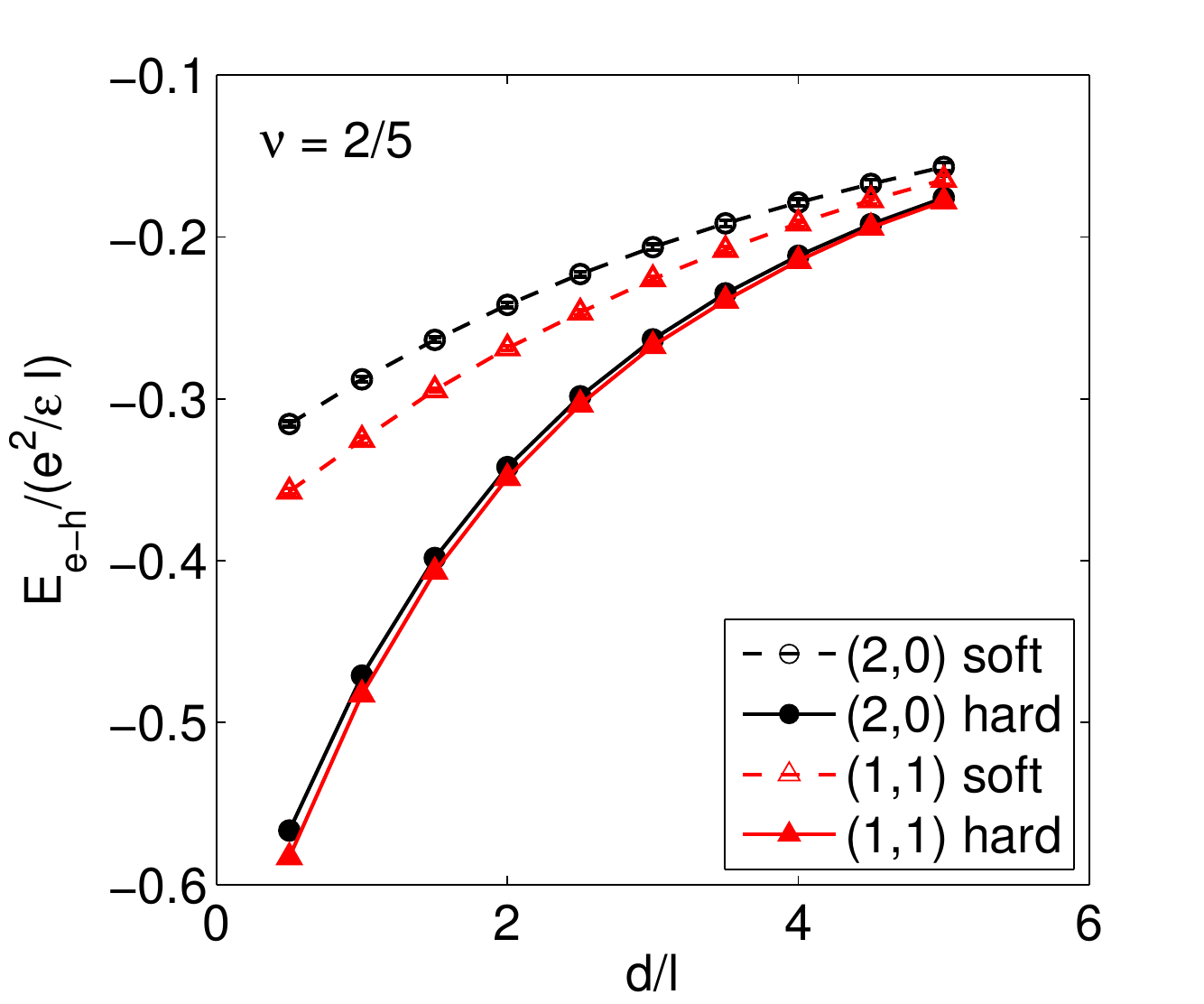}}
\hspace{-3mm}
\resizebox{0.335\textwidth}{!}{\includegraphics{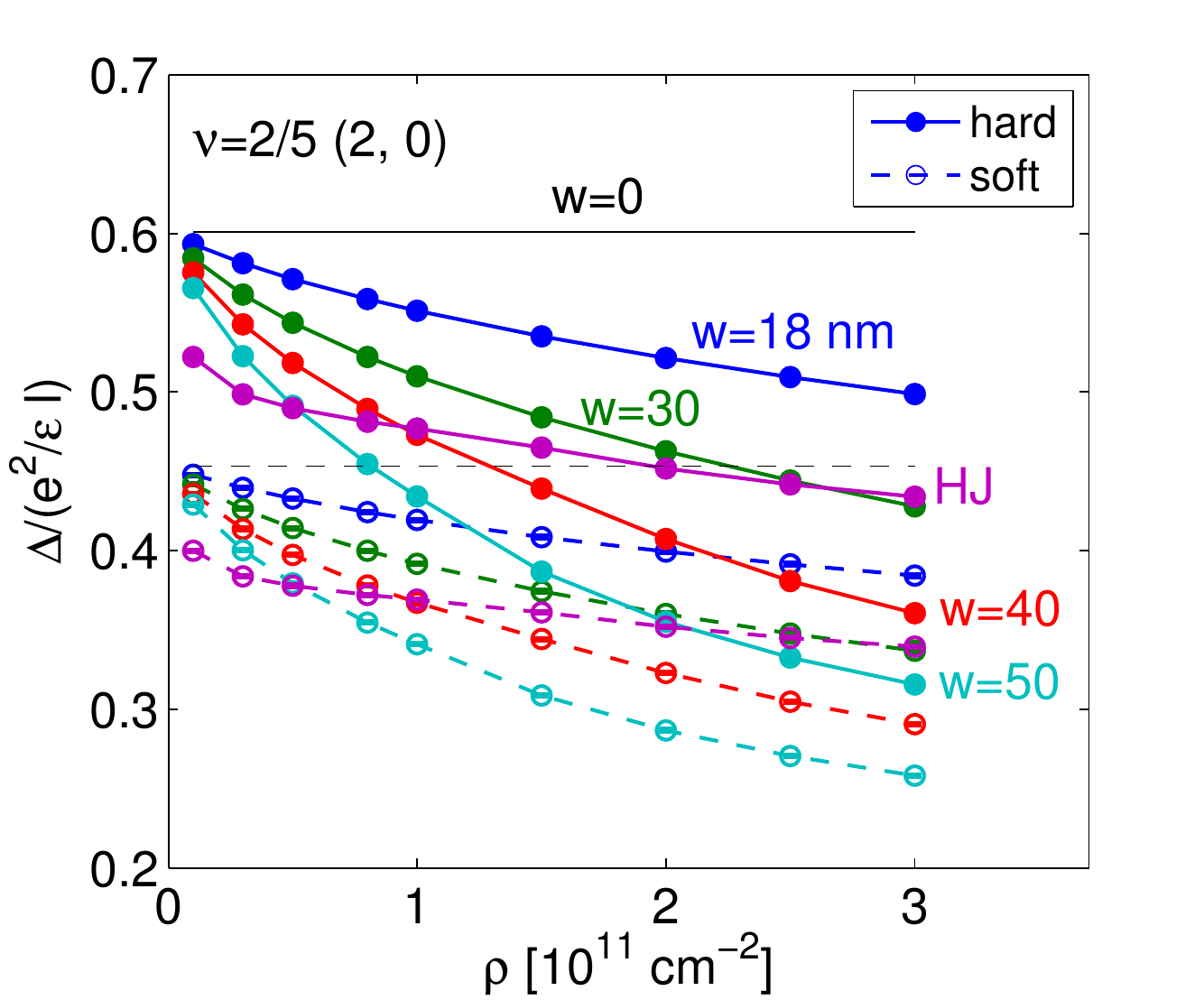}}
\hspace{-3mm}
\resizebox{0.335\textwidth}{!}{\includegraphics{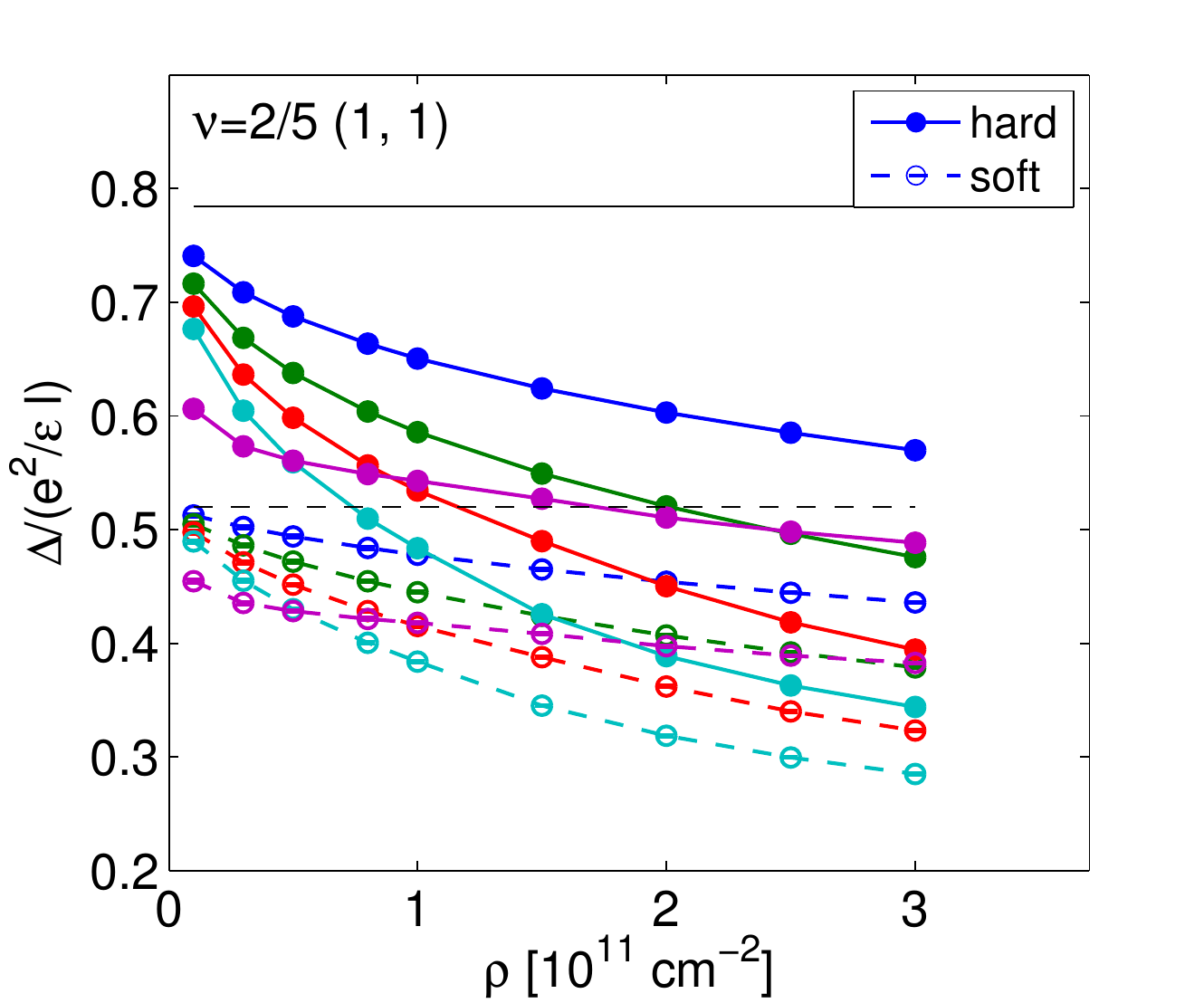}}\\
\resizebox{0.335\textwidth}{!}{\includegraphics{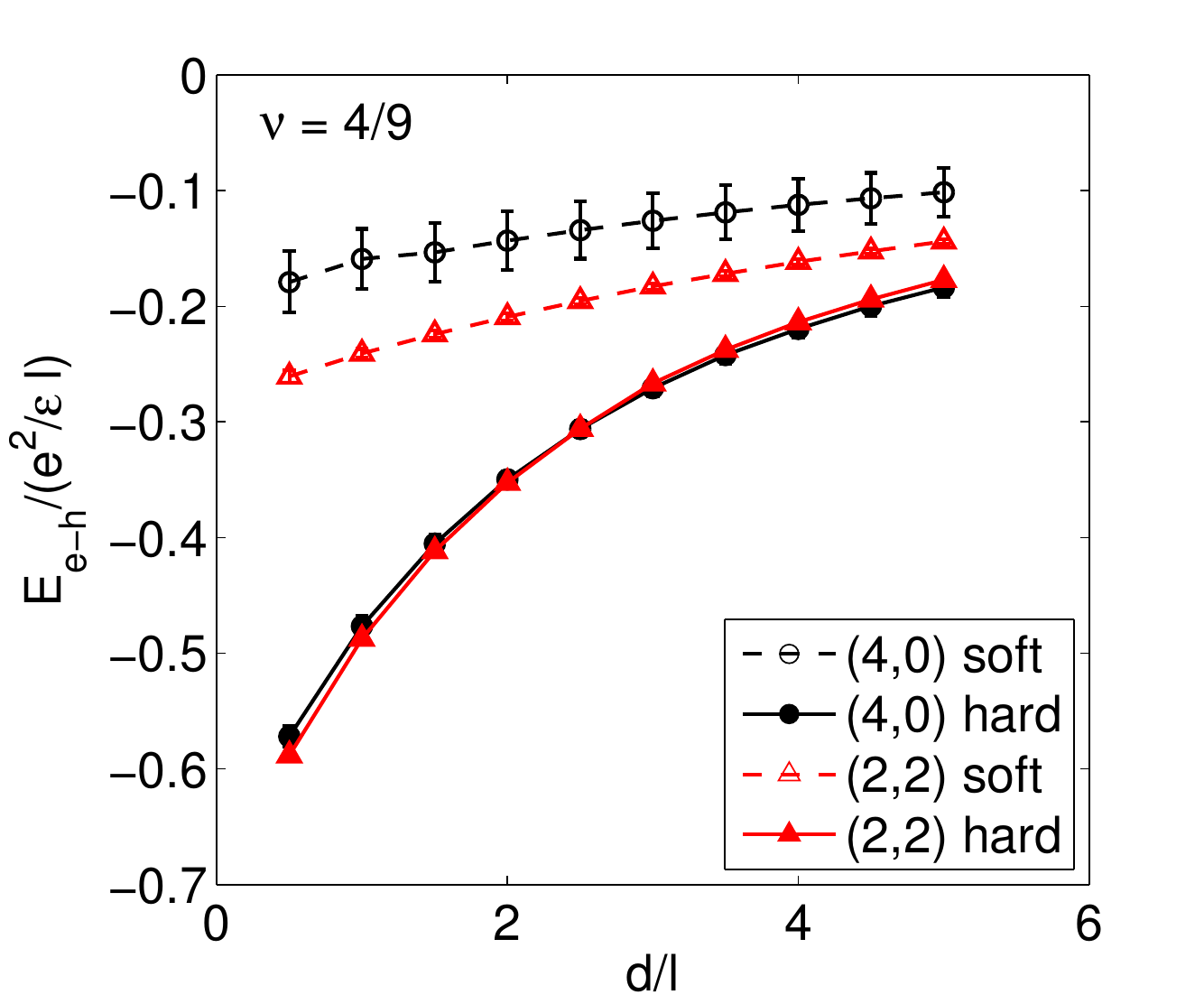}}
\hspace{-3mm}
\resizebox{0.335\textwidth}{!}{\includegraphics{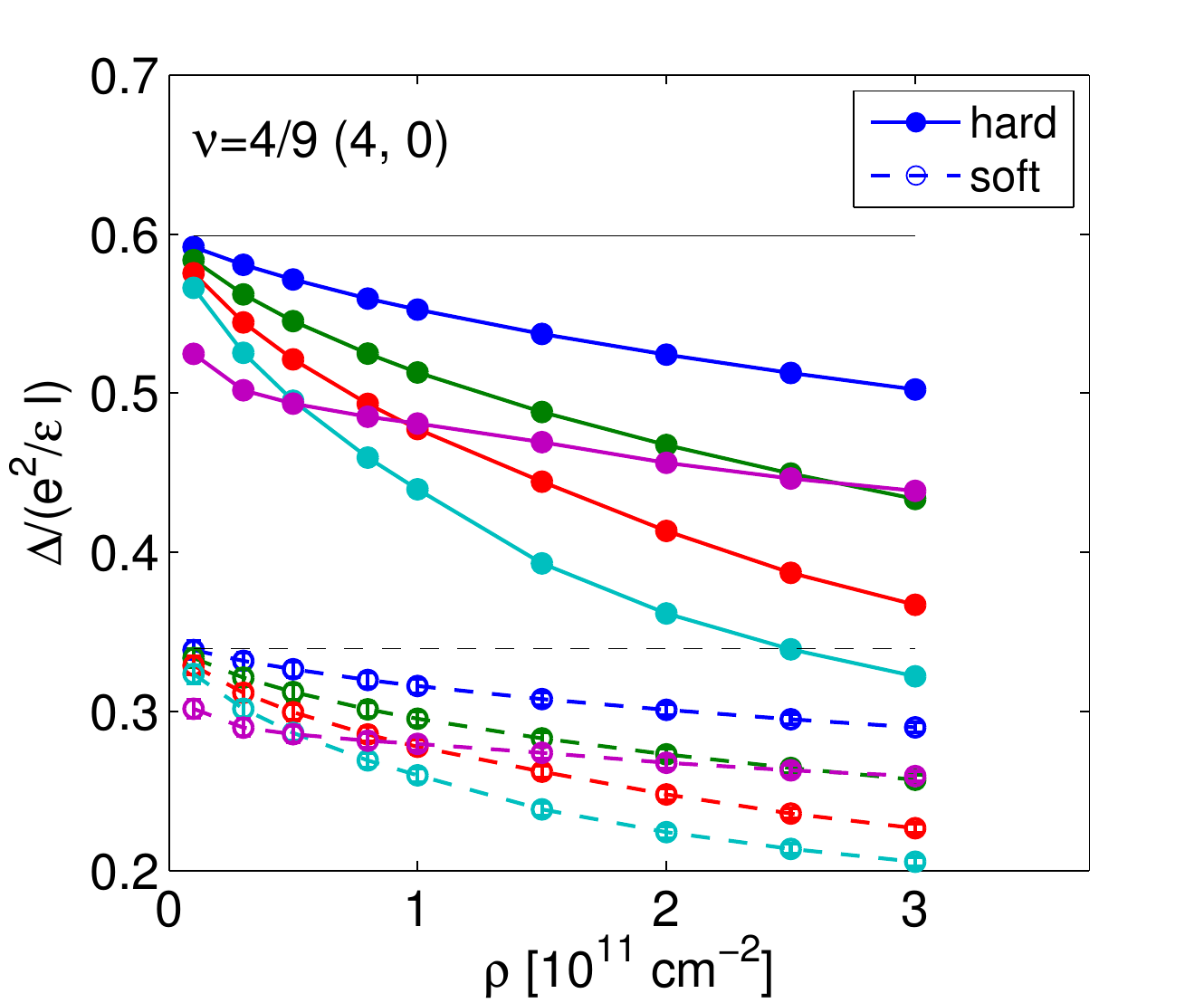}}
\hspace{-3mm}
\resizebox{0.335\textwidth}{!}{\includegraphics{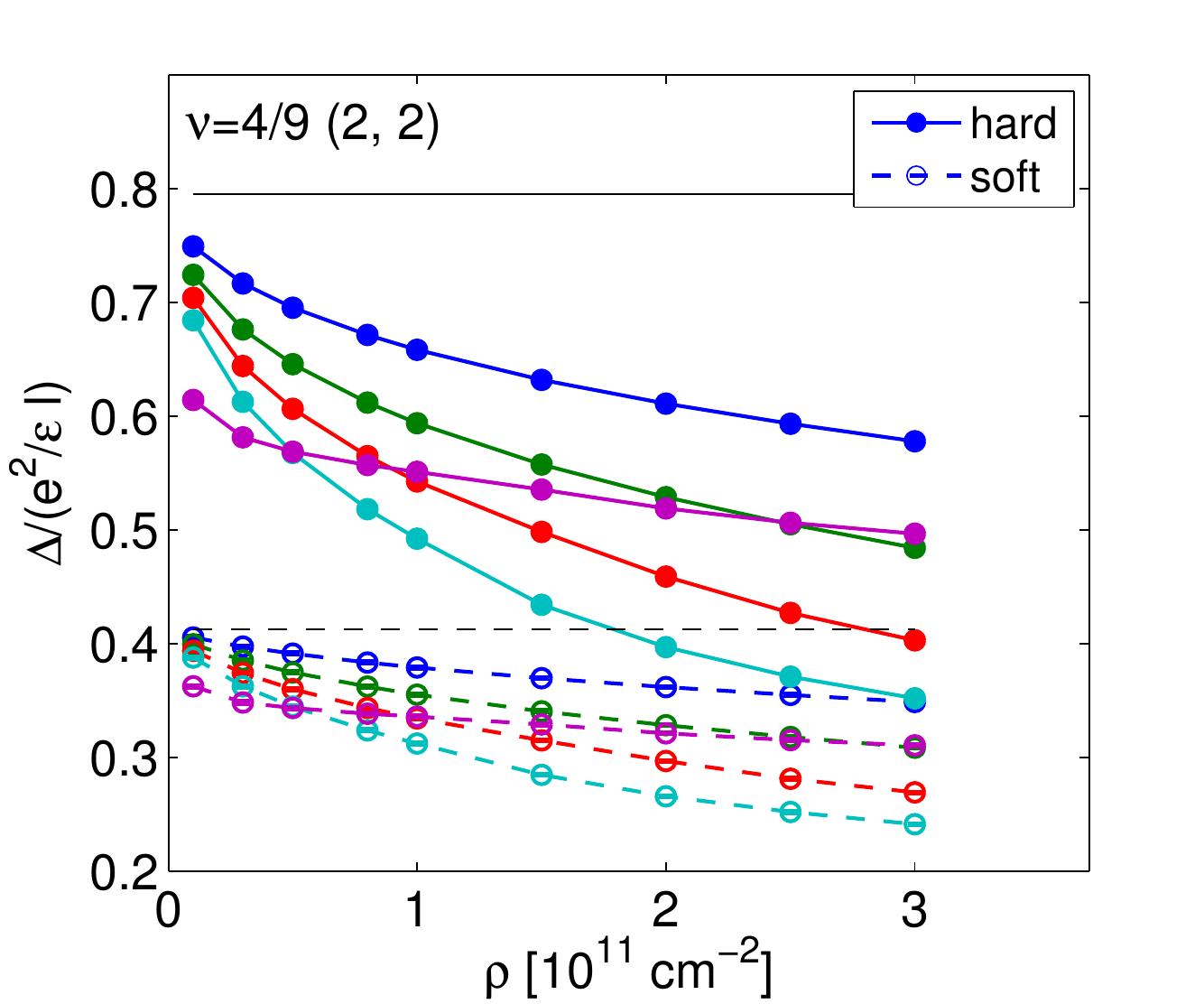}}\\
\resizebox{0.335\textwidth}{!}{\includegraphics{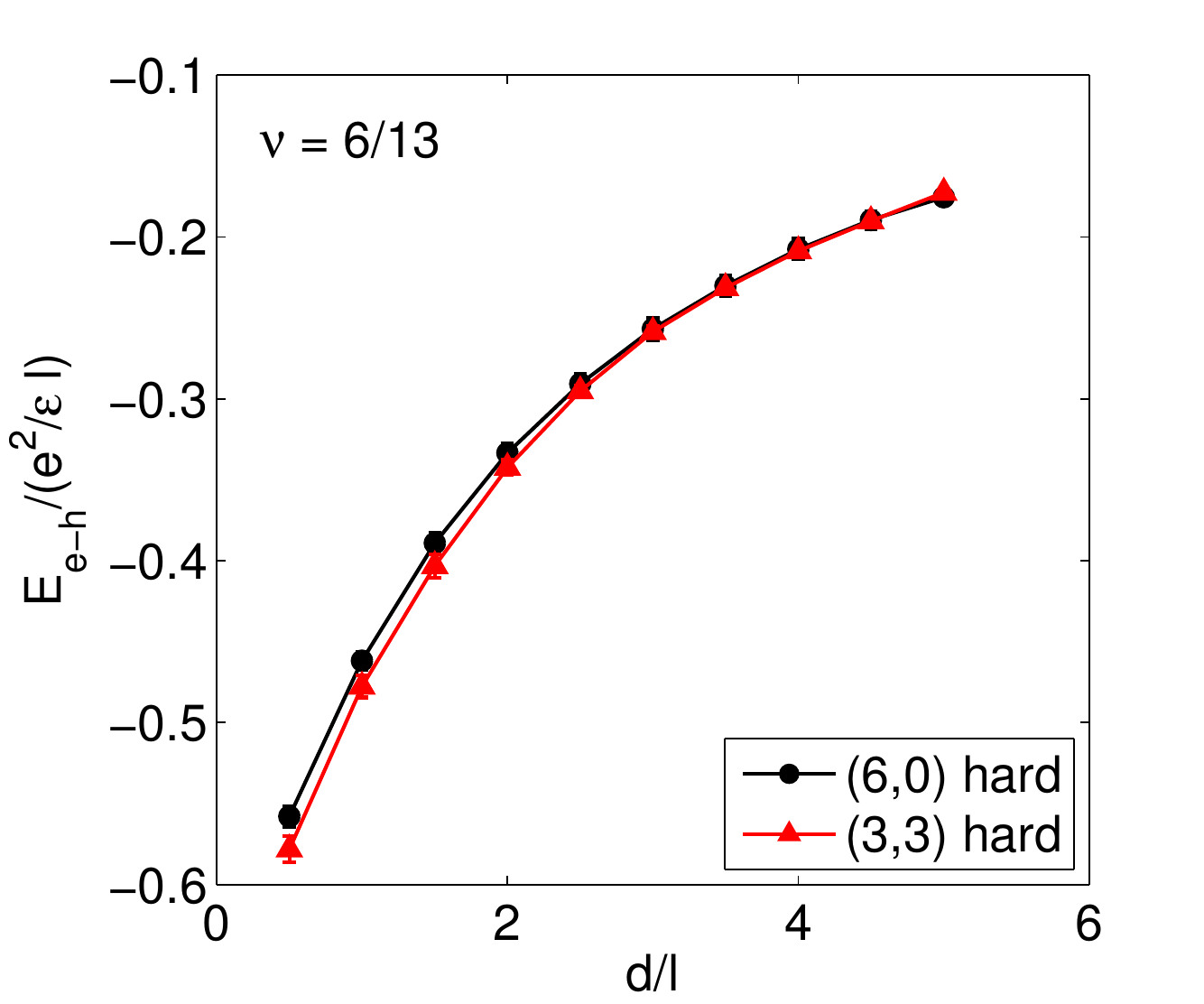}}
\hspace{-3mm}
\resizebox{0.335\textwidth}{!}{\includegraphics{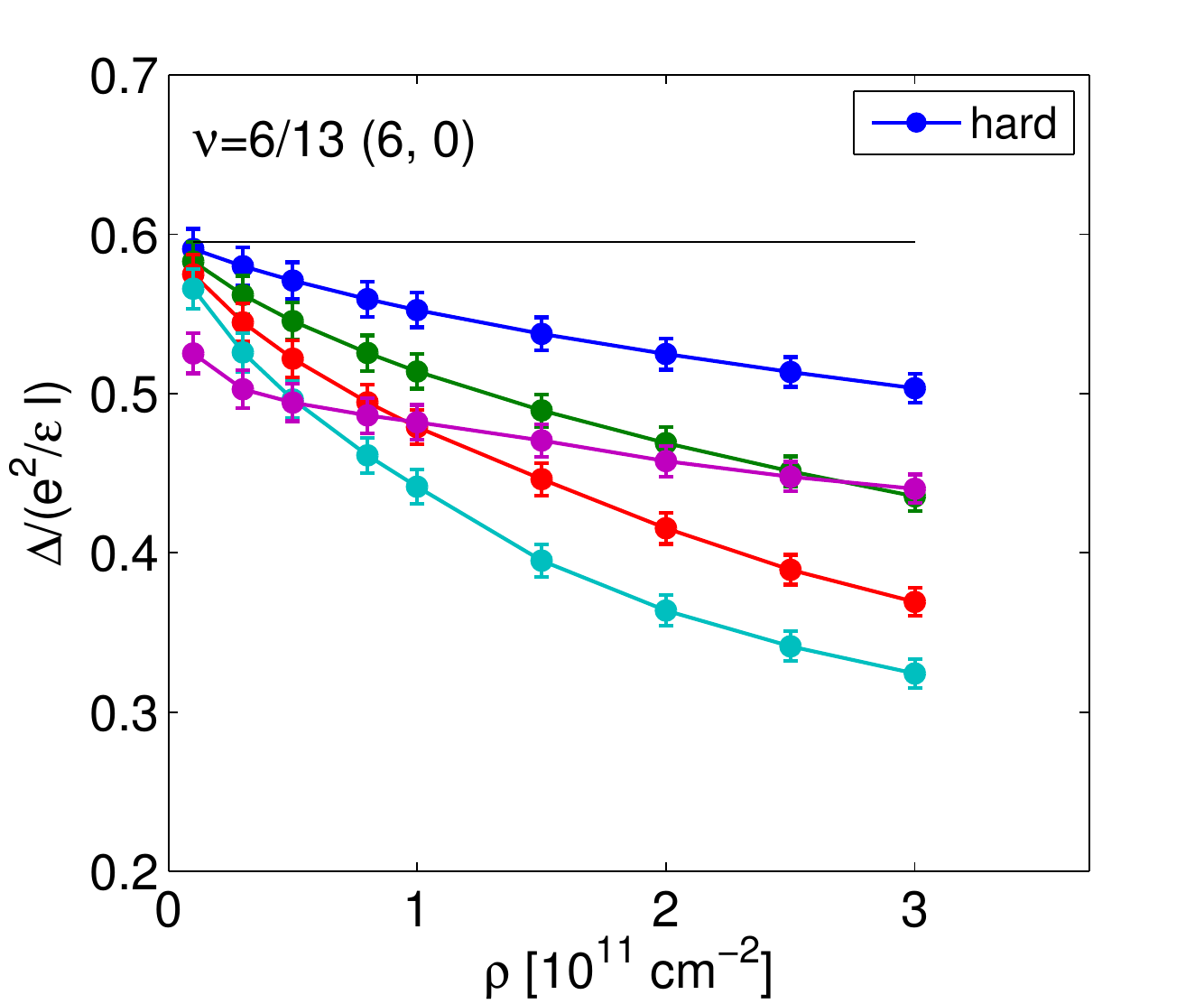}}
\hspace{-3mm}
\resizebox{0.335\textwidth}{!}{\includegraphics{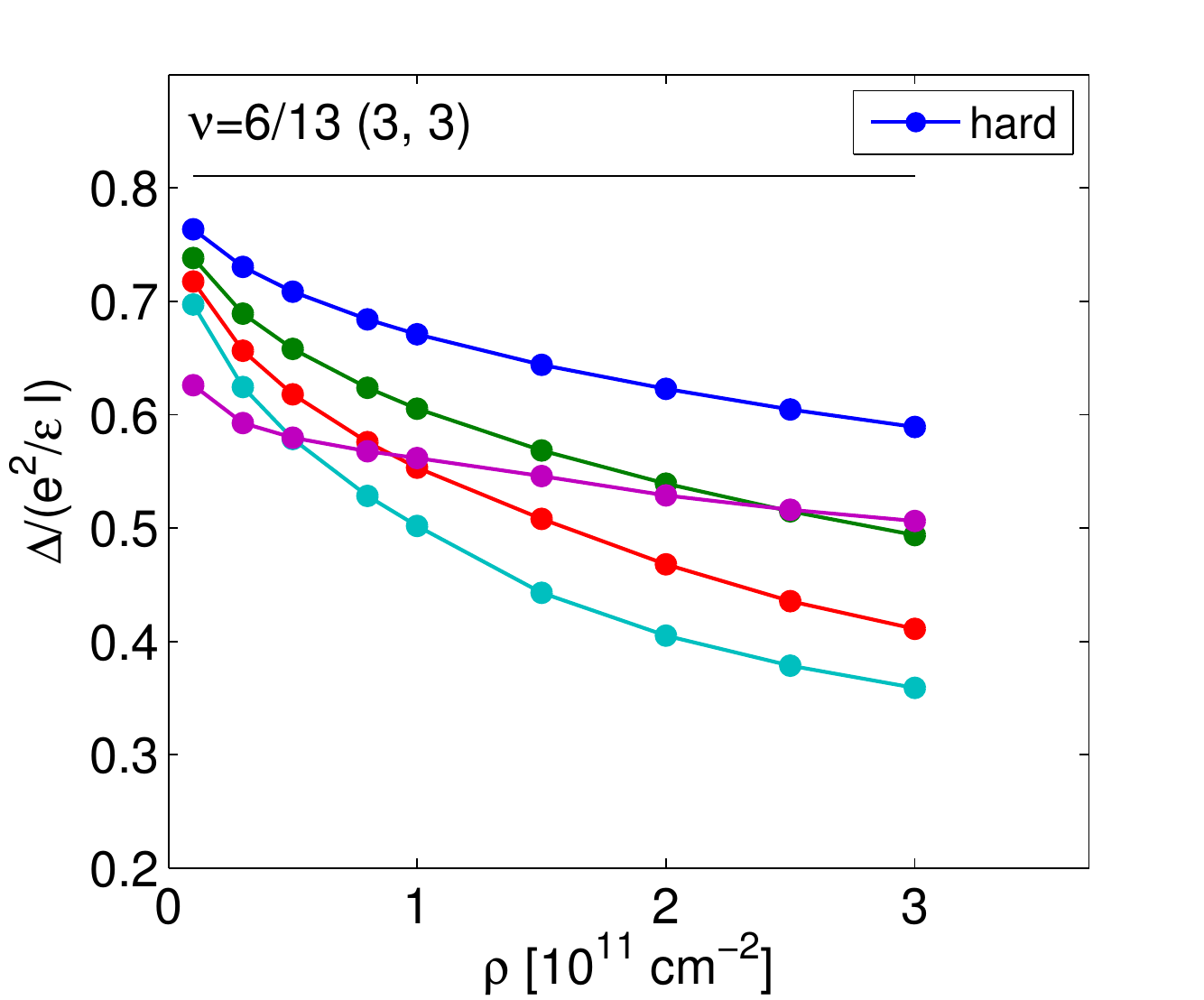}}\\
\vspace{-3mm}
\caption{Different terms contributing to the energy of the excitons of spin polarized and spin singlet states at filling factors $\nu = 2/5, 4/9, 6/13$; the spin polarized states are labeled $(2,0)$, $(4,0)$, $(6,0)$ and the spin singlet states are labeled $(1,1)$, $(2,2)$, $(3,3)$. Left panels show the hard and soft electron-hole interaction energy $E_{\rm e-h}$ as a function of the distance $d$ between two layers. Middle and right panels show the bare gap $\Delta = E_{\rm e} + E_{\rm h} - 2E_{\rm gs}$, which excludes the $E_{\rm e-h}$ term, for different quantum well widths and heterojunction (HJ) as a function of density.
}
\label{Eeh1}
\end{figure*}

\begin{figure*}
\resizebox{0.335\textwidth}{!}{\includegraphics{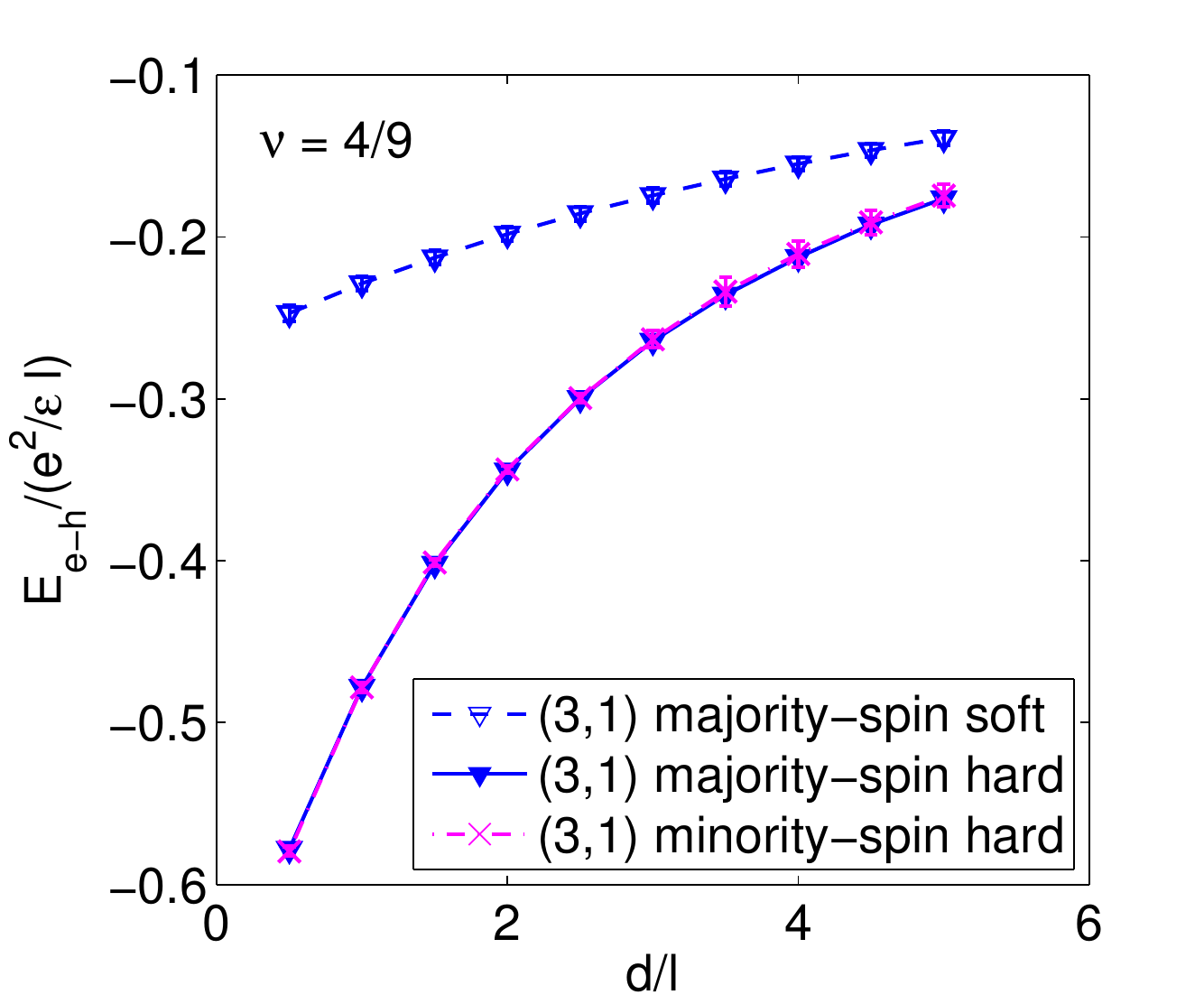}}
\hspace{-3mm}
\resizebox{0.335\textwidth}{!}{\includegraphics{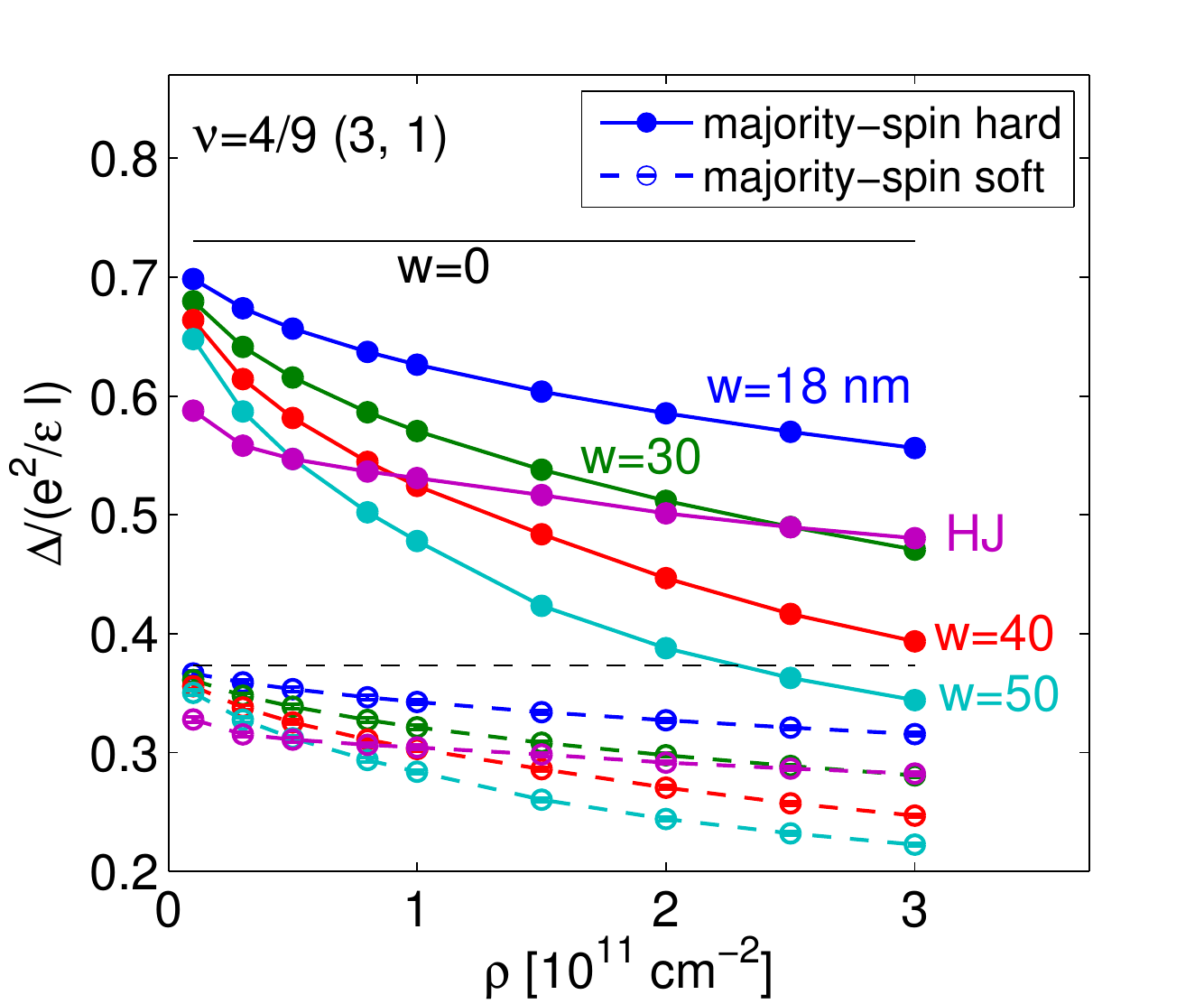}}
\hspace{-3mm}
\resizebox{0.335\textwidth}{!}{\includegraphics{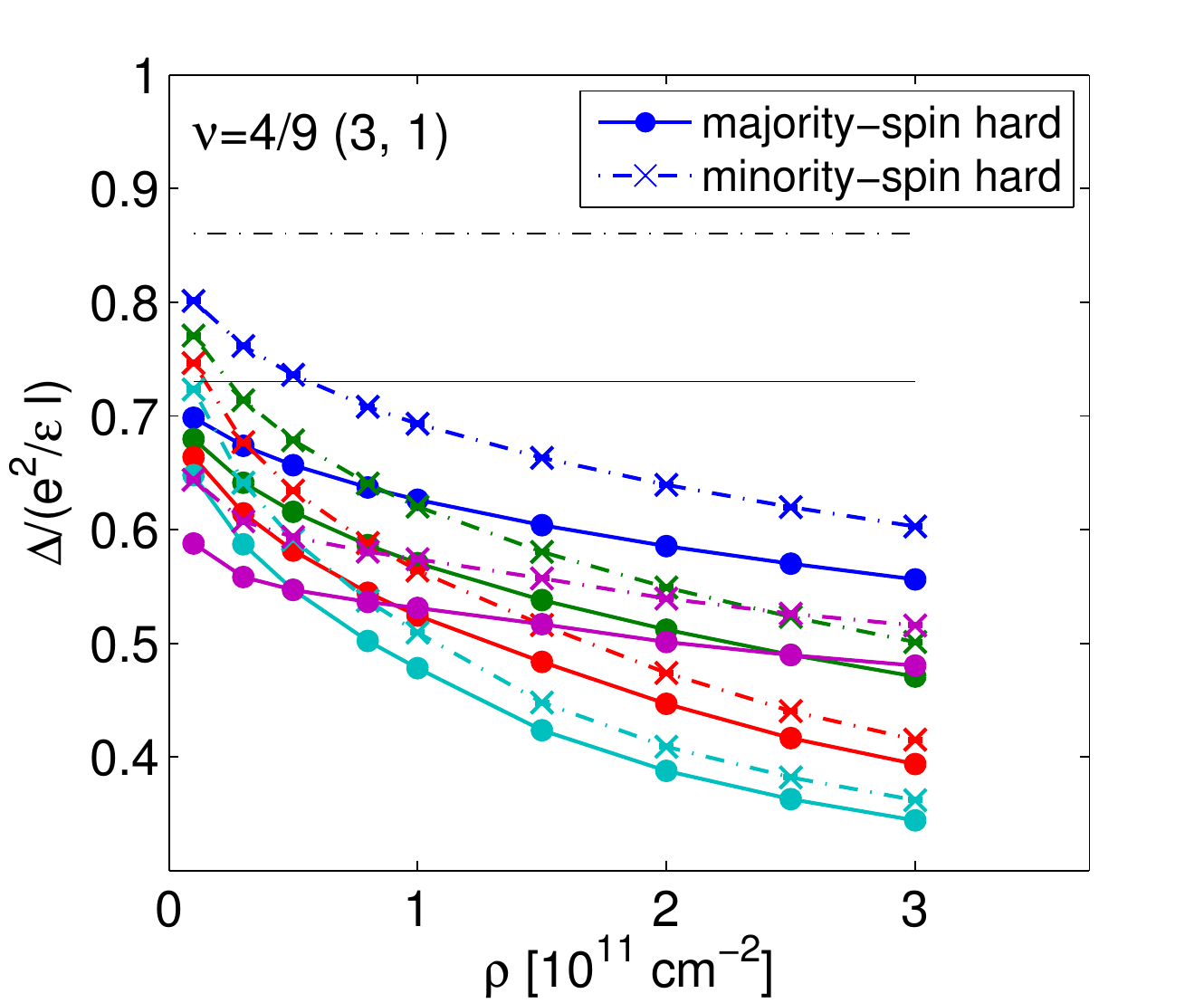}}\\
\vspace{-3mm}
\caption{The electron-hole interaction $E_{\rm e-h}$ (left panel) and bare gap $\Delta = E_{\rm e} + E_{\rm h} - 2E_{\rm gs}$ (middle and right panels) for partially polarized states at $\nu = 4/9$ labeled by $(3, 1)$. The tunneling electron can have either the majority spin or the minority spin. The right panel shows that the hard gap for a tunneling electron with minority spin is higher than that for an electron with majority spin. 
}
\label{Eeh2}
\end{figure*}

\begin{figure*}
\resizebox{0.335\textwidth}{!}{\includegraphics{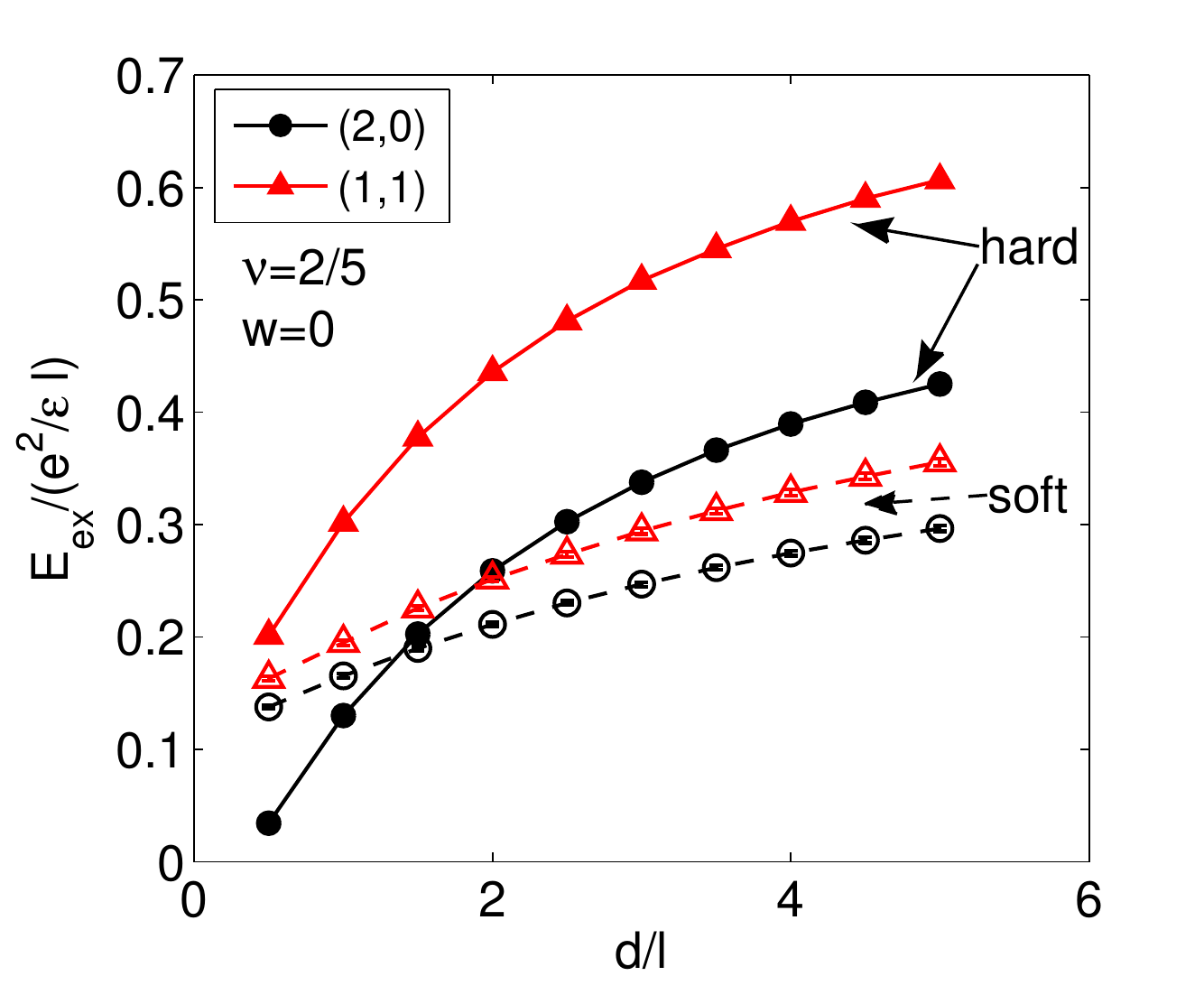}}
\hspace{-3mm}
\resizebox{0.335\textwidth}{!}{\includegraphics{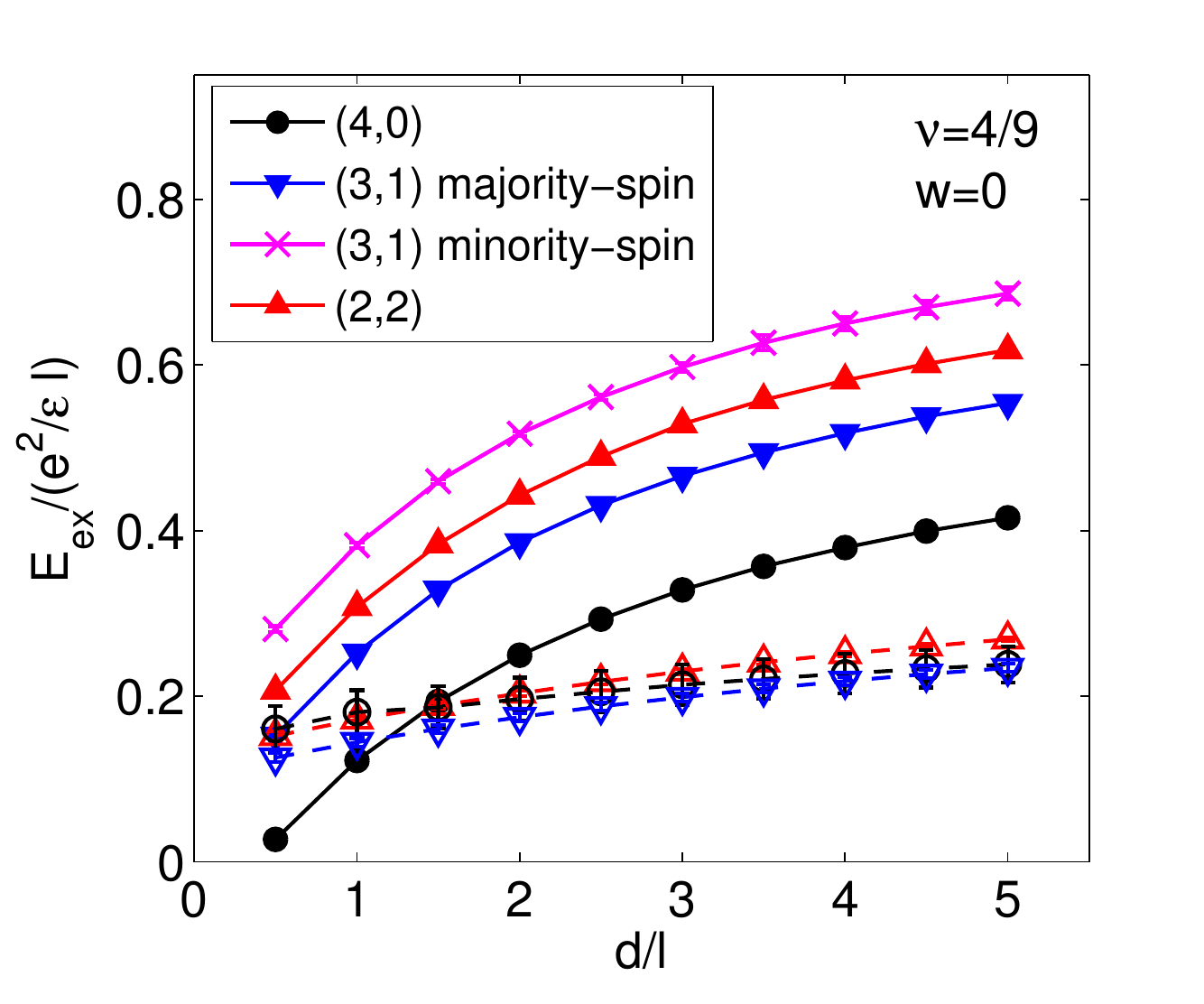}}
\hspace{-3mm}
\resizebox{0.335\textwidth}{!}{\includegraphics{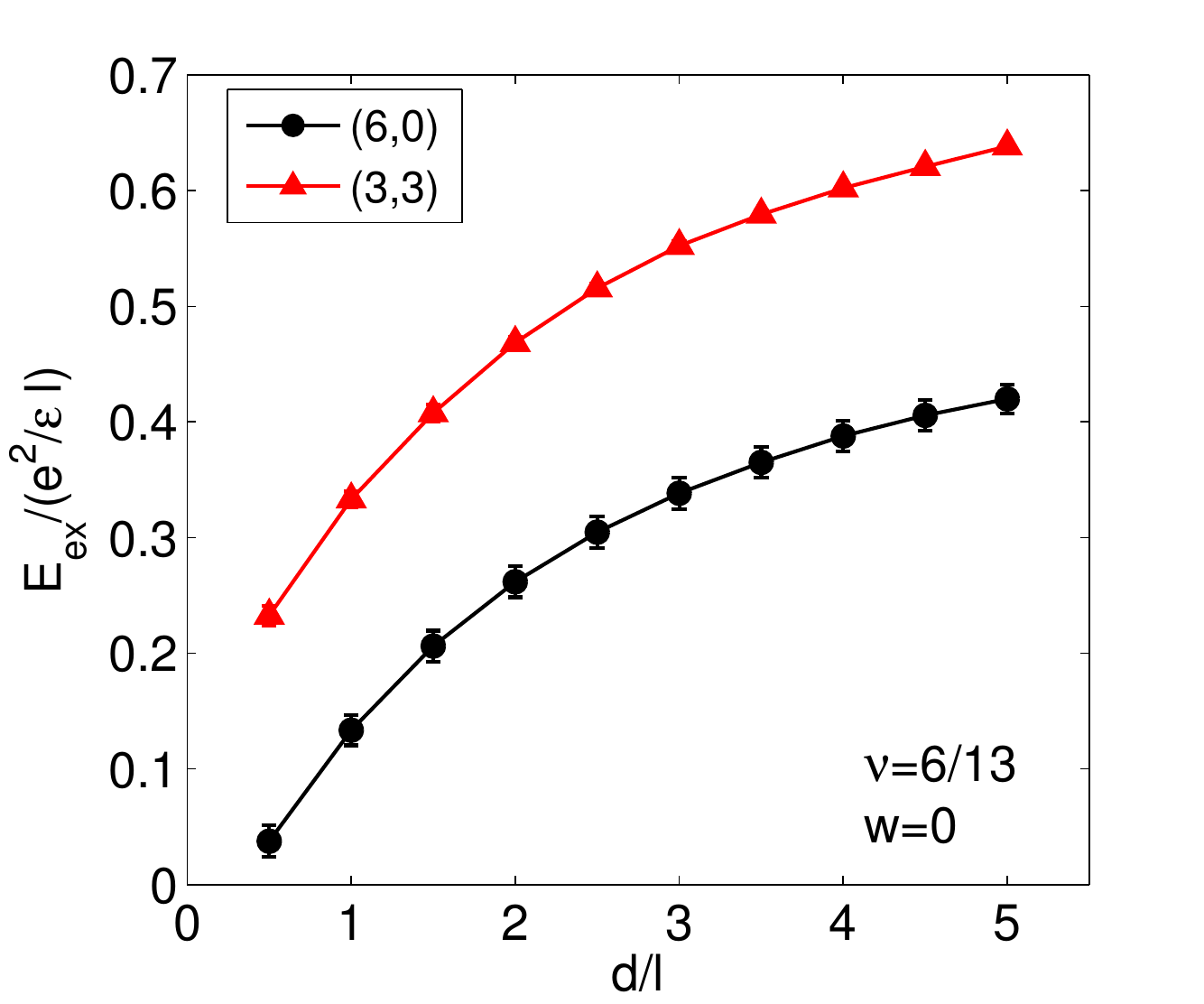}}\\
\vspace{-3mm}
\caption{The exciton energy $E_{\rm ex} = E_{\rm e-h} + \Delta$ as a function of $d$ for ideal 2D systems ($w=0$) for the hard exciton (solid symbols with solid lines) and the soft exciton (empty symbols with dashed lines). Different filling factors and spin polarizations are indicated on the figures.
}
\label{Eexw0}
\end{figure*}

\begin{figure*}
\resizebox{0.27\textwidth}{!}{\includegraphics{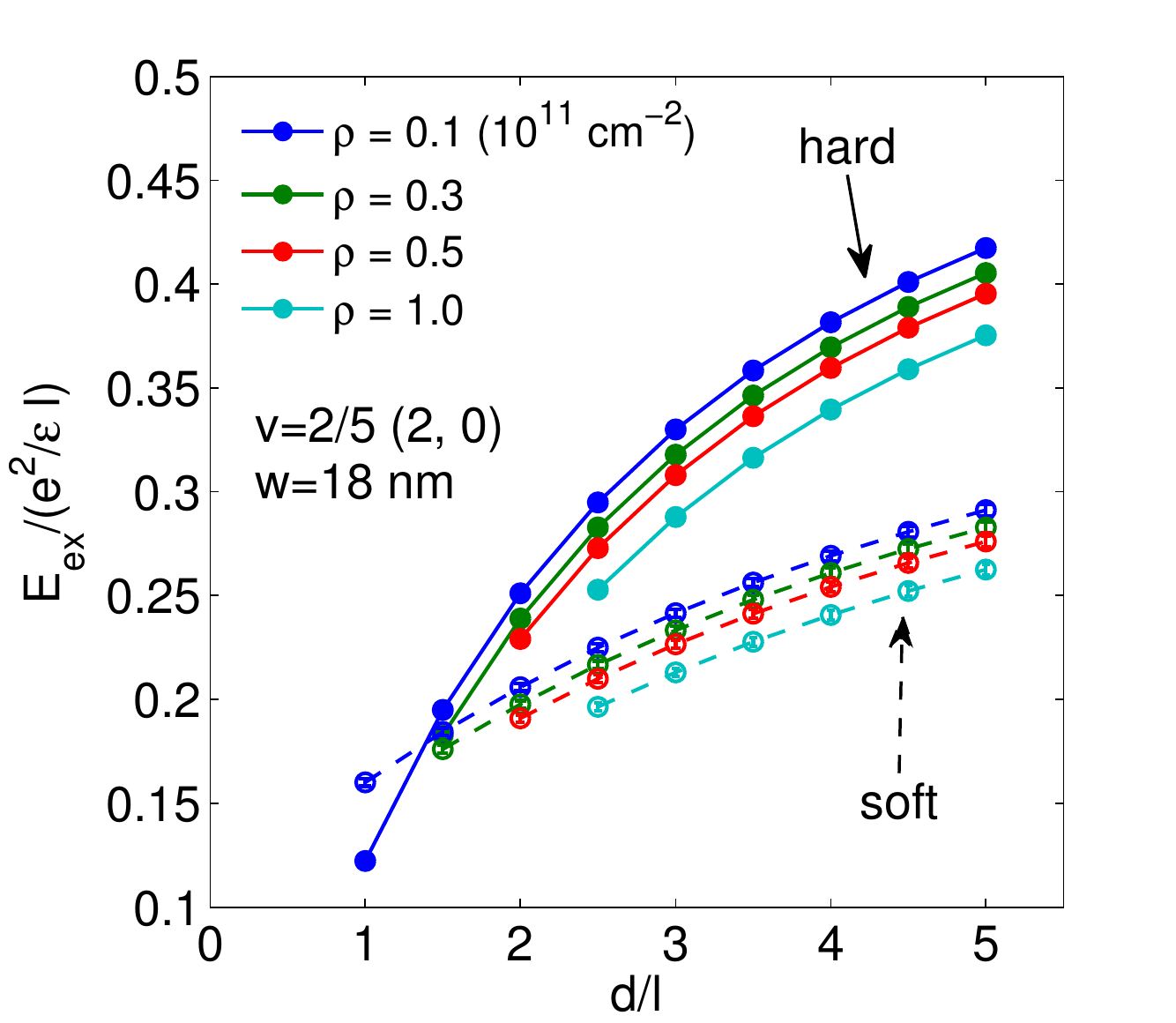}}
\hspace{-4mm}
\resizebox{0.248\textwidth}{!}{\includegraphics{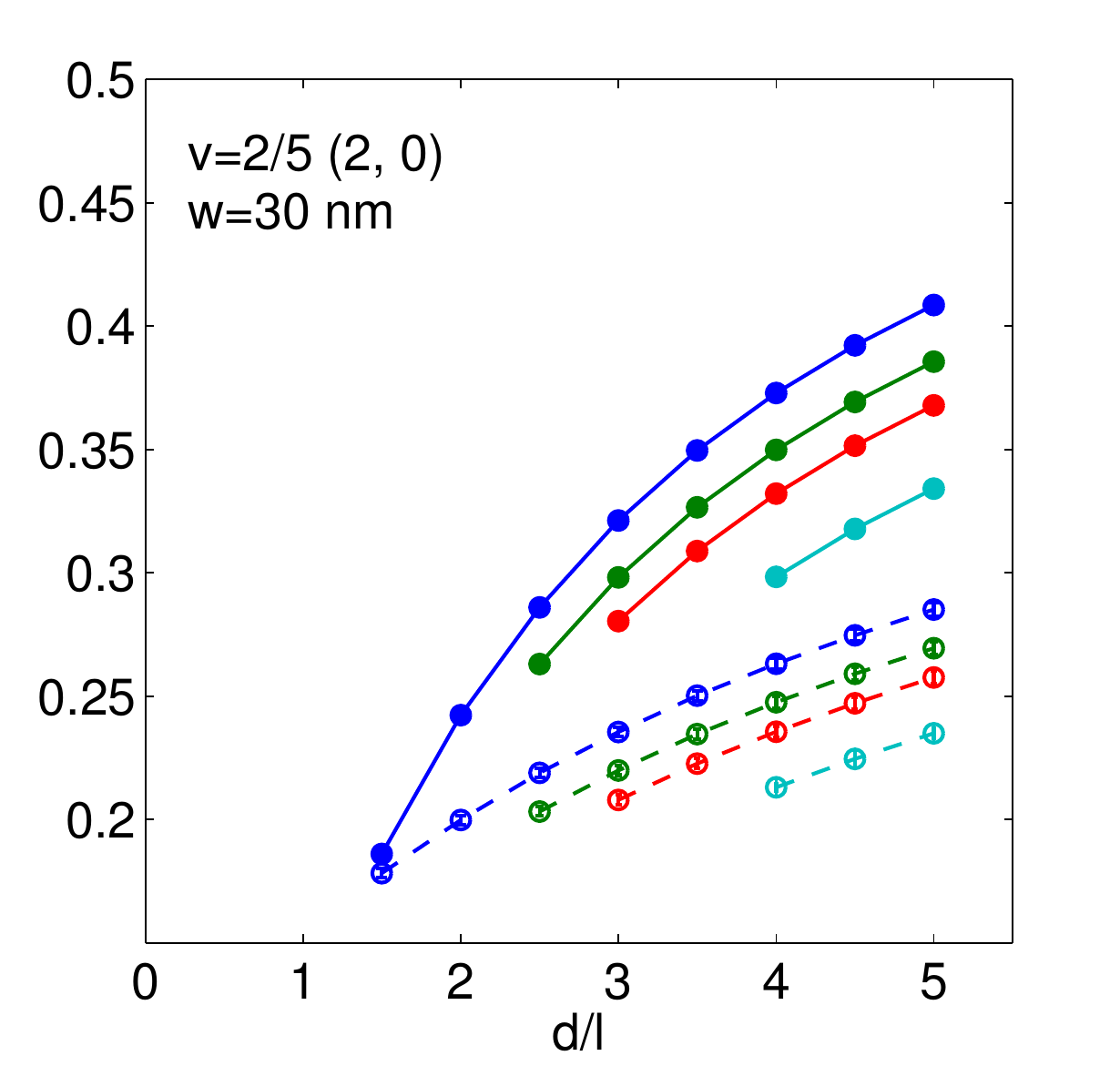}}
\hspace{-4mm}
\resizebox{0.248\textwidth}{!}{\includegraphics{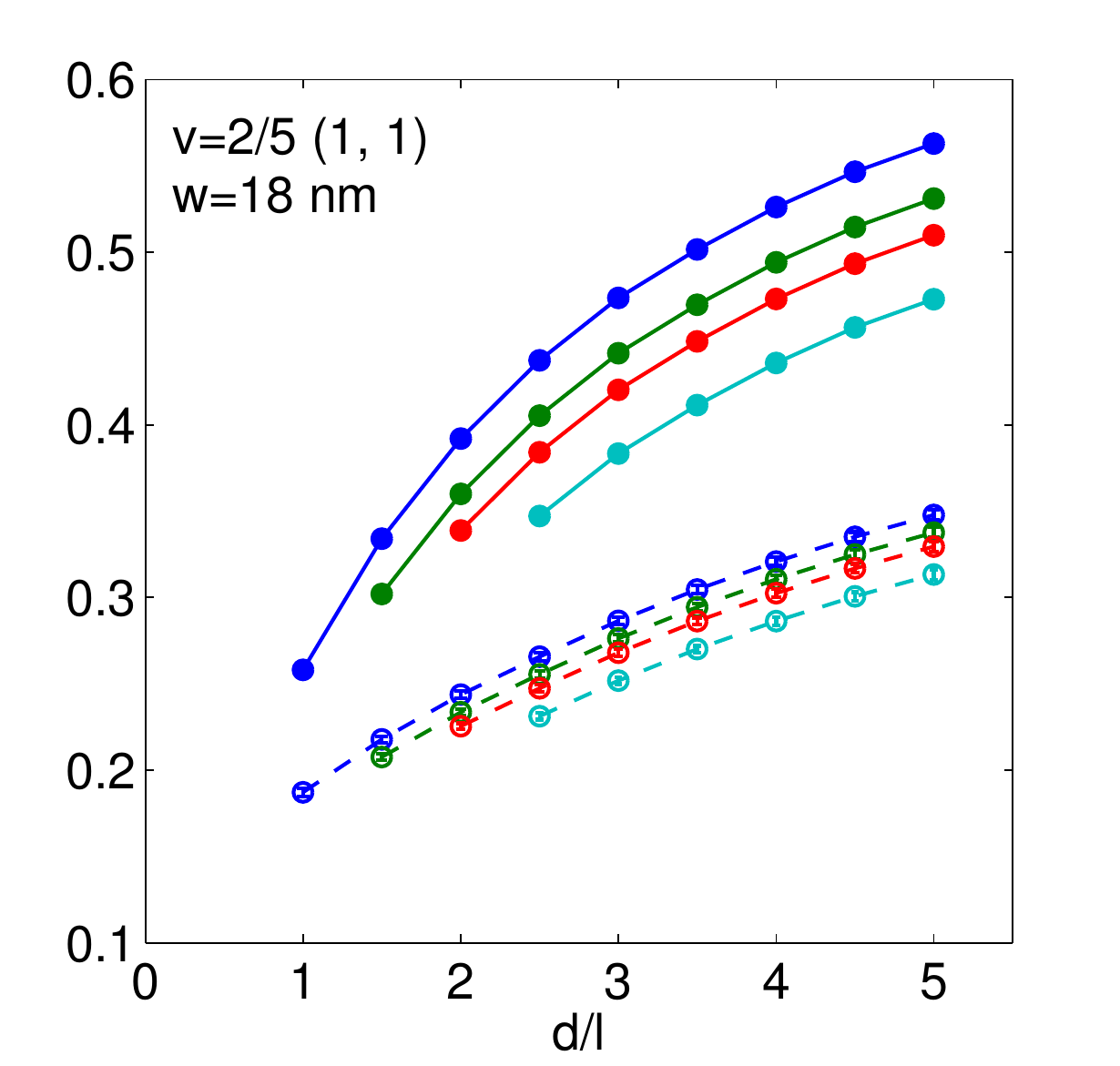}}
\hspace{-4mm}
\resizebox{0.248\textwidth}{!}{\includegraphics{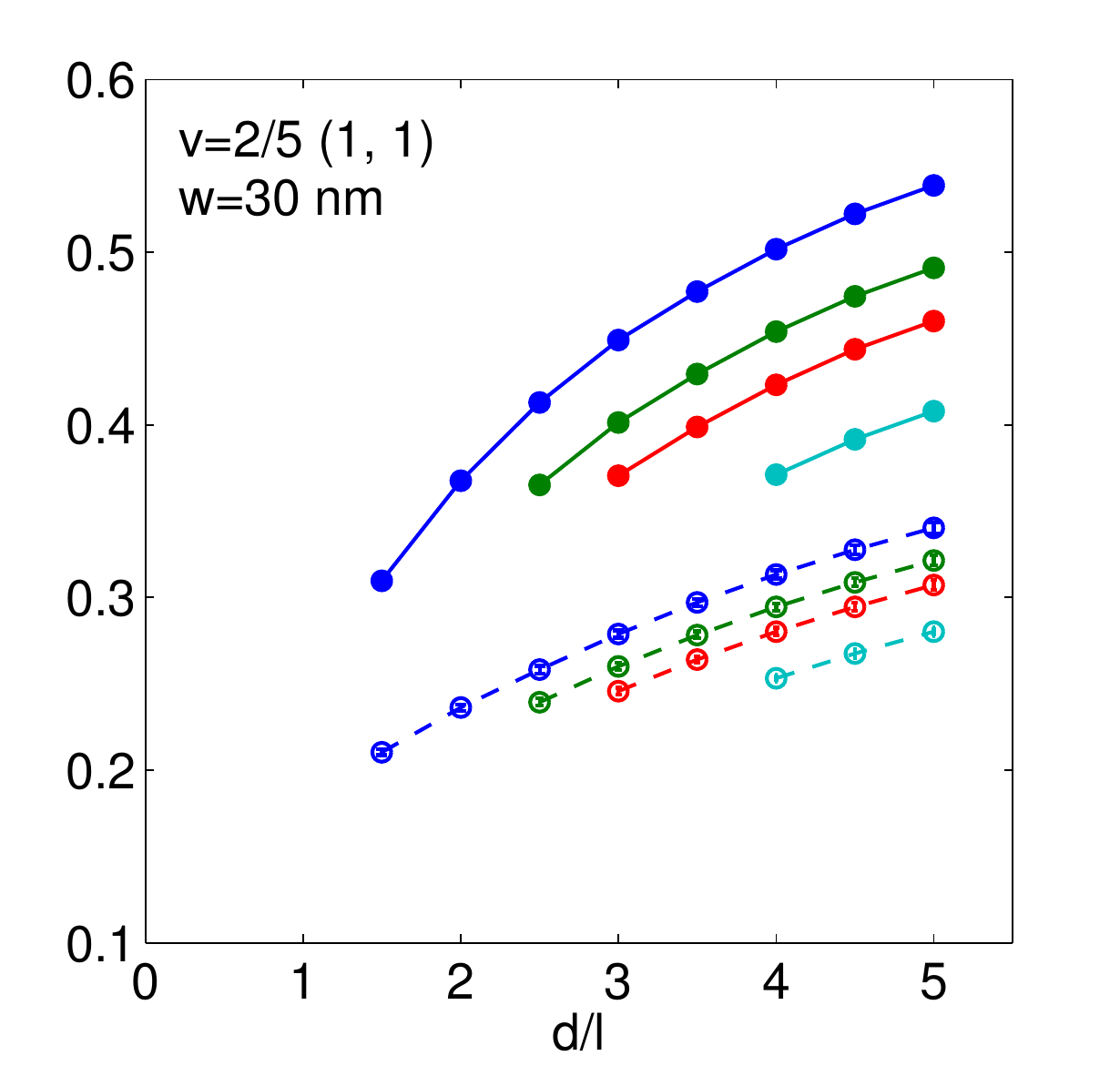}}\\
\vspace{-3mm}
\caption{The exciton energy $E_{\rm ex}$ for quantum wells of width $w=18$ and $30$ nm at different densities are shown for fully polarized and spin singlet states at $\nu = 2/5$. Energies are shown for both the hard exciton (solid symbols with solid lines) and the soft exciton (empty symbols with dashed lines). 
}
\label{Eexv25}
\end{figure*}

\begin{figure*}
\resizebox{0.268\textwidth}{!}{\includegraphics{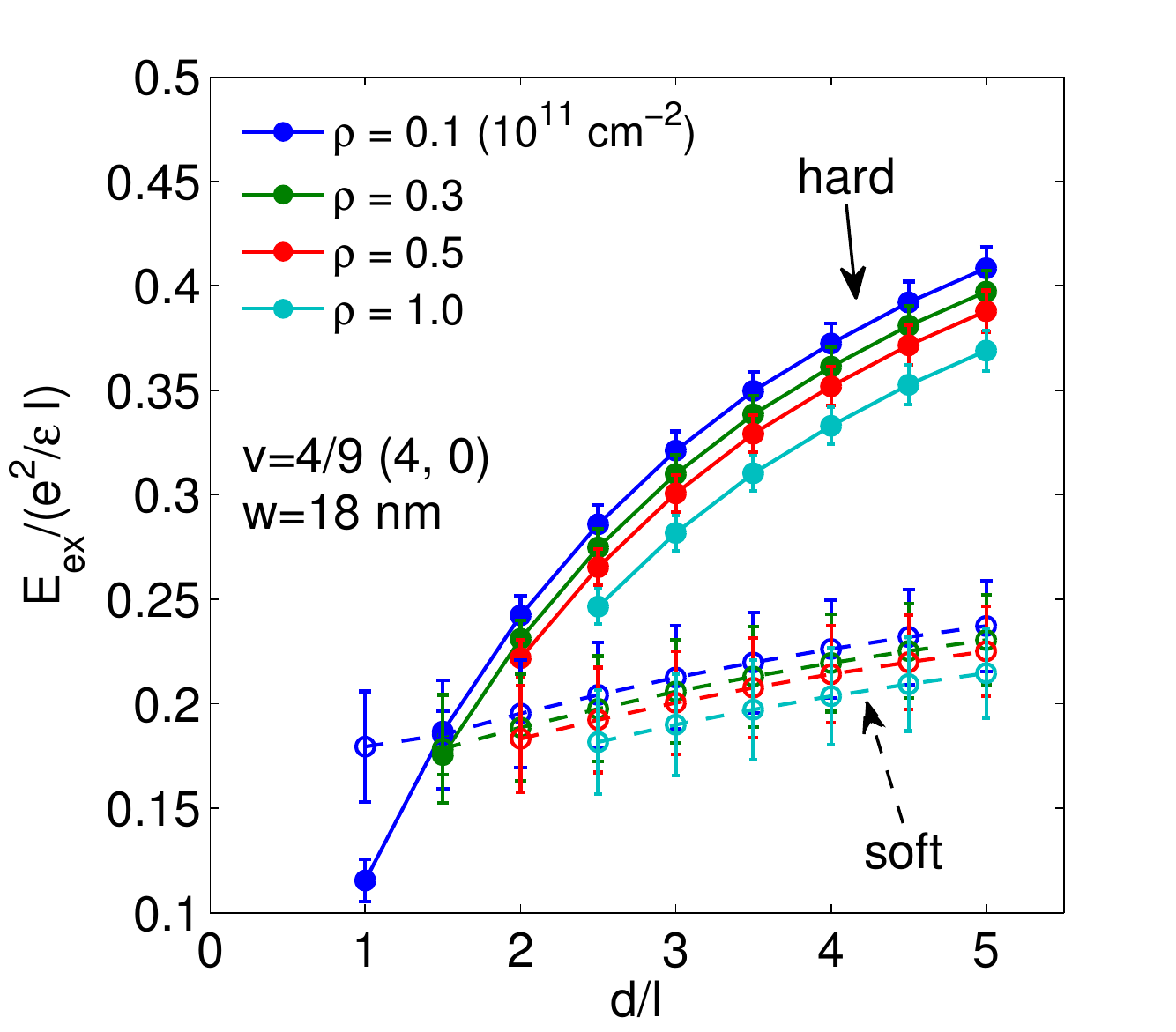}}
\hspace{-4mm}
\resizebox{0.248\textwidth}{!}{\includegraphics{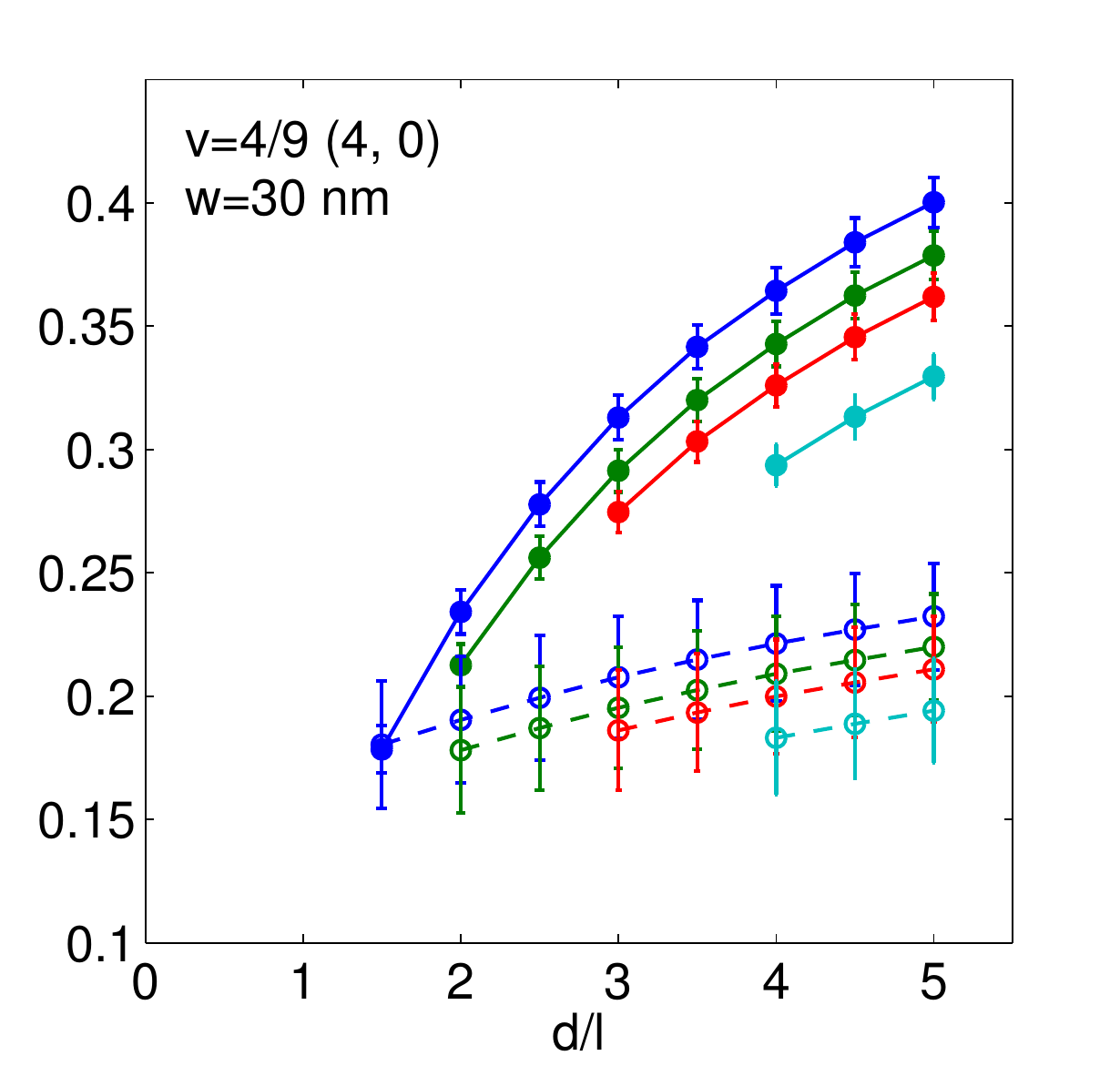}}
\hspace{-4mm}
\resizebox{0.248\textwidth}{!}{\includegraphics{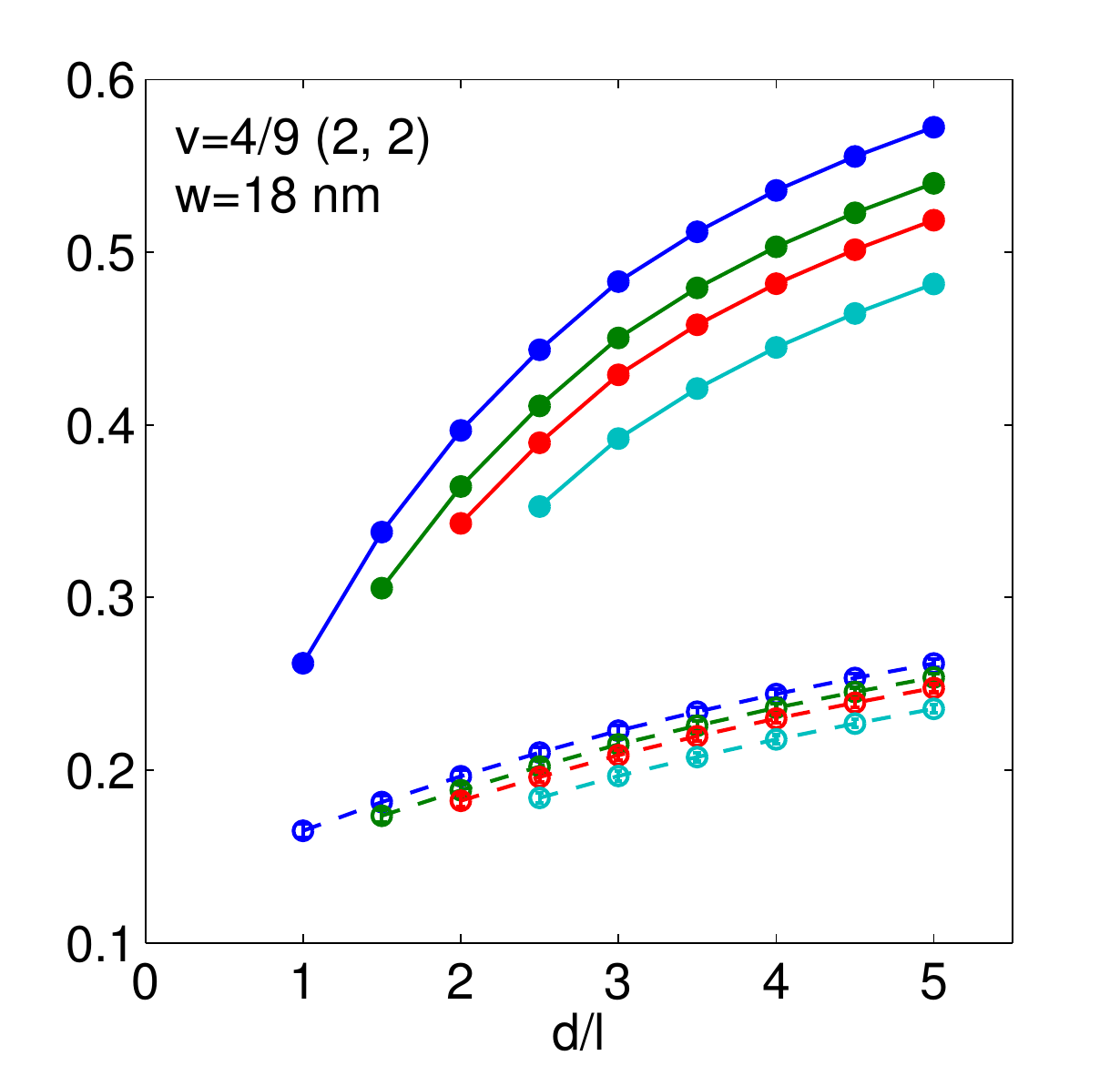}}
\hspace{-4mm}
\resizebox{0.248\textwidth}{!}{\includegraphics{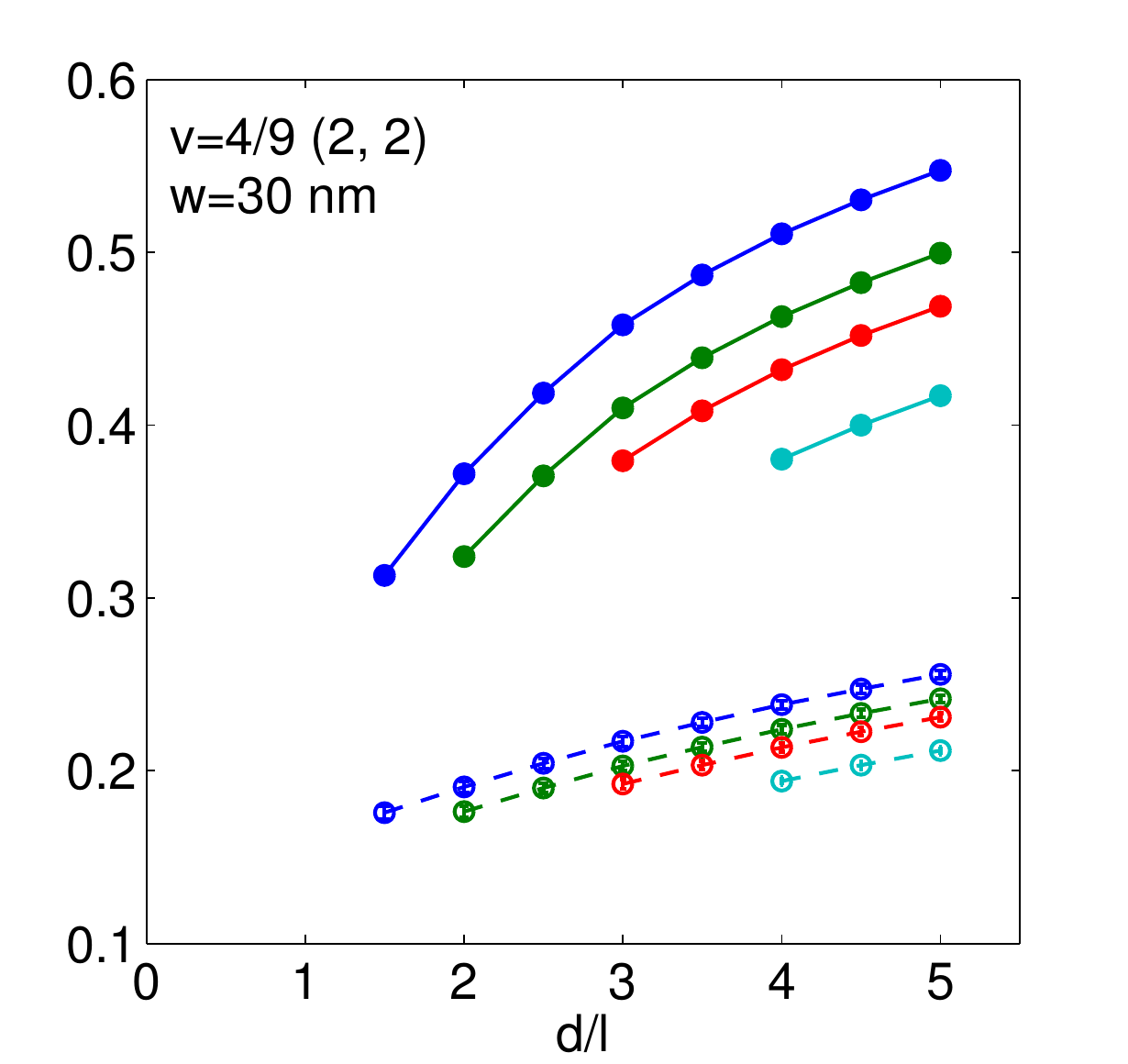}}\\
\hspace{0.5mm}
\resizebox{0.27\textwidth}{!}{\includegraphics{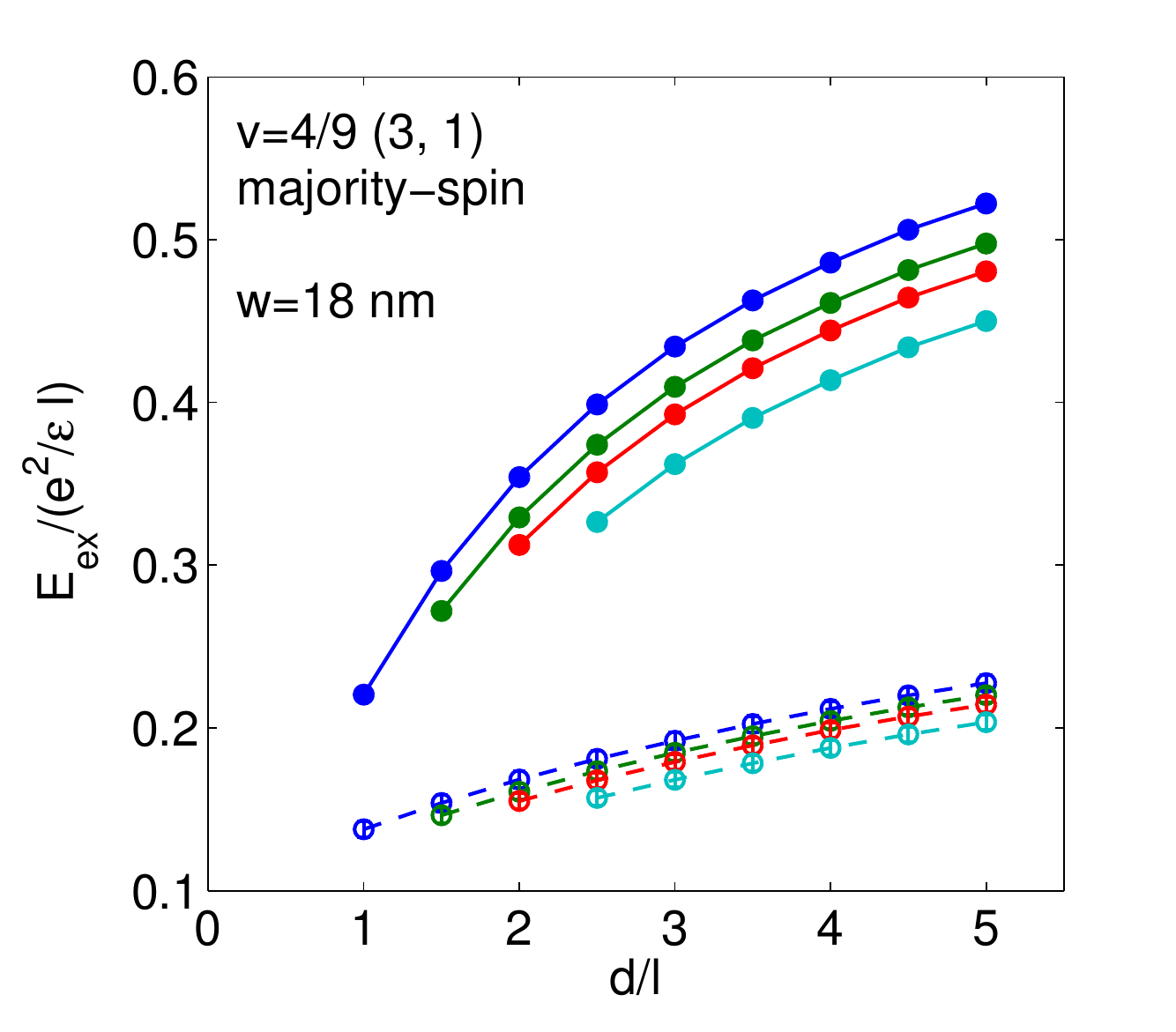}}
\hspace{-4mm}
\resizebox{0.248\textwidth}{!}{\includegraphics{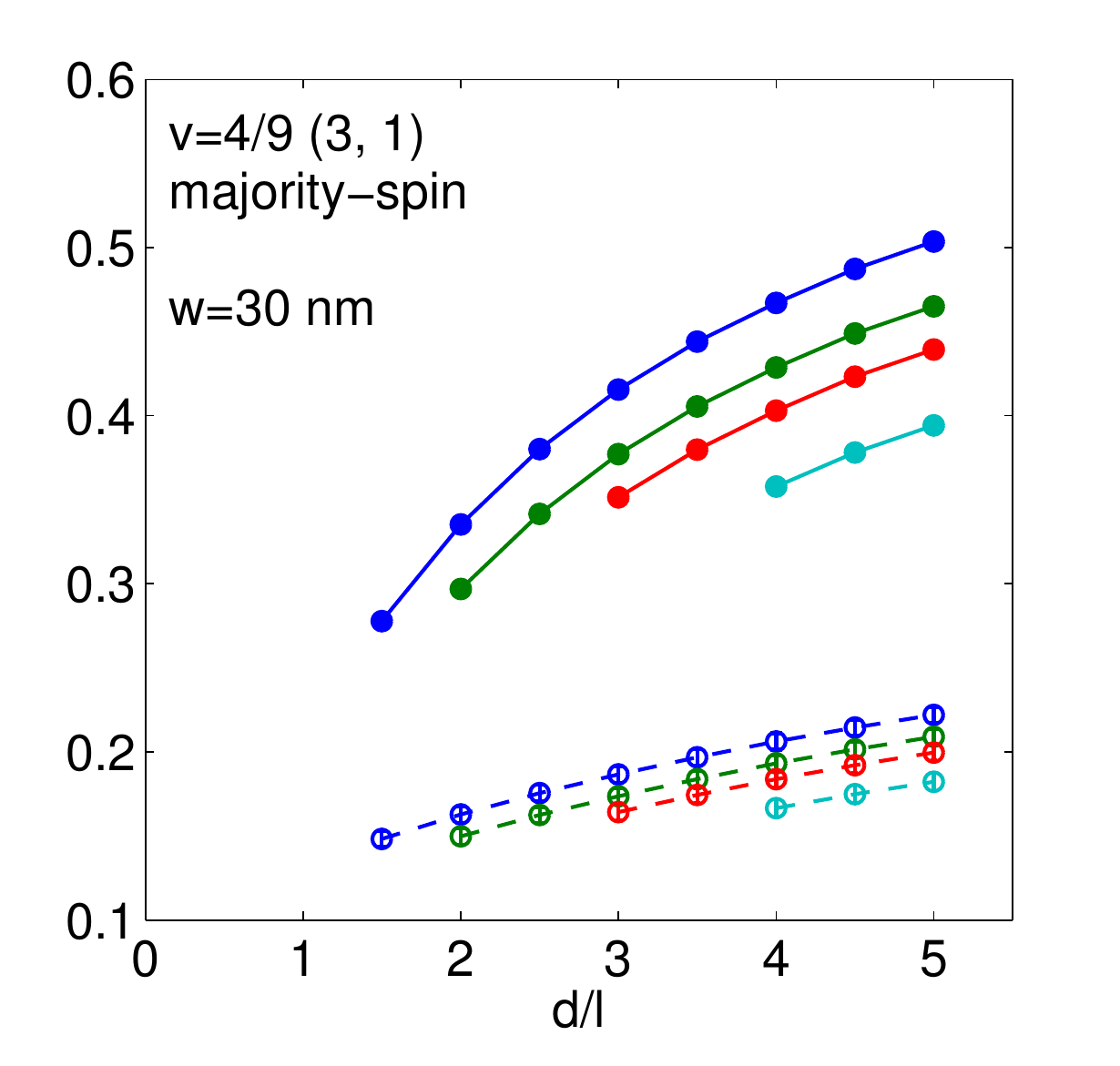}}
\hspace{-4mm}
\resizebox{0.248\textwidth}{!}{\includegraphics{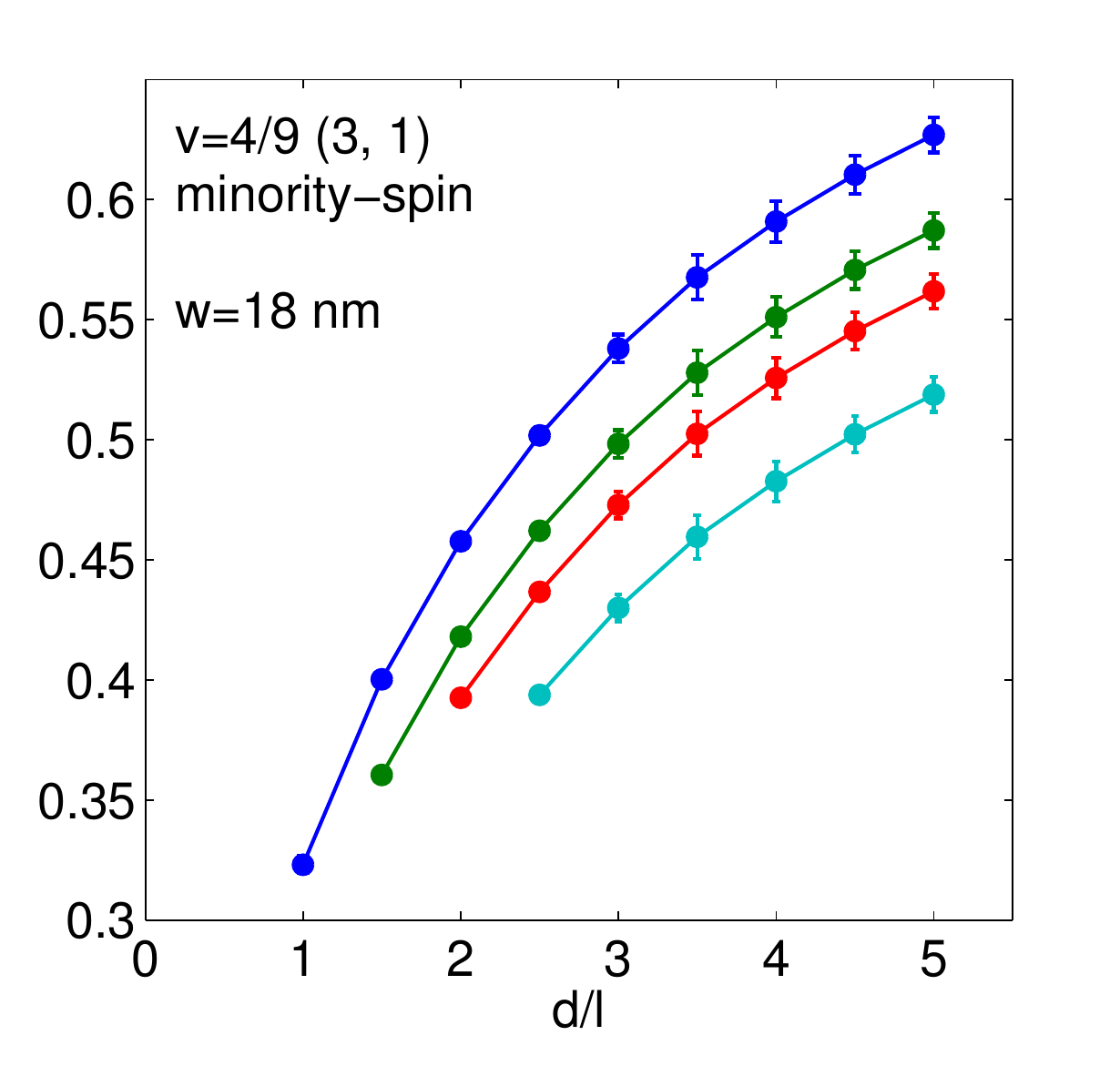}}
\hspace{-4mm}
\resizebox{0.248\textwidth}{!}{\includegraphics{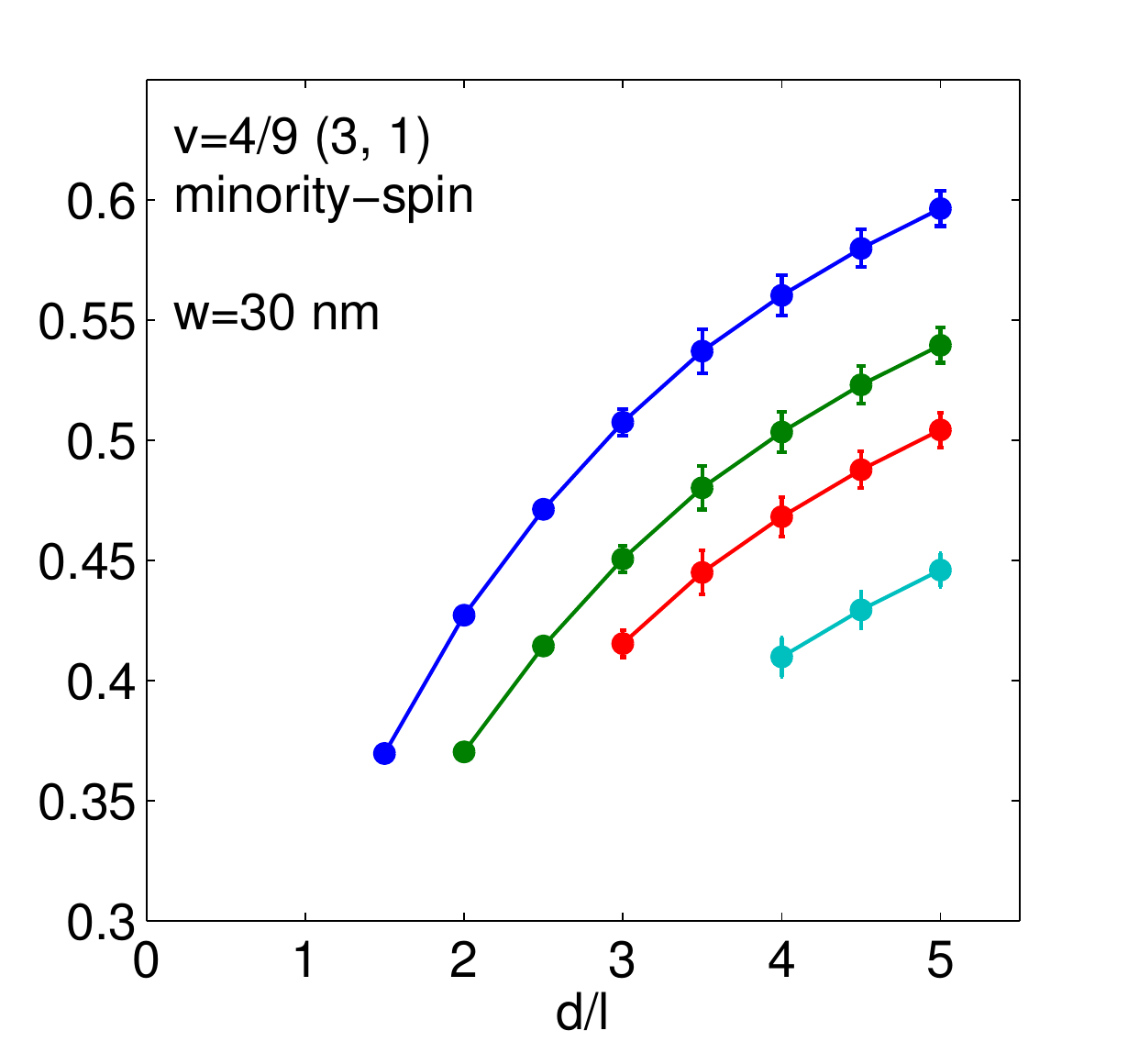}}
\vspace{-3mm}
\caption{The exciton energy $E_{\rm ex}$ for quantum wells of width $w=18$ and $30$ nm at different densities are shown for differently spin polarized states at $\nu = 4/9$. Energies are shown for both the hard exciton (solid symbols with solid lines) and the soft exciton (empty symbols with dashed lines). 
}
\label{Eexv49}
\end{figure*}

Fig.~\ref{Eeh1} shows the energies $E_{\rm e-h}$ and $\Delta=E_{\rm e}+E_{\rm h}-2E_{\rm gs}$ for both hard and the soft excitons for fully spin polarized and spin singlet FQH states at $\nu=2/5$, 4/9 and 6/13. The results are shown for quantum well widths of $w=$0, 18, 30, 40, 50 nm and for a heterojunction (HJ) as a function of density. (The width $w=18$nm is chosen to match the width of the quantum well in the experiment of Ref.~\onlinecite{Eisenstein16}.)  Fig.~\ref{Eeh2} shows the same energies for the partially polarized $(3,1)$ state at $\nu=4/9$. (We do not consider the soft exciton when the tunneling electron is of minority spin species.)
The total exciton energy $E_{\rm ex}$ for the ideal $w=0$ system is shown in Fig.~\ref{Eexw0} for several spin polarizations at $\nu=2/5$, 4/9 and 6/13 as a function of  $d/l$. (For $\nu=6/13$, we do not consider partially polarized states or the soft exciton.)  The results do not depend on density in this case. 
Figs.~\ref{Eexv25}-\ref{Eexv49} show $E_{\rm ex}$ for quantum wells of widths $w=$ 18 and 30 nm as a function of $d/l$ for several densities and spin polarizations at $\nu = 2/5$ and $4/9$.  (The data points with $w > d$ are unphysical and therefore not shown.)  For all cases, the numbers shown are obtained by a careful thermodynamic extrapolation of finite system results. For $E_{\rm e-h}$, we find that the finite width makes a negligible correction, and therefore we use the zero width results.

The following facts are evident from these results.

\begin{figure}
\resizebox{0.43\textwidth}{!}{\includegraphics{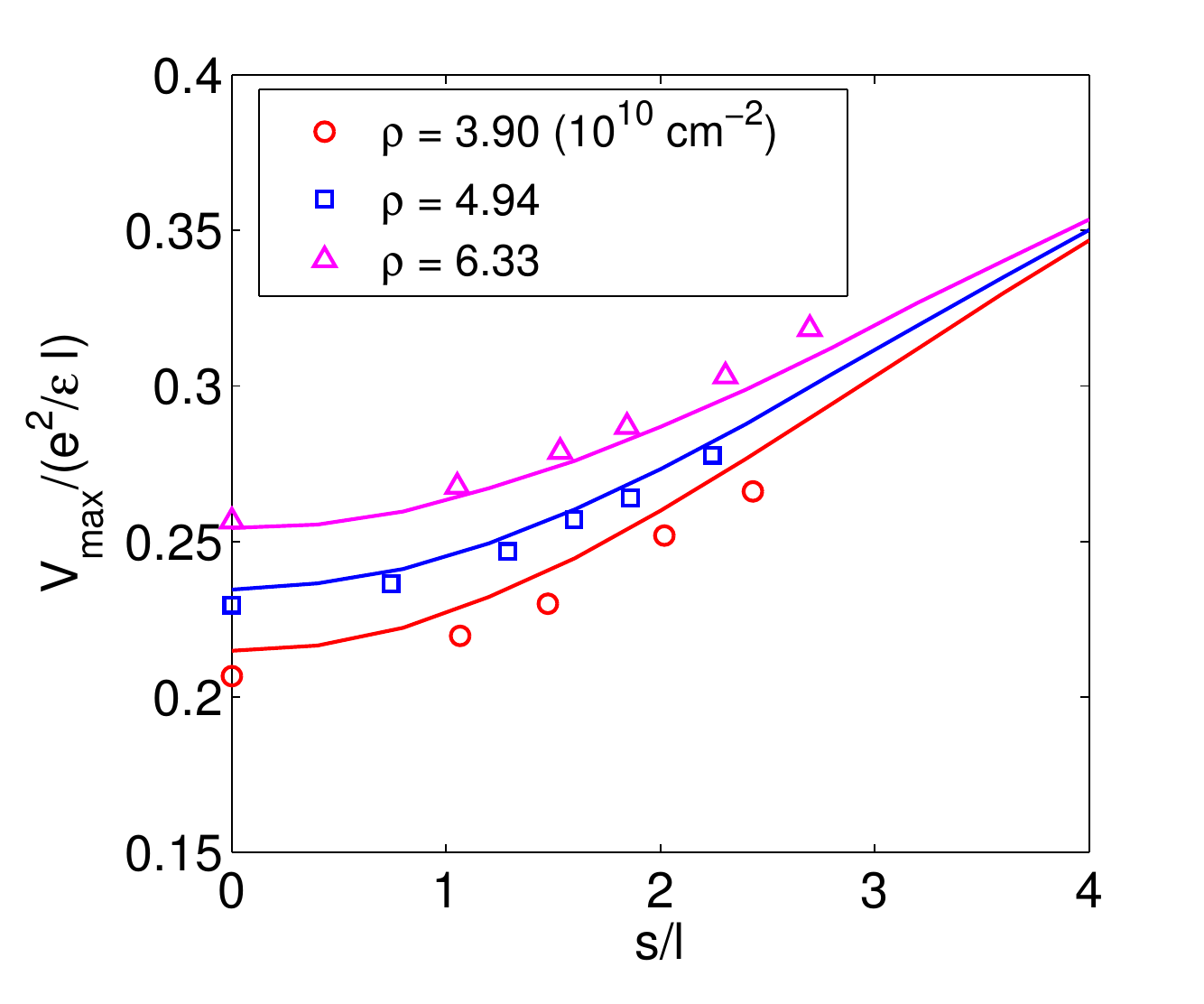}} 
\caption{Comparison of the experimentally observed $V_{\text{max}}$ in unit of $e^{2}/\epsilon l$ (symbols) with the theoretical energy of the hard exciton (solid lines) for three densities. The experiment of Eisenstein {\em et al.}\cite{Eisenstein16} is performed at $\nu=1/2$ in a system of two layers with quantum well widths of $w=18$ nm separated by a center-to-center distance of $d=28$ nm. The theoretical values are evaluated at $\nu=4/9$ but have been shown to be largely independent of the filling factor. The evaluation of the exciton-interaction energy $E_{\rm e-h}$ includes the effects of both the finite thickness of the quantum well  and a lateral shift of the tunneling electron due to the Lorentz force associated with the parallel magnetic field $B_{\parallel}$. The shift distance ($x$-axis) is equal to $s = q\ell^{2} = dB_{||}/B_{\perp}$ where $q = edB_{\parallel}/\hbar$ is the ``momentum boost'' acquired due to $B_{\parallel}$. The figure thus gives the ``dispersion" of the interlayer exciton as a function of the momentum $ql$; for large $ql$ the energy will saturate at $\Delta$. Fig.~\ref{expcompare0} shows the same results converted to $V_{\rm max}$ (mV) vs. $B_{\rm tot}$ (T).}
\label{expcompare}
\end{figure}

$\bullet$ The excitonic attraction $E_{\rm e-h}$ is substantial. This energy is given by $e^2/\epsilon d$ for $d$ large compared to the sizes of the electron and hole density profiles. However, because the interlayer separation is on the order of the electron / hole size, the energy $E_{\rm e-h}$ does not have a simple dependence on $e^2/\epsilon d$ and must be obtained from a detailed calculation that requires the knowledge of the density profiles of the electron and the hole participating in the exciton. Furthermore, the magnitude of $E_{\rm e-h}$ is much larger for the hard exciton than for the soft exciton, and brings the energy of the hard exciton below that of the soft exciton for relatively small values of $d$.

$\bullet$ Our calculation gives a quantitative account of the dependence of $E_{\rm ex}$ on the quantum well width and the density. As one might expect, the energy $E_{\rm ex}/(e^2/\epsilon l)$ goes down with increasing density and increasing width. 

$\bullet$ For the fully spin polarized state, the energy $E_{\rm ex}$ for a hard exciton is largely insensitive to the filling factor as we go from 2/5 to 4/9 to 6/13. This is evident by comparing the hard exciton energies ($E_{\rm e-h}$ and $\Delta$) for fully spin polarized states at different filling factors in Fig. \ref{Eeh1} for both the ideal zero-width and finite-width systems. Such a behavior is consistent with early experiments \cite{Eisenstein09}, and represents certain universality between all states of composite fermions carrying two vortices. We therefore conclude that we can compare our results of hard excitons at $\nu=4/9$ with the experiments performed at $\nu=1/2$. This is fortunate because while the 1/2 CF Fermi sea is convenient for experiments (because the tunneling for incompressible states is more strongly suppressed), the incompressible states are friendlier to theoretical calculations.

$\bullet$ The application of an in-plane magnetic field $B_{\parallel}$ causes the electron and the hole to be laterally offset by an amount that depends on the parallel and the perpendicular components of the magnetic field. One therefore expects that the magnitude of $E_{\rm e-h}$ decreases, and thus $E_{\rm ex}$ increases with increasing $B_{\parallel}$.

$\bullet$ In Fig.~\ref{expcompare0} we plot the energy of the hard exciton as a function of the total magnetic field (under the application of a parallel magnetic field) for parameters of the experiment of Ref.~\onlinecite{Eisenstein16} along with the experimentally observed $V_{\rm max}$. We consider the agreement to be excellent. 
In particular, the behavior as a function of B$_{\rm tot}$ is very accurately captured by theory. The excellent agreement with experiments strongly supports our assignment of $V_{\rm max}$ with the hard interlayer exciton. Same quantities are shown in Fig.~\ref{expcompare} with a different x-axis and different units. 

$\bullet$ We find that for FQH states, the energy of the hard exciton increases substantially as we reduce the spin polarization of the background incompressible state. The physical origin of this increase is clear: for partially spin polarized states the added electron does not avoid electrons of the opposite spin, thus resulting in a larger Coulomb energy. This prediction can in principle be experimentally tested by choosing parameters where spin phase transitions occur by application of a parallel field.

$\bullet$ As discussed in Ref.~\onlinecite{Eisenstein16} and in the next section, the CF Fermi sea is very likely not fully spin polarized in the entire range of $B_{\rm tot}$ shown in Fig.~\ref{expcompare0}, and comparison with our results obtained for fully spin polarized states may be questioned. However, even in the region where the CF Fermi sea is not fully polarized, it is almost fully polarized. To give a quantitative estimate, taking a model that assumes that composite fermions are noninteracting, the fraction of reversed spin, given by $0.5(1-E_{\rm Z}/E^{\rm crit}_{\rm Z})$, is less than $8\%$ even at the lowest Zeeman energies in the experiments of Ref.~\onlinecite{Eisenstein16}. This confirms that our calculation assuming a fully spin polarized Fermi sea remains a very good approximation.

$\bullet$ For partially spin polarized states we predict a split peak in I-V plot with the two maxima corresponding to the energies of the excitons resulting from the tunneling of a spin-up and a spin-down electron. For the partially polarized $\nu=4/9$ state we find a difference of $\sim 0.1$ meV between the energies of the spin up and spin down excitons (for typical experimental parameters). This matches well with the splitting seen by Eisenstein {\em et al.} \cite{Eisenstein09} at $\nu=2-4/9$. However, they also see strong splittings at $\nu=2-1/3$ and $2-2/3$, where we predict no splitting. We therefore refrain from assigning the double peak structure in terms of spin up and spin down excitons.

$\bullet$ We cannot identify any structure in experimental data that may be attributed to the soft exciton. This is not surprising, in view of our above discussion that the tunneling amplitude of the soft electron, which is a strongly correlated collective object, into a soft hole, also a strongly correlated collective object, is negligible. In particular, Fig.~\ref{expcompare0} demonstrates that the soft exciton is not relevant to the tunneling at $V_{\rm max}$.

$\bullet$ We have assumed in our discussion that no interlayer correlations are present in the ground state, i.e., the state in each layer is not affected by the other layer. FQH states in which interlayer coherence plays a crucial role can occur at $\nu=n/(2n+1)$ \cite{Halperin83,Scarola01b} as well as at $\nu=1/2$ \cite{Halperin83} for relatively small values of $d/l$ (which depends on the density and the quantum well width). The level of agreement between our theory and experiments suggests that the interlayer correlations do not substantially modify the state for the experimental parameters. 

\begin{figure}
\resizebox{0.35\textwidth}{!}{\includegraphics{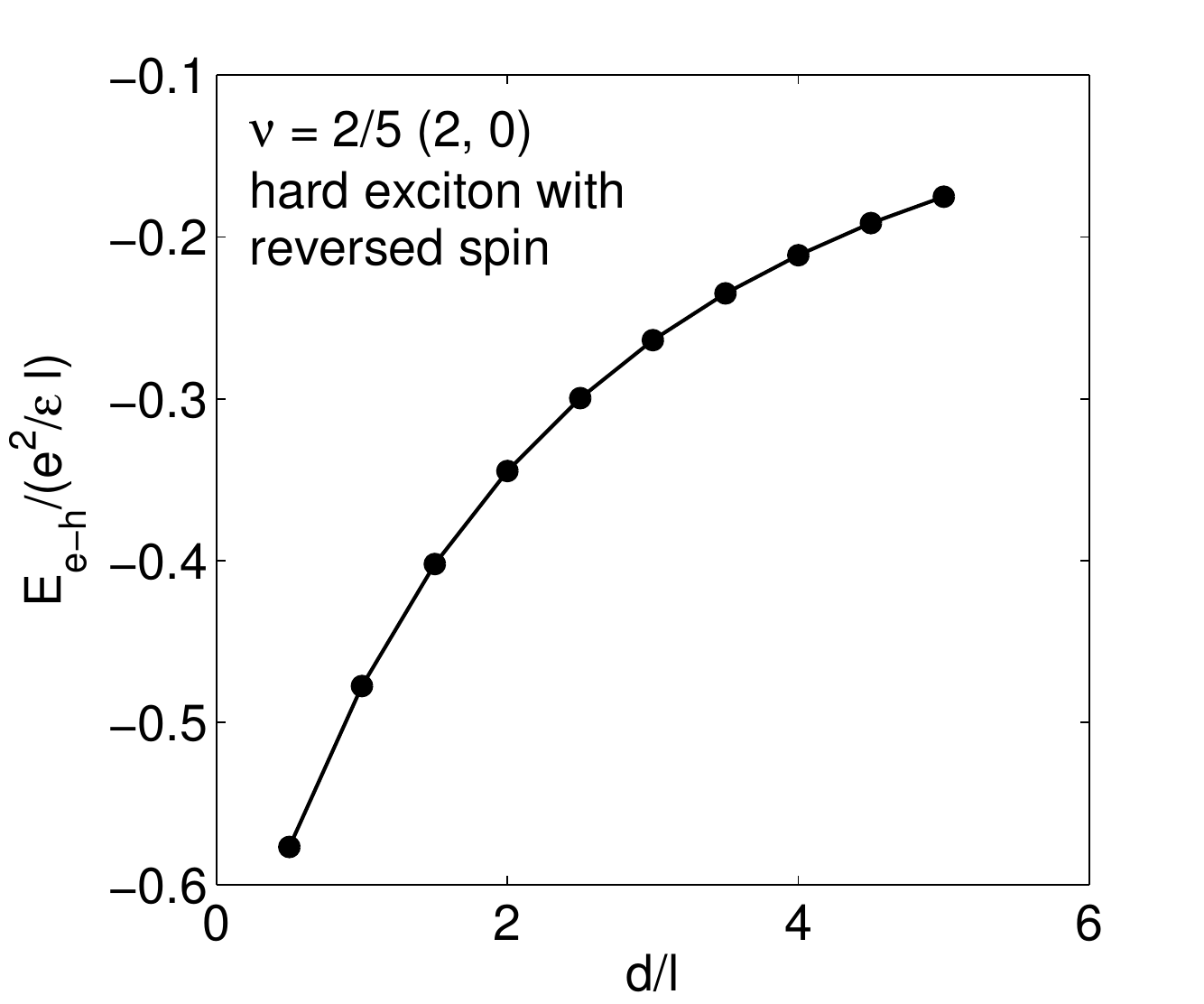}} \\
\hspace{1.6mm}
\resizebox{0.34\textwidth}{!}{\includegraphics{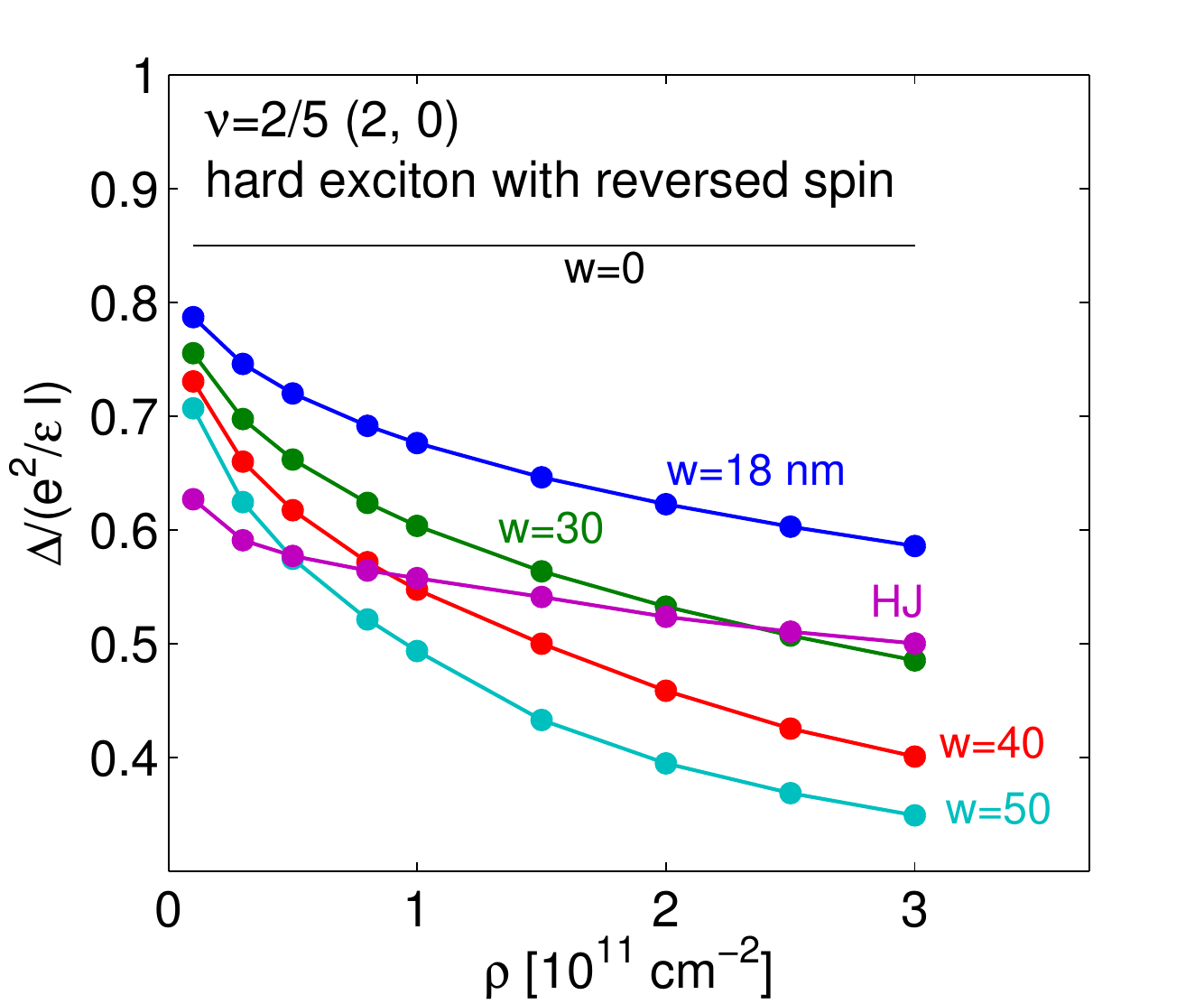}} 
\caption{The excitonic interaction $E_{\text{e-h}}$ (upper panel) and bare gap $\Delta$ (lower panel) of the spin-reversed hard exciton for the fully spin polarized 2/5 $(2, 0)$ state. The bare gap $\Delta$ of the spin-reversed hard exciton is higher than that for the spin-conserving hard exciton (Fig. \ref{Eeh1}) while the interaction energies $E_{\text{e-h}}$ for the two are about the same.}
\label{v25sf}
\end{figure}

$\bullet$ One may ask if lower energy excitons can be obtained if the electron spin is not conserved during the tunneling process. Such processes are in principle possible, because while the spin orbit coupling is very small in the usual GaAs systems, it is not zero. We show in Fig. \ref{v25sf} results for the hard exciton for the fully spin polarized 2/5 state where the added electron has a reversed spin. We find that the energy of the added spin-reversed hard electron is actually higher than that of the spin-conserving hard electron, leading to an overall increase in the exciton energy. The origin for the increase is the same as that discussed above in the context of partially spin polarized states, namely that the spin reversed electron does not Pauli-avoid the other electrons, thus resulting in a higher interaction energy.

\section{Spin polarization transition for the CF Fermi sea}
\label{SecIV}

Eisenstein {\em et al.}\cite{Eisenstein16} have measured the voltage $V$ at the onset of tunneling as a function of an additional in-plane magnetic field, and find that the behavior changes qualitatively when the total magnetic field drops below some value. They identify it with  a transition in the spin polarization of the CF Fermi sea. An earlier calculation \cite{Park98} predicted a higher value than that observed experimentally, which has motivated us to revisit this issue.

The spin phase transitions of the FQH states and the CF Fermi sea have been extensively studied both experimentally \cite{Eisenstein89, Eisenstein90, Engel92, Du95,Kang97, Kukushkin99, Yeh99,Kukushkin00, Melinte00,Freytag01, Freytag02,Tracy07,Tiemann12,Feldman13,Liu14} and theoretically\cite{Park98,Balram15a,Zhang16}. The spin transition of the CF Fermi sea has also been studied in bilayer systems \cite{Eisenstein16,Finck10, Giudici08}. 
A recent theoretical work\cite{Zhang16} treated LL mixing by a fixed phase diffusion Monte Carlo method, and found that LL mixing has a relatively large correction on the critical Zeeman energies where spin polarization transitions take place. We shall skip here the technical details of the calculation, which can be found in Ref.~\onlinecite{Ortiz93,Melik-Alaverdian97,Zhang16}, and show here results for $\nu=1/2$.

\begin{figure}[t]
\resizebox{0.45\textwidth}{!}{\includegraphics{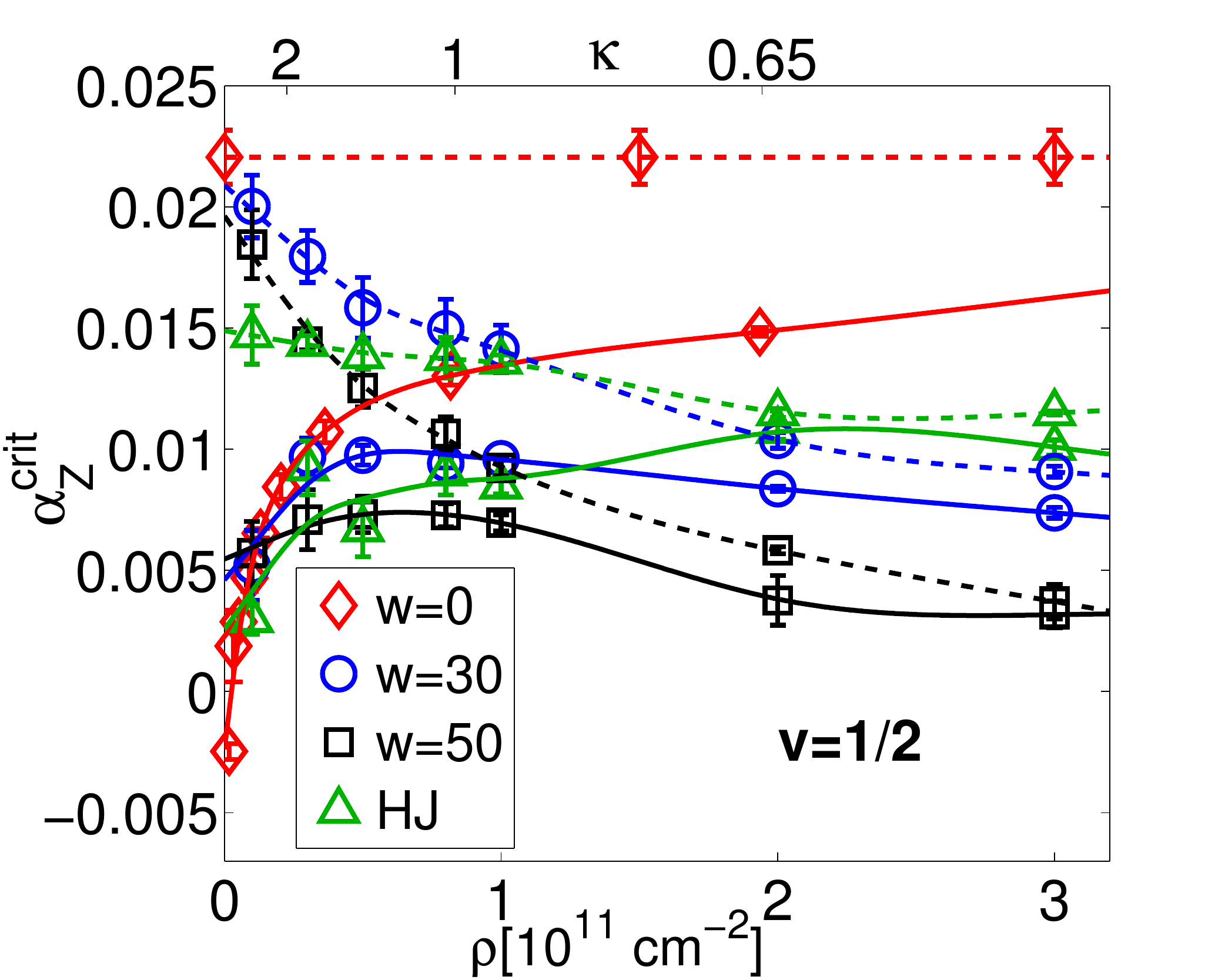}}
\vspace{-2mm}
\caption{The critical Zeeman energy $\alpha^{\rm crit}_{\rm Z}=E^{\rm crit}_{\rm Z}/(e^2/\epsilon l)$ above which the $\nu = 1/2$ CF Fermi sea is fully spin polarized.  The theoretical results are obtained from an extrapolation of the results for $\nu=n/(2n+1)$ given in Ref. \onlinecite{Zhang16} for an ideal 2D system ($w=0$), for quantum wells with width $w=30$, $50$ nm, and also for a heterojunction (HJ).  
The lower axis shows the electron density, whereas the LL mixing parameter $\kappa=(e^2/\epsilon l)/(\hbar \omega_c)$ is shown on top, assuming parameters appropriate for GaAs.
}
\label{spintransition}
\end{figure}

In Fig.~\ref{spintransition}, we show the calculated critical Zeeman energy measured in units of $e^2/\epsilon l$, i.e. $\alpha^{\rm crit}_{\rm Z}=E^{\rm crit}_{\rm Z}/(e^2/\epsilon l)$ above which the CF Fermi sea is fully spin polarized as a function of the quantum well width as well density, both indicated on the figure itself. The top axis shows the parameter $\kappa=(e^2/\epsilon l)/(\hbar \omega_c)$, where $\hbar \omega_c$ is the cyclotron energy. The horizontal dashed line at $\alpha^{\rm crit}_{\rm Z}=0.022$ is the theoretical result for an ideal 2D system with $w=0$ and no LL mixing \cite{Park98} The dashed lines include the effect of finite width but assume absence of LL mixing; these are obtained using a variational Monte Carlo (VMC) method. The solid line is calculated by a fixed phase diffusion Monte Carlo (DMC) method, and include the effect of both finite width and LL mixing. All of these results have been obtained by an extrapolation of the calculated $\alpha^{\rm crit}_{\rm Z}$ at the fractions $\nu=n/(2n+1)$ which were reported in Ref.~\onlinecite{Zhang16}. All of the calculations are performed within the CF theory.

Fig.~\ref{spintransition0} displays results for a sample width of $w=18$ nm, which can be directly compared to the critical Zeeman energies identified in the experiments of Eisenstein {\em et al.}~\cite{Eisenstein16} (magenta stars), Finck {\em et al.} \cite{Finck10} (magenta diamonds), and Giudici {\em et al.}\cite{Giudici08} (magenta square). Theoretical results are given for an ideal 2D system with zero width and no LL mixing (horizontal dashed line), for a quantum well of width $w=18$nm without LL mixing (black dashed line), and for a quantum well of width $w=18$nm including LL mixing. Inclusion of finite width and LL mixing corrections brings theoretical results into better agreement with the experiments of Eisenstein {\em et al.}~\cite{Eisenstein16}. We do not understand the origin of the larger discrepancy with the experiments in Refs.~\onlinecite{Finck10,Giudici08}.

We end this section by stressing puzzling differences between the dependencies of the onset voltage and $V_{\rm max}$ on the spin polarization of the CF Fermi sea. As noted above, the experimental plot\cite{Eisenstein16} of $V_{\rm max}$ as a function of $B_{\rm tot}$ does not show any signature of the spin transition of the CF Fermi sea, presumably due to the fact, as noted above, that the CF Fermi sea remains almost fully spin polarized in the entire parameter regime of the experiment. In contrast, the onset voltage is very sensitive to the spin polarization\cite{Eisenstein16}. Furthermore, the onset voltage {\em decreases} when the system becomes non-fully polarized, whereas, according to our calculations, $V_{\rm max}$ {\em increases} when FQH states become partially spin polarized. An explanation of these features will require a quantitative understanding of the onset voltage, which we do not currently have. We speculate that the sensitivity of the onset voltage on the spin polarization originates because when the CF Fermi sea is partially polarized, the low energy interlayer excitons can be more effectively screened due to the availability of spin flip excitations.

\section{Conclusions}
\label{SecV}

We have given a microscopic account of the energy of the inter-layer exciton that dominates the tunneling in bilayer fractional Hall systems. We find an excellent quantitative agreement with experimentally measured energy as well as its dependence on a parallel magnetic field, and identify the importance of various contributions to the energy.

\underline{Acknowledgments:}  The work at Penn State was supported in part by the US Department of Energy under Grant No. DE-SC0005042. The Caltech portion of this work was supported in part by the Institute for Quantum Information and Matter, an NSF Physics Frontiers Center with support of the Gordon and Betty Moore Foundation through Grant No. GBMF1250.

\bibliography{../../../Latex-Revtex-etc./biblio_fqhe.bib}

\begin{thebibliography}{61}%
\makeatletter
\providecommand \@ifxundefined [1]{%
 \@ifx{#1\undefined}
}%
\providecommand \@ifnum [1]{%
 \ifnum #1\expandafter \@firstoftwo
 \else \expandafter \@secondoftwo
 \fi
}%
\providecommand \@ifx [1]{%
 \ifx #1\expandafter \@firstoftwo
 \else \expandafter \@secondoftwo
 \fi
}%
\providecommand \natexlab [1]{#1}%
\providecommand \enquote  [1]{``#1''}%
\providecommand \bibnamefont  [1]{#1}%
\providecommand \bibfnamefont [1]{#1}%
\providecommand \citenamefont [1]{#1}%
\providecommand \href@noop [0]{\@secondoftwo}%
\providecommand \href [0]{\begingroup \@sanitize@url \@href}%
\providecommand \@href[1]{\@@startlink{#1}\@@href}%
\providecommand \@@href[1]{\endgroup#1\@@endlink}%
\providecommand \@sanitize@url [0]{\catcode `\\12\catcode `\$12\catcode
  `\&12\catcode `\#12\catcode `\^12\catcode `\_12\catcode `\%12\relax}%
\providecommand \@@startlink[1]{}%
\providecommand \@@endlink[0]{}%
\providecommand \url  [0]{\begingroup\@sanitize@url \@url }%
\providecommand \@url [1]{\endgroup\@href {#1}{\urlprefix }}%
\providecommand \urlprefix  [0]{URL }%
\providecommand \Eprint [0]{\href }%
\providecommand \doibase [0]{http://dx.doi.org/}%
\providecommand \selectlanguage [0]{\@gobble}%
\providecommand \bibinfo  [0]{\@secondoftwo}%
\providecommand \bibfield  [0]{\@secondoftwo}%
\providecommand \translation [1]{[#1]}%
\providecommand \BibitemOpen [0]{}%
\providecommand \bibitemStop [0]{}%
\providecommand \bibitemNoStop [0]{.\EOS\space}%
\providecommand \EOS [0]{\spacefactor3000\relax}%
\providecommand \BibitemShut  [1]{\csname bibitem#1\endcsname}%
\let\auto@bib@innerbib\@empty
\bibitem [{\citenamefont {Eisenstein}(2014)}]{Eisenstein14}%
  \BibitemOpen
  \bibfield  {author} {\bibinfo {author} {\bibfnamefont {J.}~\bibnamefont
  {Eisenstein}},\ }\href@noop {} {\bibfield  {journal} {\bibinfo  {journal}
  {Annu. Rev. Condens. Matter Phys.}\ }\textbf {\bibinfo {volume} {5}},\
  \bibinfo {pages} {159} (\bibinfo {year} {2014})}\BibitemShut {NoStop}%
\bibitem [{\citenamefont {Spielman}\ \emph {et~al.}(2000)\citenamefont
  {Spielman}, \citenamefont {Eisenstein}, \citenamefont {Pfeiffer},\ and\
  \citenamefont {West}}]{Spielman00}%
  \BibitemOpen
  \bibfield  {author} {\bibinfo {author} {\bibfnamefont {I.~B.}\ \bibnamefont
  {Spielman}}, \bibinfo {author} {\bibfnamefont {J.~P.}\ \bibnamefont
  {Eisenstein}}, \bibinfo {author} {\bibfnamefont {L.~N.}\ \bibnamefont
  {Pfeiffer}}, \ and\ \bibinfo {author} {\bibfnamefont {K.~W.}\ \bibnamefont
  {West}},\ }\href {\doibase 10.1103/PhysRevLett.84.5808} {\bibfield  {journal}
  {\bibinfo  {journal} {Phys. Rev. Lett.}\ }\textbf {\bibinfo {volume} {84}},\
  \bibinfo {pages} {5808} (\bibinfo {year} {2000})}\BibitemShut {NoStop}%
\bibitem [{\citenamefont {Kellogg}\ \emph {et~al.}(2002)\citenamefont
  {Kellogg}, \citenamefont {Spielman}, \citenamefont {Eisenstein},
  \citenamefont {Pfeiffer},\ and\ \citenamefont {West}}]{Kellogg02}%
  \BibitemOpen
  \bibfield  {author} {\bibinfo {author} {\bibfnamefont {M.}~\bibnamefont
  {Kellogg}}, \bibinfo {author} {\bibfnamefont {I.~B.}\ \bibnamefont
  {Spielman}}, \bibinfo {author} {\bibfnamefont {J.~P.}\ \bibnamefont
  {Eisenstein}}, \bibinfo {author} {\bibfnamefont {L.~N.}\ \bibnamefont
  {Pfeiffer}}, \ and\ \bibinfo {author} {\bibfnamefont {K.~W.}\ \bibnamefont
  {West}},\ }\href {\doibase 10.1103/PhysRevLett.88.126804} {\bibfield
  {journal} {\bibinfo  {journal} {Phys. Rev. Lett.}\ }\textbf {\bibinfo
  {volume} {88}},\ \bibinfo {pages} {126804} (\bibinfo {year}
  {2002})}\BibitemShut {NoStop}%
\bibitem [{\citenamefont {Kellogg}\ \emph {et~al.}(2004)\citenamefont
  {Kellogg}, \citenamefont {Eisenstein}, \citenamefont {Pfeiffer},\ and\
  \citenamefont {West}}]{Kellogg04}%
  \BibitemOpen
  \bibfield  {author} {\bibinfo {author} {\bibfnamefont {M.}~\bibnamefont
  {Kellogg}}, \bibinfo {author} {\bibfnamefont {J.~P.}\ \bibnamefont
  {Eisenstein}}, \bibinfo {author} {\bibfnamefont {L.~N.}\ \bibnamefont
  {Pfeiffer}}, \ and\ \bibinfo {author} {\bibfnamefont {K.~W.}\ \bibnamefont
  {West}},\ }\href {\doibase 10.1103/PhysRevLett.93.036801} {\bibfield
  {journal} {\bibinfo  {journal} {Phys. Rev. Lett.}\ }\textbf {\bibinfo
  {volume} {93}},\ \bibinfo {pages} {036801} (\bibinfo {year}
  {2004})}\BibitemShut {NoStop}%
\bibitem [{\citenamefont {Tutuc}\ \emph {et~al.}(2004)\citenamefont {Tutuc},
  \citenamefont {Shayegan},\ and\ \citenamefont {Huse}}]{Tutuc04}%
  \BibitemOpen
  \bibfield  {author} {\bibinfo {author} {\bibfnamefont {E.}~\bibnamefont
  {Tutuc}}, \bibinfo {author} {\bibfnamefont {M.}~\bibnamefont {Shayegan}}, \
  and\ \bibinfo {author} {\bibfnamefont {D.~A.}\ \bibnamefont {Huse}},\ }\href
  {\doibase 10.1103/PhysRevLett.93.036802} {\bibfield  {journal} {\bibinfo
  {journal} {Phys. Rev. Lett.}\ }\textbf {\bibinfo {volume} {93}},\ \bibinfo
  {pages} {036802} (\bibinfo {year} {2004})}\BibitemShut {NoStop}%
\bibitem [{\citenamefont {Wiersma}\ \emph {et~al.}(2004)\citenamefont
  {Wiersma}, \citenamefont {Lok}, \citenamefont {Kraus}, \citenamefont
  {Dietsche}, \citenamefont {von Klitzing}, \citenamefont {Schuh},
  \citenamefont {Bichler}, \citenamefont {Tranitz},\ and\ \citenamefont
  {Wegscheider}}]{Wiersma04}%
  \BibitemOpen
  \bibfield  {author} {\bibinfo {author} {\bibfnamefont {R.~D.}\ \bibnamefont
  {Wiersma}}, \bibinfo {author} {\bibfnamefont {J.~G.~S.}\ \bibnamefont {Lok}},
  \bibinfo {author} {\bibfnamefont {S.}~\bibnamefont {Kraus}}, \bibinfo
  {author} {\bibfnamefont {W.}~\bibnamefont {Dietsche}}, \bibinfo {author}
  {\bibfnamefont {K.}~\bibnamefont {von Klitzing}}, \bibinfo {author}
  {\bibfnamefont {D.}~\bibnamefont {Schuh}}, \bibinfo {author} {\bibfnamefont
  {M.}~\bibnamefont {Bichler}}, \bibinfo {author} {\bibfnamefont {H.-P.}\
  \bibnamefont {Tranitz}}, \ and\ \bibinfo {author} {\bibfnamefont
  {W.}~\bibnamefont {Wegscheider}},\ }\href {\doibase
  10.1103/PhysRevLett.93.266805} {\bibfield  {journal} {\bibinfo  {journal}
  {Phys. Rev. Lett.}\ }\textbf {\bibinfo {volume} {93}},\ \bibinfo {pages}
  {266805} (\bibinfo {year} {2004})}\BibitemShut {NoStop}%
\bibitem [{\citenamefont {Halperin}(1983)}]{Halperin83}%
  \BibitemOpen
  \bibfield  {author} {\bibinfo {author} {\bibfnamefont {B.~I.}\ \bibnamefont
  {Halperin}},\ }\href@noop {} {\bibfield  {journal} {\bibinfo  {journal}
  {Helvetica Physica Acta}\ }\textbf {\bibinfo {volume} {56}},\ \bibinfo
  {pages} {75} (\bibinfo {year} {1983})}\BibitemShut {NoStop}%
\bibitem [{\citenamefont {MacDonald}\ and\ \citenamefont
  {Rezayi}(1990)}]{MacDonald90}%
  \BibitemOpen
  \bibfield  {author} {\bibinfo {author} {\bibfnamefont {A.~H.}\ \bibnamefont
  {MacDonald}}\ and\ \bibinfo {author} {\bibfnamefont {E.~H.}\ \bibnamefont
  {Rezayi}},\ }\href {\doibase 10.1103/PhysRevB.42.3224} {\bibfield  {journal}
  {\bibinfo  {journal} {Phys. Rev. B}\ }\textbf {\bibinfo {volume} {42}},\
  \bibinfo {pages} {3224} (\bibinfo {year} {1990})}\BibitemShut {NoStop}%
\bibitem [{\citenamefont {Wen}\ and\ \citenamefont {Zee}(1992)}]{Wen92c}%
  \BibitemOpen
  \bibfield  {author} {\bibinfo {author} {\bibfnamefont {X.-G.}\ \bibnamefont
  {Wen}}\ and\ \bibinfo {author} {\bibfnamefont {A.}~\bibnamefont {Zee}},\
  }\href {\doibase 10.1103/PhysRevLett.69.1811} {\bibfield  {journal} {\bibinfo
   {journal} {Phys. Rev. Lett.}\ }\textbf {\bibinfo {volume} {69}},\ \bibinfo
  {pages} {1811} (\bibinfo {year} {1992})}\BibitemShut {NoStop}%
\bibitem [{\citenamefont {Jain}(1989)}]{Jain89}%
  \BibitemOpen
  \bibfield  {author} {\bibinfo {author} {\bibfnamefont {J.~K.}\ \bibnamefont
  {Jain}},\ }\href {\doibase 10.1103/PhysRevLett.63.199} {\bibfield  {journal}
  {\bibinfo  {journal} {Phys. Rev. Lett.}\ }\textbf {\bibinfo {volume} {63}},\
  \bibinfo {pages} {199} (\bibinfo {year} {1989})}\BibitemShut {NoStop}%
\bibitem [{\citenamefont {Halperin}\ \emph {et~al.}(1993)\citenamefont
  {Halperin}, \citenamefont {Lee},\ and\ \citenamefont {Read}}]{Halperin93}%
  \BibitemOpen
  \bibfield  {author} {\bibinfo {author} {\bibfnamefont {B.~I.}\ \bibnamefont
  {Halperin}}, \bibinfo {author} {\bibfnamefont {P.~A.}\ \bibnamefont {Lee}}, \
  and\ \bibinfo {author} {\bibfnamefont {N.}~\bibnamefont {Read}},\ }\href
  {\doibase 10.1103/PhysRevB.47.7312} {\bibfield  {journal} {\bibinfo
  {journal} {Phys. Rev. B}\ }\textbf {\bibinfo {volume} {47}},\ \bibinfo
  {pages} {7312} (\bibinfo {year} {1993})}\BibitemShut {NoStop}%
\bibitem [{\citenamefont {Jain}(2007)}]{Jain07}%
  \BibitemOpen
  \bibfield  {author} {\bibinfo {author} {\bibfnamefont {J.~K.}\ \bibnamefont
  {Jain}},\ }\href@noop {} {\emph {\bibinfo {title} {Composite Fermions}}}\
  (\bibinfo  {publisher} {Cambridge University Press, New York, US (Cambridge
  Books Online)},\ \bibinfo {year} {2007})\BibitemShut {NoStop}%
\bibitem [{\citenamefont {Ashoori}\ \emph {et~al.}(1990)\citenamefont
  {Ashoori}, \citenamefont {Lebens}, \citenamefont {Bigelow},\ and\
  \citenamefont {Silsbee}}]{Ashoori90}%
  \BibitemOpen
  \bibfield  {author} {\bibinfo {author} {\bibfnamefont {R.~C.}\ \bibnamefont
  {Ashoori}}, \bibinfo {author} {\bibfnamefont {J.~A.}\ \bibnamefont {Lebens}},
  \bibinfo {author} {\bibfnamefont {N.~P.}\ \bibnamefont {Bigelow}}, \ and\
  \bibinfo {author} {\bibfnamefont {R.~H.}\ \bibnamefont {Silsbee}},\ }\href
  {\doibase 10.1103/PhysRevLett.64.681} {\bibfield  {journal} {\bibinfo
  {journal} {Phys. Rev. Lett.}\ }\textbf {\bibinfo {volume} {64}},\ \bibinfo
  {pages} {681} (\bibinfo {year} {1990})}\BibitemShut {NoStop}%
\bibitem [{\citenamefont {Eisenstein}\ \emph {et~al.}(1992)\citenamefont
  {Eisenstein}, \citenamefont {Pfeiffer},\ and\ \citenamefont
  {West}}]{Eisenstein92b}%
  \BibitemOpen
  \bibfield  {author} {\bibinfo {author} {\bibfnamefont {J.~P.}\ \bibnamefont
  {Eisenstein}}, \bibinfo {author} {\bibfnamefont {L.~N.}\ \bibnamefont
  {Pfeiffer}}, \ and\ \bibinfo {author} {\bibfnamefont {K.~W.}\ \bibnamefont
  {West}},\ }\href {\doibase 10.1103/PhysRevLett.69.3804} {\bibfield  {journal}
  {\bibinfo  {journal} {Phys. Rev. Lett.}\ }\textbf {\bibinfo {volume} {69}},\
  \bibinfo {pages} {3804} (\bibinfo {year} {1992})}\BibitemShut {NoStop}%
\bibitem [{\citenamefont {Brown}\ \emph {et~al.}(1994)\citenamefont {Brown},
  \citenamefont {Turner}, \citenamefont {Nicholls}, \citenamefont {Linfield},
  \citenamefont {Pepper}, \citenamefont {Ritchie},\ and\ \citenamefont
  {Jones}}]{Brown94}%
  \BibitemOpen
  \bibfield  {author} {\bibinfo {author} {\bibfnamefont {K.~M.}\ \bibnamefont
  {Brown}}, \bibinfo {author} {\bibfnamefont {N.}~\bibnamefont {Turner}},
  \bibinfo {author} {\bibfnamefont {J.~T.}\ \bibnamefont {Nicholls}}, \bibinfo
  {author} {\bibfnamefont {E.~H.}\ \bibnamefont {Linfield}}, \bibinfo {author}
  {\bibfnamefont {M.}~\bibnamefont {Pepper}}, \bibinfo {author} {\bibfnamefont
  {D.~A.}\ \bibnamefont {Ritchie}}, \ and\ \bibinfo {author} {\bibfnamefont
  {G.~A.~C.}\ \bibnamefont {Jones}},\ }\href {\doibase
  10.1103/PhysRevB.50.15465} {\bibfield  {journal} {\bibinfo  {journal} {Phys.
  Rev. B}\ }\textbf {\bibinfo {volume} {50}},\ \bibinfo {pages} {15465}
  (\bibinfo {year} {1994})}\BibitemShut {NoStop}%
\bibitem [{\citenamefont {Eisenstein}\ \emph {et~al.}(1995)\citenamefont
  {Eisenstein}, \citenamefont {Pfeiffer},\ and\ \citenamefont
  {West}}]{Eisenstein95}%
  \BibitemOpen
  \bibfield  {author} {\bibinfo {author} {\bibfnamefont {J.~P.}\ \bibnamefont
  {Eisenstein}}, \bibinfo {author} {\bibfnamefont {L.~N.}\ \bibnamefont
  {Pfeiffer}}, \ and\ \bibinfo {author} {\bibfnamefont {K.~W.}\ \bibnamefont
  {West}},\ }\href {\doibase 10.1103/PhysRevLett.74.1419} {\bibfield  {journal}
  {\bibinfo  {journal} {Phys. Rev. Lett.}\ }\textbf {\bibinfo {volume} {74}},\
  \bibinfo {pages} {1419} (\bibinfo {year} {1995})}\BibitemShut {NoStop}%
\bibitem [{\citenamefont {Eisenstein}\ \emph {et~al.}(2009)\citenamefont
  {Eisenstein}, \citenamefont {Pfeiffer},\ and\ \citenamefont
  {West}}]{Eisenstein09}%
  \BibitemOpen
  \bibfield  {author} {\bibinfo {author} {\bibfnamefont {J.}~\bibnamefont
  {Eisenstein}}, \bibinfo {author} {\bibfnamefont {L.}~\bibnamefont
  {Pfeiffer}}, \ and\ \bibinfo {author} {\bibfnamefont {K.}~\bibnamefont
  {West}},\ }\href {\doibase http://dx.doi.org/10.1016/j.ssc.2009.08.004}
  {\bibfield  {journal} {\bibinfo  {journal} {Solid State Communications}\
  }\textbf {\bibinfo {volume} {149}},\ \bibinfo {pages} {1867 } (\bibinfo
  {year} {2009})}\BibitemShut {NoStop}%
\bibitem [{\citenamefont {Eisenstein}\ \emph {et~al.}(2016)\citenamefont
  {Eisenstein}, \citenamefont {Khaire}, \citenamefont {Nandi}, \citenamefont
  {Finck}, \citenamefont {Pfeiffer},\ and\ \citenamefont
  {West}}]{Eisenstein16}%
  \BibitemOpen
  \bibfield  {author} {\bibinfo {author} {\bibfnamefont {J.~P.}\ \bibnamefont
  {Eisenstein}}, \bibinfo {author} {\bibfnamefont {T.}~\bibnamefont {Khaire}},
  \bibinfo {author} {\bibfnamefont {D.}~\bibnamefont {Nandi}}, \bibinfo
  {author} {\bibfnamefont {A.~D.~K.}\ \bibnamefont {Finck}}, \bibinfo {author}
  {\bibfnamefont {L.~N.}\ \bibnamefont {Pfeiffer}}, \ and\ \bibinfo {author}
  {\bibfnamefont {K.~W.}\ \bibnamefont {West}},\ }\href {\doibase
  10.1103/PhysRevB.94.125409} {\bibfield  {journal} {\bibinfo  {journal} {Phys.
  Rev. B}\ }\textbf {\bibinfo {volume} {94}},\ \bibinfo {pages} {125409}
  (\bibinfo {year} {2016})}\BibitemShut {NoStop}%
\bibitem [{\citenamefont {He}\ \emph {et~al.}(1993)\citenamefont {He},
  \citenamefont {Platzman},\ and\ \citenamefont {Halperin}}]{He93b}%
  \BibitemOpen
  \bibfield  {author} {\bibinfo {author} {\bibfnamefont {S.}~\bibnamefont
  {He}}, \bibinfo {author} {\bibfnamefont {P.~M.}\ \bibnamefont {Platzman}}, \
  and\ \bibinfo {author} {\bibfnamefont {B.~I.}\ \bibnamefont {Halperin}},\
  }\href {\doibase 10.1103/PhysRevLett.71.777} {\bibfield  {journal} {\bibinfo
  {journal} {Phys. Rev. Lett.}\ }\textbf {\bibinfo {volume} {71}},\ \bibinfo
  {pages} {777} (\bibinfo {year} {1993})}\BibitemShut {NoStop}%
\bibitem [{\citenamefont {Haussmann}\ \emph {et~al.}(1996)\citenamefont
  {Haussmann}, \citenamefont {Mori},\ and\ \citenamefont
  {MacDonald}}]{Haussmann96}%
  \BibitemOpen
  \bibfield  {author} {\bibinfo {author} {\bibfnamefont {R.}~\bibnamefont
  {Haussmann}}, \bibinfo {author} {\bibfnamefont {H.}~\bibnamefont {Mori}}, \
  and\ \bibinfo {author} {\bibfnamefont {A.~H.}\ \bibnamefont {MacDonald}},\
  }\href {\doibase 10.1103/PhysRevLett.76.979} {\bibfield  {journal} {\bibinfo
  {journal} {Phys. Rev. Lett.}\ }\textbf {\bibinfo {volume} {76}},\ \bibinfo
  {pages} {979} (\bibinfo {year} {1996})}\BibitemShut {NoStop}%
\bibitem [{\citenamefont {Hatsugai}\ \emph {et~al.}(1993)\citenamefont
  {Hatsugai}, \citenamefont {Bares},\ and\ \citenamefont {Wen}}]{Hatsugai93}%
  \BibitemOpen
  \bibfield  {author} {\bibinfo {author} {\bibfnamefont {Y.}~\bibnamefont
  {Hatsugai}}, \bibinfo {author} {\bibfnamefont {P.-A.}\ \bibnamefont {Bares}},
  \ and\ \bibinfo {author} {\bibfnamefont {X.~G.}\ \bibnamefont {Wen}},\ }\href
  {\doibase 10.1103/PhysRevLett.71.424} {\bibfield  {journal} {\bibinfo
  {journal} {Phys. Rev. Lett.}\ }\textbf {\bibinfo {volume} {71}},\ \bibinfo
  {pages} {424} (\bibinfo {year} {1993})}\BibitemShut {NoStop}%
\bibitem [{\citenamefont {Efros}\ and\ \citenamefont {Pikus}(1993)}]{Efros93}%
  \BibitemOpen
  \bibfield  {author} {\bibinfo {author} {\bibfnamefont {A.~L.}\ \bibnamefont
  {Efros}}\ and\ \bibinfo {author} {\bibfnamefont {F.~G.}\ \bibnamefont
  {Pikus}},\ }\href {\doibase 10.1103/PhysRevB.48.14694} {\bibfield  {journal}
  {\bibinfo  {journal} {Phys. Rev. B}\ }\textbf {\bibinfo {volume} {48}},\
  \bibinfo {pages} {14694} (\bibinfo {year} {1993})}\BibitemShut {NoStop}%
\bibitem [{\citenamefont {Levitov}\ and\ \citenamefont
  {Shytov}(1997)}]{Levitov97}%
  \BibitemOpen
  \bibfield  {author} {\bibinfo {author} {\bibfnamefont {S.}~\bibnamefont
  {Levitov}}\ and\ \bibinfo {author} {\bibfnamefont {A.~V.}\ \bibnamefont
  {Shytov}},\ }\href {\doibase 10.1134/1.567489} {\bibfield  {journal}
  {\bibinfo  {journal} {Journal of Experimental and Theoretical Physics
  Letters}\ }\textbf {\bibinfo {volume} {66}},\ \bibinfo {pages} {214}
  (\bibinfo {year} {1997})}\BibitemShut {NoStop}%
\bibitem [{\citenamefont {Johansson}\ and\ \citenamefont
  {Kinaret}(1993)}]{Johansson93}%
  \BibitemOpen
  \bibfield  {author} {\bibinfo {author} {\bibfnamefont {P.}~\bibnamefont
  {Johansson}}\ and\ \bibinfo {author} {\bibfnamefont {J.~M.}\ \bibnamefont
  {Kinaret}},\ }\href {\doibase 10.1103/PhysRevLett.71.1435} {\bibfield
  {journal} {\bibinfo  {journal} {Phys. Rev. Lett.}\ }\textbf {\bibinfo
  {volume} {71}},\ \bibinfo {pages} {1435} (\bibinfo {year}
  {1993})}\BibitemShut {NoStop}%
\bibitem [{\citenamefont {Finck}\ \emph {et~al.}(2010)\citenamefont {Finck},
  \citenamefont {Eisenstein}, \citenamefont {Pfeiffer},\ and\ \citenamefont
  {West}}]{Finck10}%
  \BibitemOpen
  \bibfield  {author} {\bibinfo {author} {\bibfnamefont {A.~D.~K.}\
  \bibnamefont {Finck}}, \bibinfo {author} {\bibfnamefont {J.~P.}\ \bibnamefont
  {Eisenstein}}, \bibinfo {author} {\bibfnamefont {L.~N.}\ \bibnamefont
  {Pfeiffer}}, \ and\ \bibinfo {author} {\bibfnamefont {K.~W.}\ \bibnamefont
  {West}},\ }\href {\doibase 10.1103/PhysRevLett.104.016801} {\bibfield
  {journal} {\bibinfo  {journal} {Phys. Rev. Lett.}\ }\textbf {\bibinfo
  {volume} {104}},\ \bibinfo {pages} {016801} (\bibinfo {year}
  {2010})}\BibitemShut {NoStop}%
\bibitem [{\citenamefont {Giudici}\ \emph {et~al.}(2008)\citenamefont
  {Giudici}, \citenamefont {Muraki}, \citenamefont {Kumada}, \citenamefont
  {Hirayama},\ and\ \citenamefont {Fujisawa}}]{Giudici08}%
  \BibitemOpen
  \bibfield  {author} {\bibinfo {author} {\bibfnamefont {P.}~\bibnamefont
  {Giudici}}, \bibinfo {author} {\bibfnamefont {K.}~\bibnamefont {Muraki}},
  \bibinfo {author} {\bibfnamefont {N.}~\bibnamefont {Kumada}}, \bibinfo
  {author} {\bibfnamefont {Y.}~\bibnamefont {Hirayama}}, \ and\ \bibinfo
  {author} {\bibfnamefont {T.}~\bibnamefont {Fujisawa}},\ }\href {\doibase
  10.1103/PhysRevLett.100.106803} {\bibfield  {journal} {\bibinfo  {journal}
  {Phys. Rev. Lett.}\ }\textbf {\bibinfo {volume} {100}},\ \bibinfo {pages}
  {106803} (\bibinfo {year} {2008})}\BibitemShut {NoStop}%
\bibitem [{\citenamefont {Park}\ and\ \citenamefont {Jain}(1998)}]{Park98}%
  \BibitemOpen
  \bibfield  {author} {\bibinfo {author} {\bibfnamefont {K.}~\bibnamefont
  {Park}}\ and\ \bibinfo {author} {\bibfnamefont {J.~K.}\ \bibnamefont
  {Jain}},\ }\href {\doibase 10.1103/PhysRevLett.80.4237} {\bibfield  {journal}
  {\bibinfo  {journal} {Phys. Rev. Lett.}\ }\textbf {\bibinfo {volume} {80}},\
  \bibinfo {pages} {4237} (\bibinfo {year} {1998})}\BibitemShut {NoStop}%
\bibitem [{\citenamefont {Giuliani}\ and\ \citenamefont
  {Vignale}(2008)}]{Giuliani08}%
  \BibitemOpen
  \bibfield  {author} {\bibinfo {author} {\bibfnamefont {G.}~\bibnamefont
  {Giuliani}}\ and\ \bibinfo {author} {\bibfnamefont {G.}~\bibnamefont
  {Vignale}},\ }\href@noop {} {\emph {\bibinfo {title} {Quantum Theory of the
  Electron Liquid}}}\ (\bibinfo  {publisher} {Cambridge University Press, The
  Edinburgh Building, Cambridge CB2 2RU, UK},\ \bibinfo {year}
  {2008})\BibitemShut {NoStop}%
\bibitem [{\citenamefont {Mahan}(2000)}]{Mahan00}%
  \BibitemOpen
  \bibfield  {author} {\bibinfo {author} {\bibfnamefont {G.~D.}\ \bibnamefont
  {Mahan}},\ }\href@noop {} {\emph {\bibinfo {title} {Many Particle Physics}}}\
  (\bibinfo  {publisher} {Kluwer Academic / Plenum Publishers, New York},\
  \bibinfo {year} {2000})\BibitemShut {NoStop}%
\bibitem [{\citenamefont {Scarola}\ and\ \citenamefont
  {Jain}(2001)}]{Scarola01b}%
  \BibitemOpen
  \bibfield  {author} {\bibinfo {author} {\bibfnamefont {V.~W.}\ \bibnamefont
  {Scarola}}\ and\ \bibinfo {author} {\bibfnamefont {J.~K.}\ \bibnamefont
  {Jain}},\ }\href {\doibase 10.1103/PhysRevB.64.085313} {\bibfield  {journal}
  {\bibinfo  {journal} {Phys. Rev. B}\ }\textbf {\bibinfo {volume} {64}},\
  \bibinfo {pages} {085313} (\bibinfo {year} {2001})}\BibitemShut {NoStop}%
\bibitem [{\citenamefont {Haldane}(1983)}]{Haldane83}%
  \BibitemOpen
  \bibfield  {author} {\bibinfo {author} {\bibfnamefont {F.~D.~M.}\
  \bibnamefont {Haldane}},\ }\href {\doibase 10.1103/PhysRevLett.51.605}
  {\bibfield  {journal} {\bibinfo  {journal} {Phys. Rev. Lett.}\ }\textbf
  {\bibinfo {volume} {51}},\ \bibinfo {pages} {605} (\bibinfo {year}
  {1983})}\BibitemShut {NoStop}%
\bibitem [{\citenamefont {Wu}\ \emph {et~al.}(1993)\citenamefont {Wu},
  \citenamefont {Dev},\ and\ \citenamefont {Jain}}]{Wu93}%
  \BibitemOpen
  \bibfield  {author} {\bibinfo {author} {\bibfnamefont {X.~G.}\ \bibnamefont
  {Wu}}, \bibinfo {author} {\bibfnamefont {G.}~\bibnamefont {Dev}}, \ and\
  \bibinfo {author} {\bibfnamefont {J.~K.}\ \bibnamefont {Jain}},\ }\href
  {\doibase 10.1103/PhysRevLett.71.153} {\bibfield  {journal} {\bibinfo
  {journal} {Phys. Rev. Lett.}\ }\textbf {\bibinfo {volume} {71}},\ \bibinfo
  {pages} {153} (\bibinfo {year} {1993})}\BibitemShut {NoStop}%
\bibitem [{\citenamefont {Jain}\ and\ \citenamefont
  {Kamilla}(1997{\natexlab{a}})}]{Jain97}%
  \BibitemOpen
  \bibfield  {author} {\bibinfo {author} {\bibfnamefont {J.~K.}\ \bibnamefont
  {Jain}}\ and\ \bibinfo {author} {\bibfnamefont {R.~K.}\ \bibnamefont
  {Kamilla}},\ }\href {\doibase 10.1142/S0217979297001301} {\bibfield
  {journal} {\bibinfo  {journal} {Int. J. Mod. Phys. B}\ }\textbf {\bibinfo
  {volume} {11}},\ \bibinfo {pages} {2621} (\bibinfo {year}
  {1997}{\natexlab{a}})}\BibitemShut {NoStop}%
\bibitem [{\citenamefont {Jain}\ and\ \citenamefont
  {Kamilla}(1997{\natexlab{b}})}]{Jain97b}%
  \BibitemOpen
  \bibfield  {author} {\bibinfo {author} {\bibfnamefont {J.~K.}\ \bibnamefont
  {Jain}}\ and\ \bibinfo {author} {\bibfnamefont {R.~K.}\ \bibnamefont
  {Kamilla}},\ }\href {\doibase 10.1103/PhysRevB.55.R4895} {\bibfield
  {journal} {\bibinfo  {journal} {Phys. Rev. B}\ }\textbf {\bibinfo {volume}
  {55}},\ \bibinfo {pages} {R4895} (\bibinfo {year}
  {1997}{\natexlab{b}})}\BibitemShut {NoStop}%
\bibitem [{\citenamefont {Foulkes}\ \emph {et~al.}(2001)\citenamefont
  {Foulkes}, \citenamefont {Mitas}, \citenamefont {Needs},\ and\ \citenamefont
  {Rajagopal}}]{Foulkes01}%
  \BibitemOpen
  \bibfield  {author} {\bibinfo {author} {\bibfnamefont {W.~M.~C.}\
  \bibnamefont {Foulkes}}, \bibinfo {author} {\bibfnamefont {L.}~\bibnamefont
  {Mitas}}, \bibinfo {author} {\bibfnamefont {R.~J.}\ \bibnamefont {Needs}}, \
  and\ \bibinfo {author} {\bibfnamefont {G.}~\bibnamefont {Rajagopal}},\ }\href
  {\doibase 10.1103/RevModPhys.73.33} {\bibfield  {journal} {\bibinfo
  {journal} {Rev. Mod. Phys.}\ }\textbf {\bibinfo {volume} {73}},\ \bibinfo
  {pages} {33} (\bibinfo {year} {2001})}\BibitemShut {NoStop}%
\bibitem [{\citenamefont {Jain}\ and\ \citenamefont
  {Peterson}(2005)}]{Peterson05}%
  \BibitemOpen
  \bibfield  {author} {\bibinfo {author} {\bibfnamefont {J.~K.}\ \bibnamefont
  {Jain}}\ and\ \bibinfo {author} {\bibfnamefont {M.~R.}\ \bibnamefont
  {Peterson}},\ }\href {\doibase 10.1103/PhysRevLett.94.186808} {\bibfield
  {journal} {\bibinfo  {journal} {Phys. Rev. Lett.}\ }\textbf {\bibinfo
  {volume} {94}},\ \bibinfo {pages} {186808} (\bibinfo {year}
  {2005})}\BibitemShut {NoStop}%
\bibitem [{\citenamefont {Ortalano}\ \emph {et~al.}(1997)\citenamefont
  {Ortalano}, \citenamefont {He},\ and\ \citenamefont
  {Das~Sarma}}]{Ortalano97}%
  \BibitemOpen
  \bibfield  {author} {\bibinfo {author} {\bibfnamefont {M.~W.}\ \bibnamefont
  {Ortalano}}, \bibinfo {author} {\bibfnamefont {S.}~\bibnamefont {He}}, \ and\
  \bibinfo {author} {\bibfnamefont {S.}~\bibnamefont {Das~Sarma}},\ }\href
  {\doibase 10.1103/PhysRevB.55.7702} {\bibfield  {journal} {\bibinfo
  {journal} {Phys. Rev. B}\ }\textbf {\bibinfo {volume} {55}},\ \bibinfo
  {pages} {7702} (\bibinfo {year} {1997})}\BibitemShut {NoStop}%
\bibitem [{\citenamefont {Mueed}\ \emph {et~al.}(2015)\citenamefont {Mueed},
  \citenamefont {Kamburov}, \citenamefont {Shayegan}, \citenamefont {Pfeiffer},
  \citenamefont {West}, \citenamefont {Baldwin},\ and\ \citenamefont
  {Winkler}}]{Mueed15a}%
  \BibitemOpen
  \bibfield  {author} {\bibinfo {author} {\bibfnamefont {M.~A.}\ \bibnamefont
  {Mueed}}, \bibinfo {author} {\bibfnamefont {D.}~\bibnamefont {Kamburov}},
  \bibinfo {author} {\bibfnamefont {M.}~\bibnamefont {Shayegan}}, \bibinfo
  {author} {\bibfnamefont {L.~N.}\ \bibnamefont {Pfeiffer}}, \bibinfo {author}
  {\bibfnamefont {K.~W.}\ \bibnamefont {West}}, \bibinfo {author}
  {\bibfnamefont {K.~W.}\ \bibnamefont {Baldwin}}, \ and\ \bibinfo {author}
  {\bibfnamefont {R.}~\bibnamefont {Winkler}},\ }\href {\doibase
  10.1103/PhysRevLett.114.236404} {\bibfield  {journal} {\bibinfo  {journal}
  {Phys. Rev. Lett.}\ }\textbf {\bibinfo {volume} {114}},\ \bibinfo {pages}
  {236404} (\bibinfo {year} {2015})}\BibitemShut {NoStop}%
\bibitem [{\citenamefont {Kamburov}\ \emph {et~al.}(2014)\citenamefont
  {Kamburov}, \citenamefont {Mueed}, \citenamefont {Shayegan}, \citenamefont
  {Pfeiffer}, \citenamefont {West}, \citenamefont {Baldwin}, \citenamefont
  {Lee},\ and\ \citenamefont {Winkler}}]{Kamburov14}%
  \BibitemOpen
  \bibfield  {author} {\bibinfo {author} {\bibfnamefont {D.}~\bibnamefont
  {Kamburov}}, \bibinfo {author} {\bibfnamefont {M.~A.}\ \bibnamefont {Mueed}},
  \bibinfo {author} {\bibfnamefont {M.}~\bibnamefont {Shayegan}}, \bibinfo
  {author} {\bibfnamefont {L.~N.}\ \bibnamefont {Pfeiffer}}, \bibinfo {author}
  {\bibfnamefont {K.~W.}\ \bibnamefont {West}}, \bibinfo {author}
  {\bibfnamefont {K.~W.}\ \bibnamefont {Baldwin}}, \bibinfo {author}
  {\bibfnamefont {J.~J.~D.}\ \bibnamefont {Lee}}, \ and\ \bibinfo {author}
  {\bibfnamefont {R.}~\bibnamefont {Winkler}},\ }\href {\doibase
  10.1103/PhysRevB.89.085304} {\bibfield  {journal} {\bibinfo  {journal} {Phys.
  Rev. B}\ }\textbf {\bibinfo {volume} {89}},\ \bibinfo {pages} {085304}
  (\bibinfo {year} {2014})}\BibitemShut {NoStop}%
\bibitem [{\citenamefont {Balram}\ and\ \citenamefont {Jain}(2016)}]{Balram16}%
  \BibitemOpen
  \bibfield  {author} {\bibinfo {author} {\bibfnamefont {A.~C.}\ \bibnamefont
  {Balram}}\ and\ \bibinfo {author} {\bibfnamefont {J.~K.}\ \bibnamefont
  {Jain}},\ }\href {\doibase 10.1103/PhysRevB.93.075121} {\bibfield  {journal}
  {\bibinfo  {journal} {Phys. Rev. B}\ }\textbf {\bibinfo {volume} {93}},\
  \bibinfo {pages} {075121} (\bibinfo {year} {2016})}\BibitemShut {NoStop}%
\bibitem [{\citenamefont {Melik-Alaverdian}\ and\ \citenamefont
  {Bonesteel}(1995)}]{Melik-Alaverdian95}%
  \BibitemOpen
  \bibfield  {author} {\bibinfo {author} {\bibfnamefont {V.}~\bibnamefont
  {Melik-Alaverdian}}\ and\ \bibinfo {author} {\bibfnamefont {N.~E.}\
  \bibnamefont {Bonesteel}},\ }\href {\doibase 10.1103/PhysRevB.52.R17032}
  {\bibfield  {journal} {\bibinfo  {journal} {Phys. Rev. B}\ }\textbf {\bibinfo
  {volume} {52}},\ \bibinfo {pages} {R17032} (\bibinfo {year}
  {1995})}\BibitemShut {NoStop}%
\bibitem [{\citenamefont {Melik-Alaverdian}\ \emph {et~al.}(1997)\citenamefont
  {Melik-Alaverdian}, \citenamefont {Bonesteel},\ and\ \citenamefont
  {Ortiz}}]{Melik-Alaverdian97}%
  \BibitemOpen
  \bibfield  {author} {\bibinfo {author} {\bibfnamefont {V.}~\bibnamefont
  {Melik-Alaverdian}}, \bibinfo {author} {\bibfnamefont {N.~E.}\ \bibnamefont
  {Bonesteel}}, \ and\ \bibinfo {author} {\bibfnamefont {G.}~\bibnamefont
  {Ortiz}},\ }\href {\doibase 10.1103/PhysRevLett.79.5286} {\bibfield
  {journal} {\bibinfo  {journal} {Phys. Rev. Lett.}\ }\textbf {\bibinfo
  {volume} {79}},\ \bibinfo {pages} {5286} (\bibinfo {year}
  {1997})}\BibitemShut {NoStop}%
\bibitem [{\citenamefont {Scarola}\ \emph {et~al.}(2000)\citenamefont
  {Scarola}, \citenamefont {Park},\ and\ \citenamefont {Jain}}]{Scarola00}%
  \BibitemOpen
  \bibfield  {author} {\bibinfo {author} {\bibfnamefont {V.~W.}\ \bibnamefont
  {Scarola}}, \bibinfo {author} {\bibfnamefont {K.}~\bibnamefont {Park}}, \
  and\ \bibinfo {author} {\bibfnamefont {J.~K.}\ \bibnamefont {Jain}},\ }\href
  {\doibase 10.1103/PhysRevB.61.13064} {\bibfield  {journal} {\bibinfo
  {journal} {Phys. Rev. B}\ }\textbf {\bibinfo {volume} {61}},\ \bibinfo
  {pages} {13064} (\bibinfo {year} {2000})}\BibitemShut {NoStop}%
\bibitem [{\citenamefont {Zhang}\ \emph {et~al.}(2016)\citenamefont {Zhang},
  \citenamefont {W\'ojs},\ and\ \citenamefont {Jain}}]{Zhang16}%
  \BibitemOpen
  \bibfield  {author} {\bibinfo {author} {\bibfnamefont {Y.}~\bibnamefont
  {Zhang}}, \bibinfo {author} {\bibfnamefont {A.}~\bibnamefont {W\'ojs}}, \
  and\ \bibinfo {author} {\bibfnamefont {J.~K.}\ \bibnamefont {Jain}},\ }\href
  {\doibase 10.1103/PhysRevLett.117.116803} {\bibfield  {journal} {\bibinfo
  {journal} {Phys. Rev. Lett.}\ }\textbf {\bibinfo {volume} {117}},\ \bibinfo
  {pages} {116803} (\bibinfo {year} {2016})}\BibitemShut {NoStop}%
\bibitem [{\citenamefont {Eisenstein}\ \emph {et~al.}(1989)\citenamefont
  {Eisenstein}, \citenamefont {Stormer}, \citenamefont {Pfeiffer},\ and\
  \citenamefont {West}}]{Eisenstein89}%
  \BibitemOpen
  \bibfield  {author} {\bibinfo {author} {\bibfnamefont {J.~P.}\ \bibnamefont
  {Eisenstein}}, \bibinfo {author} {\bibfnamefont {H.~L.}\ \bibnamefont
  {Stormer}}, \bibinfo {author} {\bibfnamefont {L.}~\bibnamefont {Pfeiffer}}, \
  and\ \bibinfo {author} {\bibfnamefont {K.~W.}\ \bibnamefont {West}},\ }\href
  {\doibase 10.1103/PhysRevLett.62.1540} {\bibfield  {journal} {\bibinfo
  {journal} {Phys. Rev. Lett.}\ }\textbf {\bibinfo {volume} {62}},\ \bibinfo
  {pages} {1540} (\bibinfo {year} {1989})}\BibitemShut {NoStop}%
\bibitem [{\citenamefont {Eisenstein}\ \emph {et~al.}(1990)\citenamefont
  {Eisenstein}, \citenamefont {Stormer}, \citenamefont {Pfeiffer},\ and\
  \citenamefont {West}}]{Eisenstein90}%
  \BibitemOpen
  \bibfield  {author} {\bibinfo {author} {\bibfnamefont {J.~P.}\ \bibnamefont
  {Eisenstein}}, \bibinfo {author} {\bibfnamefont {H.~L.}\ \bibnamefont
  {Stormer}}, \bibinfo {author} {\bibfnamefont {L.~N.}\ \bibnamefont
  {Pfeiffer}}, \ and\ \bibinfo {author} {\bibfnamefont {K.~W.}\ \bibnamefont
  {West}},\ }\href {\doibase 10.1103/PhysRevB.41.7910} {\bibfield  {journal}
  {\bibinfo  {journal} {Phys. Rev. B}\ }\textbf {\bibinfo {volume} {41}},\
  \bibinfo {pages} {7910} (\bibinfo {year} {1990})}\BibitemShut {NoStop}%
\bibitem [{\citenamefont {Engel}\ \emph {et~al.}(1992)\citenamefont {Engel},
  \citenamefont {Hwang}, \citenamefont {Sajoto}, \citenamefont {Tsui},\ and\
  \citenamefont {Shayegan}}]{Engel92}%
  \BibitemOpen
  \bibfield  {author} {\bibinfo {author} {\bibfnamefont {L.~W.}\ \bibnamefont
  {Engel}}, \bibinfo {author} {\bibfnamefont {S.~W.}\ \bibnamefont {Hwang}},
  \bibinfo {author} {\bibfnamefont {T.}~\bibnamefont {Sajoto}}, \bibinfo
  {author} {\bibfnamefont {D.~C.}\ \bibnamefont {Tsui}}, \ and\ \bibinfo
  {author} {\bibfnamefont {M.}~\bibnamefont {Shayegan}},\ }\href {\doibase
  10.1103/PhysRevB.45.3418} {\bibfield  {journal} {\bibinfo  {journal} {Phys.
  Rev. B}\ }\textbf {\bibinfo {volume} {45}},\ \bibinfo {pages} {3418}
  (\bibinfo {year} {1992})}\BibitemShut {NoStop}%
\bibitem [{\citenamefont {Du}\ \emph {et~al.}(1995)\citenamefont {Du},
  \citenamefont {Yeh}, \citenamefont {Stormer}, \citenamefont {Tsui},
  \citenamefont {Pfeiffer},\ and\ \citenamefont {West}}]{Du95}%
  \BibitemOpen
  \bibfield  {author} {\bibinfo {author} {\bibfnamefont {R.~R.}\ \bibnamefont
  {Du}}, \bibinfo {author} {\bibfnamefont {A.~S.}\ \bibnamefont {Yeh}},
  \bibinfo {author} {\bibfnamefont {H.~L.}\ \bibnamefont {Stormer}}, \bibinfo
  {author} {\bibfnamefont {D.~C.}\ \bibnamefont {Tsui}}, \bibinfo {author}
  {\bibfnamefont {L.~N.}\ \bibnamefont {Pfeiffer}}, \ and\ \bibinfo {author}
  {\bibfnamefont {K.~W.}\ \bibnamefont {West}},\ }\href {\doibase
  10.1103/PhysRevLett.75.3926} {\bibfield  {journal} {\bibinfo  {journal}
  {Phys. Rev. Lett.}\ }\textbf {\bibinfo {volume} {75}},\ \bibinfo {pages}
  {3926} (\bibinfo {year} {1995})}\BibitemShut {NoStop}%
\bibitem [{\citenamefont {Kang}\ \emph {et~al.}(1997)\citenamefont {Kang},
  \citenamefont {Young}, \citenamefont {Hannahs}, \citenamefont {Palm},
  \citenamefont {Campman},\ and\ \citenamefont {Gossard}}]{Kang97}%
  \BibitemOpen
  \bibfield  {author} {\bibinfo {author} {\bibfnamefont {W.}~\bibnamefont
  {Kang}}, \bibinfo {author} {\bibfnamefont {J.~B.}\ \bibnamefont {Young}},
  \bibinfo {author} {\bibfnamefont {S.~T.}\ \bibnamefont {Hannahs}}, \bibinfo
  {author} {\bibfnamefont {E.}~\bibnamefont {Palm}}, \bibinfo {author}
  {\bibfnamefont {K.~L.}\ \bibnamefont {Campman}}, \ and\ \bibinfo {author}
  {\bibfnamefont {A.~C.}\ \bibnamefont {Gossard}},\ }\href {\doibase
  10.1103/PhysRevB.56.R12776} {\bibfield  {journal} {\bibinfo  {journal} {Phys.
  Rev. B}\ }\textbf {\bibinfo {volume} {56}},\ \bibinfo {pages} {R12776}
  (\bibinfo {year} {1997})}\BibitemShut {NoStop}%
\bibitem [{\citenamefont {Kukushkin}\ \emph {et~al.}(1999)\citenamefont
  {Kukushkin}, \citenamefont {v.~Klitzing},\ and\ \citenamefont
  {Eberl}}]{Kukushkin99}%
  \BibitemOpen
  \bibfield  {author} {\bibinfo {author} {\bibfnamefont {I.~V.}\ \bibnamefont
  {Kukushkin}}, \bibinfo {author} {\bibfnamefont {K.}~\bibnamefont
  {v.~Klitzing}}, \ and\ \bibinfo {author} {\bibfnamefont {K.}~\bibnamefont
  {Eberl}},\ }\href {\doibase 10.1103/PhysRevLett.82.3665} {\bibfield
  {journal} {\bibinfo  {journal} {Phys. Rev. Lett.}\ }\textbf {\bibinfo
  {volume} {82}},\ \bibinfo {pages} {3665} (\bibinfo {year}
  {1999})}\BibitemShut {NoStop}%
\bibitem [{\citenamefont {Yeh}\ \emph {et~al.}(1999)\citenamefont {Yeh},
  \citenamefont {Stormer}, \citenamefont {Tsui}, \citenamefont {Pfeiffer},
  \citenamefont {Baldwin},\ and\ \citenamefont {West}}]{Yeh99}%
  \BibitemOpen
  \bibfield  {author} {\bibinfo {author} {\bibfnamefont {A.~S.}\ \bibnamefont
  {Yeh}}, \bibinfo {author} {\bibfnamefont {H.~L.}\ \bibnamefont {Stormer}},
  \bibinfo {author} {\bibfnamefont {D.~C.}\ \bibnamefont {Tsui}}, \bibinfo
  {author} {\bibfnamefont {L.~N.}\ \bibnamefont {Pfeiffer}}, \bibinfo {author}
  {\bibfnamefont {K.~W.}\ \bibnamefont {Baldwin}}, \ and\ \bibinfo {author}
  {\bibfnamefont {K.~W.}\ \bibnamefont {West}},\ }\href {\doibase
  10.1103/PhysRevLett.82.592} {\bibfield  {journal} {\bibinfo  {journal} {Phys.
  Rev. Lett.}\ }\textbf {\bibinfo {volume} {82}},\ \bibinfo {pages} {592}
  (\bibinfo {year} {1999})}\BibitemShut {NoStop}%
\bibitem [{\citenamefont {Kukushkin}\ \emph {et~al.}(2000)\citenamefont
  {Kukushkin}, \citenamefont {Smet}, \citenamefont {von Klitzing},\ and\
  \citenamefont {Eberl}}]{Kukushkin00}%
  \BibitemOpen
  \bibfield  {author} {\bibinfo {author} {\bibfnamefont {I.~V.}\ \bibnamefont
  {Kukushkin}}, \bibinfo {author} {\bibfnamefont {J.~H.}\ \bibnamefont {Smet}},
  \bibinfo {author} {\bibfnamefont {K.}~\bibnamefont {von Klitzing}}, \ and\
  \bibinfo {author} {\bibfnamefont {K.}~\bibnamefont {Eberl}},\ }\href
  {\doibase 10.1103/PhysRevLett.85.3688} {\bibfield  {journal} {\bibinfo
  {journal} {Phys. Rev. Lett.}\ }\textbf {\bibinfo {volume} {85}},\ \bibinfo
  {pages} {3688} (\bibinfo {year} {2000})}\BibitemShut {NoStop}%
\bibitem [{\citenamefont {Melinte}\ \emph {et~al.}(2000)\citenamefont
  {Melinte}, \citenamefont {Freytag}, \citenamefont {Horvatic}, \citenamefont
  {Berthier}, \citenamefont {L\'evy}, \citenamefont {Bayot},\ and\
  \citenamefont {Shayegan}}]{Melinte00}%
  \BibitemOpen
  \bibfield  {author} {\bibinfo {author} {\bibfnamefont {S.}~\bibnamefont
  {Melinte}}, \bibinfo {author} {\bibfnamefont {N.}~\bibnamefont {Freytag}},
  \bibinfo {author} {\bibfnamefont {M.}~\bibnamefont {Horvatic}}, \bibinfo
  {author} {\bibfnamefont {C.}~\bibnamefont {Berthier}}, \bibinfo {author}
  {\bibfnamefont {L.~P.}\ \bibnamefont {L\'evy}}, \bibinfo {author}
  {\bibfnamefont {V.}~\bibnamefont {Bayot}}, \ and\ \bibinfo {author}
  {\bibfnamefont {M.}~\bibnamefont {Shayegan}},\ }\href {\doibase
  10.1103/PhysRevLett.84.354} {\bibfield  {journal} {\bibinfo  {journal} {Phys.
  Rev. Lett.}\ }\textbf {\bibinfo {volume} {84}},\ \bibinfo {pages} {354}
  (\bibinfo {year} {2000})}\BibitemShut {NoStop}%
\bibitem [{\citenamefont {Freytag}\ \emph {et~al.}(2001)\citenamefont
  {Freytag}, \citenamefont {Tokunaga}, \citenamefont {Horvati\'{c}},
  \citenamefont {Berthier}, \citenamefont {Shayegan},\ and\ \citenamefont
  {L\'evy}}]{Freytag01}%
  \BibitemOpen
  \bibfield  {author} {\bibinfo {author} {\bibfnamefont {N.}~\bibnamefont
  {Freytag}}, \bibinfo {author} {\bibfnamefont {Y.}~\bibnamefont {Tokunaga}},
  \bibinfo {author} {\bibfnamefont {M.}~\bibnamefont {Horvati\'{c}}}, \bibinfo
  {author} {\bibfnamefont {C.}~\bibnamefont {Berthier}}, \bibinfo {author}
  {\bibfnamefont {M.}~\bibnamefont {Shayegan}}, \ and\ \bibinfo {author}
  {\bibfnamefont {L.~P.}\ \bibnamefont {L\'evy}},\ }\href {\doibase
  10.1103/PhysRevLett.87.136801} {\bibfield  {journal} {\bibinfo  {journal}
  {Phys. Rev. Lett.}\ }\textbf {\bibinfo {volume} {87}},\ \bibinfo {pages}
  {136801} (\bibinfo {year} {2001})}\BibitemShut {NoStop}%
\bibitem [{\citenamefont {Freytag}\ \emph {et~al.}(2002)\citenamefont
  {Freytag}, \citenamefont {Horvatic}, \citenamefont {Berthier}, \citenamefont
  {Shayegan},\ and\ \citenamefont {L\'evy}}]{Freytag02}%
  \BibitemOpen
  \bibfield  {author} {\bibinfo {author} {\bibfnamefont {N.}~\bibnamefont
  {Freytag}}, \bibinfo {author} {\bibfnamefont {M.}~\bibnamefont {Horvatic}},
  \bibinfo {author} {\bibfnamefont {C.}~\bibnamefont {Berthier}}, \bibinfo
  {author} {\bibfnamefont {M.}~\bibnamefont {Shayegan}}, \ and\ \bibinfo
  {author} {\bibfnamefont {L.~P.}\ \bibnamefont {L\'evy}},\ }\href {\doibase
  10.1103/PhysRevLett.89.246804} {\bibfield  {journal} {\bibinfo  {journal}
  {Phys. Rev. Lett.}\ }\textbf {\bibinfo {volume} {89}},\ \bibinfo {pages}
  {246804} (\bibinfo {year} {2002})}\BibitemShut {NoStop}%
\bibitem [{\citenamefont {Tracy}\ \emph {et~al.}(2007)\citenamefont {Tracy},
  \citenamefont {Eisenstein}, \citenamefont {Pfeiffer},\ and\ \citenamefont
  {West}}]{Tracy07}%
  \BibitemOpen
  \bibfield  {author} {\bibinfo {author} {\bibfnamefont {L.~A.}\ \bibnamefont
  {Tracy}}, \bibinfo {author} {\bibfnamefont {J.~P.}\ \bibnamefont
  {Eisenstein}}, \bibinfo {author} {\bibfnamefont {L.~N.}\ \bibnamefont
  {Pfeiffer}}, \ and\ \bibinfo {author} {\bibfnamefont {K.~W.}\ \bibnamefont
  {West}},\ }\href {\doibase 10.1103/PhysRevLett.98.086801} {\bibfield
  {journal} {\bibinfo  {journal} {Phys. Rev. Lett.}\ }\textbf {\bibinfo
  {volume} {98}},\ \bibinfo {pages} {086801} (\bibinfo {year}
  {2007})}\BibitemShut {NoStop}%
\bibitem [{\citenamefont {Tiemann}\ \emph {et~al.}(2012)\citenamefont
  {Tiemann}, \citenamefont {Gamez}, \citenamefont {Kumada},\ and\ \citenamefont
  {Muraki}}]{Tiemann12}%
  \BibitemOpen
  \bibfield  {author} {\bibinfo {author} {\bibfnamefont {L.}~\bibnamefont
  {Tiemann}}, \bibinfo {author} {\bibfnamefont {G.}~\bibnamefont {Gamez}},
  \bibinfo {author} {\bibfnamefont {N.}~\bibnamefont {Kumada}}, \ and\ \bibinfo
  {author} {\bibfnamefont {K.}~\bibnamefont {Muraki}},\ }\href {\doibase
  10.1126/science.1216697} {\bibfield  {journal} {\bibinfo  {journal}
  {Science}\ }\textbf {\bibinfo {volume} {335}},\ \bibinfo {pages} {828}
  (\bibinfo {year} {2012})},\ \Eprint
  {http://arxiv.org/abs/http://www.sciencemag.org/content/335/6070/828.full.pdf}
  {http://www.sciencemag.org/content/335/6070/828.full.pdf} \BibitemShut
  {NoStop}%
\bibitem [{\citenamefont {Feldman}\ \emph {et~al.}(2013)\citenamefont
  {Feldman}, \citenamefont {Levin}, \citenamefont {Krauss}, \citenamefont
  {Abanin}, \citenamefont {Halperin}, \citenamefont {Smet},\ and\ \citenamefont
  {Yacoby}}]{Feldman13}%
  \BibitemOpen
  \bibfield  {author} {\bibinfo {author} {\bibfnamefont {B.~E.}\ \bibnamefont
  {Feldman}}, \bibinfo {author} {\bibfnamefont {A.~J.}\ \bibnamefont {Levin}},
  \bibinfo {author} {\bibfnamefont {B.}~\bibnamefont {Krauss}}, \bibinfo
  {author} {\bibfnamefont {D.~A.}\ \bibnamefont {Abanin}}, \bibinfo {author}
  {\bibfnamefont {B.~I.}\ \bibnamefont {Halperin}}, \bibinfo {author}
  {\bibfnamefont {J.~H.}\ \bibnamefont {Smet}}, \ and\ \bibinfo {author}
  {\bibfnamefont {A.}~\bibnamefont {Yacoby}},\ }\href {\doibase
  10.1103/PhysRevLett.111.076802} {\bibfield  {journal} {\bibinfo  {journal}
  {Phys. Rev. Lett.}\ }\textbf {\bibinfo {volume} {111}},\ \bibinfo {pages}
  {076802} (\bibinfo {year} {2013})}\BibitemShut {NoStop}%
\bibitem [{\citenamefont {Liu}\ \emph {et~al.}(2014)\citenamefont {Liu},
  \citenamefont {Hasdemir}, \citenamefont {W\'ojs}, \citenamefont {Jain},
  \citenamefont {Pfeiffer}, \citenamefont {West}, \citenamefont {Baldwin},\
  and\ \citenamefont {Shayegan}}]{Liu14}%
  \BibitemOpen
  \bibfield  {author} {\bibinfo {author} {\bibfnamefont {Y.}~\bibnamefont
  {Liu}}, \bibinfo {author} {\bibfnamefont {S.}~\bibnamefont {Hasdemir}},
  \bibinfo {author} {\bibfnamefont {A.}~\bibnamefont {W\'ojs}}, \bibinfo
  {author} {\bibfnamefont {J.~K.}\ \bibnamefont {Jain}}, \bibinfo {author}
  {\bibfnamefont {L.~N.}\ \bibnamefont {Pfeiffer}}, \bibinfo {author}
  {\bibfnamefont {K.~W.}\ \bibnamefont {West}}, \bibinfo {author}
  {\bibfnamefont {K.~W.}\ \bibnamefont {Baldwin}}, \ and\ \bibinfo {author}
  {\bibfnamefont {M.}~\bibnamefont {Shayegan}},\ }\href {\doibase
  10.1103/PhysRevB.90.085301} {\bibfield  {journal} {\bibinfo  {journal} {Phys.
  Rev. B}\ }\textbf {\bibinfo {volume} {90}},\ \bibinfo {pages} {085301}
  (\bibinfo {year} {2014})}\BibitemShut {NoStop}%
\bibitem [{\citenamefont {Balram}\ \emph {et~al.}(2015)\citenamefont {Balram},
  \citenamefont {T\"oke}, \citenamefont {W\'ojs},\ and\ \citenamefont
  {Jain}}]{Balram15a}%
  \BibitemOpen
  \bibfield  {author} {\bibinfo {author} {\bibfnamefont {A.~C.}\ \bibnamefont
  {Balram}}, \bibinfo {author} {\bibfnamefont {C.}~\bibnamefont {T\"oke}},
  \bibinfo {author} {\bibfnamefont {A.}~\bibnamefont {W\'ojs}}, \ and\ \bibinfo
  {author} {\bibfnamefont {J.~K.}\ \bibnamefont {Jain}},\ }\href {\doibase
  10.1103/PhysRevB.92.075410} {\bibfield  {journal} {\bibinfo  {journal} {Phys.
  Rev. B}\ }\textbf {\bibinfo {volume} {92}},\ \bibinfo {pages} {075410}
  (\bibinfo {year} {2015})}\BibitemShut {NoStop}%
\bibitem [{\citenamefont {Ortiz}\ \emph {et~al.}(1993)\citenamefont {Ortiz},
  \citenamefont {Ceperley},\ and\ \citenamefont {Martin}}]{Ortiz93}%
  \BibitemOpen
  \bibfield  {author} {\bibinfo {author} {\bibfnamefont {G.}~\bibnamefont
  {Ortiz}}, \bibinfo {author} {\bibfnamefont {D.~M.}\ \bibnamefont {Ceperley}},
  \ and\ \bibinfo {author} {\bibfnamefont {R.~M.}\ \bibnamefont {Martin}},\
  }\href {\doibase 10.1103/PhysRevLett.71.2777} {\bibfield  {journal} {\bibinfo
   {journal} {Phys. Rev. Lett.}\ }\textbf {\bibinfo {volume} {71}},\ \bibinfo
  {pages} {2777} (\bibinfo {year} {1993})}\BibitemShut {NoStop}%
\end{thebibliography}%

\end{document}